\documentclass[aps,twocolumn,prd,longbibliography,nofootinbib,showpacs,preprintnumbers,superscriptaddress]{revtex4-2}

\usepackage{tikz-cd}

\usepackage{psfrag}
\usepackage{mathrsfs}

\usepackage{amsfonts}
\usepackage{amsmath} 
\usepackage{amsthm}
\usepackage{amssymb} 

\usepackage{xparse} 
\usepackage{empheq}		
\usepackage{cancel}     

\usepackage[toc,page,titletoc]{appendix}

\usepackage{bbm} 

\usepackage{commath}

\usepackage{graphicx}
\usepackage{epsfig}
\usepackage{float}
\usepackage{color,array,subfigure}
\usepackage{epstopdf}
\usepackage{esint}

\usepackage{geometry}
\usepackage{mathdesign}

\usepackage[utf8]{inputenc}

\RequirePackage{CJKutf8}

\usepackage[colorlinks=true,  
            citecolor=red,
            urlcolor= blue,                     %
            filecolor=black,
            linktocpage=true,  
            linkcolor=blue,
             ]{hyperref}

\usepackage{dcolumn}
\usepackage{bm}

\newcommand{\beq}{\begin{equation}}
\newcommand{\eeq}{\end{equation}}

\newcommand{\ba}{\begin{array}}
\newcommand{\ea}{\end{array}}

\newcommand{\beqa}{\begin{eqnarray}}
\newcommand{\eeqa}{\end{eqnarray}}
\newcommand{\bit}{\begin{itemize}}
\newcommand{\eit}{\end{itemize}}

\newcommand{\bsp}{\begin{split}}
\newcommand{\esp}{\end{split}}
\newcommand{\bpm}{\begin{pmatrix}}
\newcommand{\epm}{\end{pmatrix}}
\newcommand{\bbm}{\begin{bmatrix}}
\newcommand{\ebm}{\end{bmatrix}}
\newcommand{\bBm}{\begin{Bmatrix}}
\newcommand{\eBm}{\end{Bmatrix}}
\newcommand{\bvm}{\begin{vmatrix}}
\newcommand{\evm}{\end{vmatrix}}
\newcommand{\bVm}{\begin{Vmatrix}}
\newcommand{\eVm}{\end{Vmatrix}}

\newcommand{\bsel}{{\begin{subequations}\begin{empheq}}}
\newcommand{\bse}{{\begin{subequations}\begin{empheq}[left={\ii}\empheqlbrace]{align}}}
\newcommand{\ese}{{\end{empheq}\end{subequations}}}
\def\bc{\begin{center}}
\def\ec{\end{center}}
\def\bnum{\begin{enumerate} }
\def\enum{\end{enumerate}}
\def\nn{\nonumber}
\def\ii{\!\!\!\!\!\!}  
\def\3i{\!\!\!}
\def\2i{\!\!}

\def\ea{{e_a}}

\def\ec{{e_c}}

\def\log{\ln}

\def\nn{\nonumber}

\def\({\left(}
\def\){\right)}

\def\abs#1{\left|#1\right|}
\def\vev#1{\left\langle#1\right\rangle}
\def\>{\rightarrow}

\def\Diracslash#1{\not{\hbox{\kern-4pt $#1$}}}

\def\Dslash{\not{\hbox{\kern-4pt $D$}}}
\def\pslash{\not{\hbox{\kern-4pt $p$}}}
\def\qslash{\not{\hbox{\kern-4pt $q$}}}

\def\lv{\not{\hbox{\kern-4pt $L$}}}
\def\lsim{\mathrel{\raise.3ex\hbox{$<$\kern-.75em\lower1ex\hbox{$\sim$}}}}
\def\gsim{\mathrel{\raise.3ex\hbox{$>$\kern-.75em\lower1ex\hbox{$\sim$}}}}
\def\ifmath#1{\relax\ifmmode #1\else $#1$\fi}

\evensidemargin=\oddsidemargin

\usepackage[normalem]{ulem}  

\renewcommand\sout{\bgroup \color{red} \ULdepth=-.5ex \ULset}

\usepackage[capitalise,nameinlink]{cleveref}

\crefname{supp}{Supplement}{Supplements}

\makeatletter
\renewcommand{\appendix}{\par
  \setcounter{section}{0}%
  }
  \makeatother

\crefname{paragraph}{paragraph}{paragraphs}
\Crefname{paragraph}{Paragraph}{Paragraphs}

\begin{document}

\begin{CJK}{UTF8}{gbsn}

\title{Superradiance in the Kerr-Taub-NUT spacetime}

\author{Bum-Hoon Lee}
\email{bhl@sogang.ac.kr} 

\affiliation{
Center for Quantum Spacetime, Sogang University, Seoul 04107, Korea
}

\affiliation{
Department of Physics, Sogang University, Seoul 04107, Korea
}

\author{Wonwoo Lee}
\email{warrior@sogang.ac.kr} 

\affiliation{
Center for Quantum Spacetime, Sogang University, Seoul 04107, Korea
}

\author{Yong-Hui~Qi}
\email{yhqipku@gmail.com}

\affiliation{
Center for Quantum Spacetime, Sogang University, Seoul 04107, Korea
}

\affiliation{
School of Physics and Astronomy, Sun Yat-sen University, Zhuhai 519082, China
}

\begin{abstract}
Superradiance is the effect of field waves being amplified during reflection from a charged or rotating black hole.
In this paper, we study the low-energy dynamics of super-radiant scattering of massive scalar and massless higher spin field perturbations in a generic axisymmetric stationary Ker-Taub-NUT (Newman-Unti-Tamburino) spacetime, which represents sources with both gravitomagnetic monopole moment (magnetic mass) and gravitomagnetic dipole moment (angular momentum).
We obtain a generalized Teukolsky master equation for all spin perturbation fields.
The equations are separated into their angular and radial parts.
The angular equations lead to spin-weighted spheroidal harmonic functions that generalize those in Kerr spacetime. We identify an effective spin as a coupling between frequency (or energy) and the NUT parameter.
%
%
The behaviors of the radial wave function near the horizon and at the infinite boundary are studied.
We provide analytical expressions for low-energy observables such as emission rates and cross sections of all massless fields with spin, including scalar, neutrino, electromagnetic, Rarita-Schwinger, and gravitational waves.
\end{abstract}

\maketitle


\tableofcontents

\section{Introduction}

The origins of energy in the universe, such as the quasi-stellar radio source, are intriguing.
The violent release of enormous amounts of energy might result from the gravitational collapse. A natural consequence is the formation of compact objects without spherical symmetry, such as neutron stars or black hole spacetime with physical singularities~\cite{Penrose:1964wq}.
The Penrose process corresponds to a radiation enhancement process that permits continuously extracting the Coulomb energy or rotational energy of a charged or rotating black hole, through its absorption of particles with negative energies or angular momentum~\cite{Penrose:1969pc,Penrose:1971uk}.
Superradiance is a field theory generalization of the Penrose process that takes into account the intrinsic properties of the point particles, such as spins. It is an enhanced radiating effect with monochromatic amplification of scattering waves~\cite{Zel'Dovich:1971,Misner:1972kx,Teukolsky:1972my,Bardeen:1972fi,Brill:1972xj,Fackerell:1972hg,Starobinskii:1973hgd}.
The superradiant scattering of spin particles in the Kerr spacetime has been studied. They include electromagnetic and gravitational waves~\cite{Zel'Dovich:1972,Starobinskil:1974nkd,Cohen:1974cm,Chrzanowski:1975wv,Chrzanowski:1976jy,Kegeles:1979an}, electrons and neutrino waves~\cite{Unruh:1974bw,Chandrasekhar:1976ap,Page:1976jj,Gueven:1977dq,Martellini:1977qf,Chandrasekhar:1985kt}, the Rarita-Schwinger field~\cite{Gueven:1980be,Kamran:1985tp,TorresdelCastillo1989}, and the gravitational waves~\cite{Ryan:1974nt,Chrzanowski:1974nr,Chrzanowski:1976jb,Matzner:1977dn,Handler:1980un,Futterman:1988ni}.
There is no superadiance for the Dirac and Rarita-Schwinger fields.
The super-radiant scattering of scalar waves in the Kerr spacetime is described by a relativistic Klein-Gordon equation, which turns out to be separable~\cite{Detweiler:1980uk,Detweiler:1983zz}. 
%
%
Recently, the superradiance instability of the Kerr and Kerr-Newman black holes has been revisited~\cite{Guo:2021xao,Myung:2022biw}.


The perturbation equation for all spin particles can also be derived in the Newman-Penrose formalism~\cite{Newman:1961qr,Newman:1965ik,Frolov1979,Penrose:1985bww}.
The electromagnetic and gravitational perturbation equations in the Kerr-Newman spacetime are not decoupled~\cite{Lee1976,Chitre:1976bb}.
%
When derive the perturbation equation for all spin particles in Kerr spacetime in terms of the Teukolsky equation~\cite{Teukolsky:1973ha,Press:1973zz,Teukolsky:1974yv}, it is common to adopt the Kinnersley tetrad~\cite{Kinnersley1969}.
The real null tetrad lies in the two repeated principal null directions. Thus, the Kerr spacetime belongs to Petrov type D~\cite{Petrov:2000bs,Pirani1965}, according to the Goldberg-Sachs theorem~\cite{Goldberg2009}.
The Teukolsky equation describes the dynamics of superradiance for spin particles in rotating spacetime.
The equation turns out to be separable in the Kerr-Taub-NUT spacetime with a symmetric Misner string~\cite{Bini:2003sy}.
The solutions to the equations in the Kerr spacetime turn out to be exactly solvable and can be expressed in terms of confluent Heun equations~\cite{Fiziev:2009ud,Cook:2014cta}.
The solution to the angular equation can be expressed in terms of spin-weighted spheroidal harmonics~\cite{Goldberg1967,Breuer:1977,Li1998,Falloon2003,Casals:2004zq,Berti:2005gp}.
The radial equation can be recasted in Schr\"{o}dinger equation form with an effective potential, named the Regge-Wheeler-Zerilli equation ~\cite{Regge:1957td,Vishveshwara:1970cc,Zerilli:1970wzz,Vishveshwara:1970cc}.
The perturbation equation due to source can be constructed~\cite{Merlin:2016boc}.
%
Higher-order Teukolsky perturbation equations in the Kerr spacetime have been studied~\cite{Loutrel:2020wbw,Ripley:2020xby,Cano:2023tmv}.

%
The equilibrium state of the superradiance process can be understood through black hole thermodynamics~\cite{Bekenstein:1973mi,Bekenstein:1973ur,Hawking:1976de,Bekenstein:1998nt}.
Combining with quantum mechanics methods, it is found that a black hole can create and emit particles, as if it were a macroscopic thermal object with temperature. The radiation would carry away energy, and the mass of the black hole would decrease, thus causing an evaporation process~\cite{Hawking:1975vcx,Page:1976df,Damour:1976jd}.
The process can be understood by considering the vacuum fluctuations outside of the horizon, which create virtual photon pairs, and the tidal forces pulling them apart and converting them into real ones. If the tidal forces get strong enough, even massive virtual particle pairs such as electrons and positions, etc., can also be pulled apart. 
The analytical properties of Hawking radiation can be described by transition coefficients and energy flux~\cite{Koga:1995bs}.
The low-energy dynamics of the rotating spacetime can be probed and inferred by emission rates, including super-radiant emission and graybody factors~\cite{Maldacena:1996ix,Maldacena:1997ih}.
The separability of the equations for radiative higher-spin fields is obvious in spherical, symmetric spacetime~\cite{Gubser:1997cm,Cvetic:1997ap,Kanti:2002ge}.
Recently, Teukolsky equations for all spin fields in symmetric spacetime were obtained~\cite{Arbey:2021jif}.

The Superradiance of the massless field in different rotating spacetimes, such as dilaton gravity, high-dimensional gravity, brane gravity and gauged supergravity, has also been studied~\cite{Kanti:2002nr,Ida:2002ez,Sakalli:2016fif}. 
The conformal symmetry of the scalar wave function at low frequencies is emergent in the extremal limit of the Kerr black hole and has raised interest in the conjecture called Kerr/CFT correspondence~\cite{Bredberg:2009pv,Castro:2010fd}.
If a light boson, e.g,. an axion, exists with proper mass, gravitational bound states are formed around rotating charged black holes. These bound states could continuously extract electromagnetic or rotational energy from black holes~\cite{Arvanitaki:2010sy}.
Moreover, black hole evaporation can be explored by considering a massive charged scalar field in Kerr-Newman spacetime~\cite{Hod:2014baa,Benone:2014ssa}.
In particular, motivated by that scalar as a dark matter candidate around the galaxy, the superradiance phenomenon of a scalar boson field in a Kerr black hole has been studied recently~\cite{Benone:2019all,Hui:2019aqm,Hui:2022sri}.

%
The angular momentum per unit mass can be viewed as the gravitomagnetic dipole moment of spacetime; it would be interesting to also consider the gravitomagnetic monopole of spacetime, namely, the Taub-NUT spacetime~\cite{Taub:1950ez,Newman:1963yy}.
Compared to Schwarzschild spacetime, there is an additional NUT parameter. 
The NUT parameter has the interpretation of a gravitomagnetic (monopole) charge. It may also be interpreted as the twist~\cite{Nouri-Zonoz:1997mnd} or vorticity of holographic fluid~\cite{Leigh:2011au}.
There are two unique properties of the Taub-NUT metric. 
One is that spacetime is intrinsically rotating due to the magnetic mass source, namely, the NUT parameter.
The other is that it is not asymptotically flat at infinity in the large distance limit.
The first property can be understood by noticing a non-vanishing $g_{t \phi }$ component of the metric. It's also worth noticing that the component is not asymptotically flat at infinity. This represents a topological line defect, i.e., a wire singularity of spacetime, in terms of ``Misner string'', which is symmetrically distributed along the polar axis~\cite{Misner:1965zz,Miller:1971em,Misner-Taub:1969}. There are conical singularities on its axis of symmetry that can be removed by imposing an appropriate periodic condition on the time coordinate. It results in the Misner condition, the gravitomagnetic analog of Dirac's string quantization conditions~\cite{Misner:1963fr,Misner-Taub:1969}. The periodic condition implies a finite temperature~\cite{Chamblin:1998pz,Emparan:1999pm,Hawking:1973uf}.
The position of the Misner string as the singularity on the axis of the respective Taub-NUT spacetime can be described by the Manko-Ruiz (MR) parameter~\cite{Manko_2005}, which is present in the non-diagonal elements of the metric $g_{t\phi}$. 
The metric is related to closed time-like lines~\cite{Misner-Taub:1969}, time machines, and wormholes~\cite{Clement:2015aka}. 

%
In this paper, we study the super-radiant scattering of all spin fields in the Kerr-Taub-NUT spacetime~\cite{Newman:1965tw,Newman:1965my}. The spacetime is a local analytic axial-symmetric stationary solution of the vacuum Einstein-Maxwell equations~\cite{Miller1973}. It represents a source with a mass $M$, a NUT parameter $n$, and an angular momentum per unit mass $a$.
The metric is the nature generalization of the Kerr spacetime, in which $a$ can be interpreted as the gravitomagnetic dipole moment~\cite{Kerr:1963ud}. The NUT parameter $n$ has an interpretation of gravitomagnetic monopole moment or magnetic mass, which is dual to the gravitoelectric mass $M$~\cite{Lynden-Bell:1996dpw,Hawking:1998ct}.
The metric with electric charge $Q$ is the the most general stationary solution of the Einstein-Maxwell equation, in terms of the Ker-Newman-Taub-NUT spacetime~\cite{Demianski-Newman,Aliev:2007fy}.
It is known that given a spherical, symmetric metric, we can construct its rotating counterpart by using the extended Newman-Jain algorithm (ENJA) with a noncomplexification procedure~\cite{Drake:1998gf,Azreg-Ainou:2014pra,Azreg-Ainou:2014aqa}.
In this paper, we generalize the method from a spherical, symmetric spacetime to a generic axial, symmetric stationary spacetime.

The paper is organized as follows: In Sec.~\ref{sec:KTN}, we give the derivation of an axial symmetric stationary rotating metric. The method is generic, and we obtain the Kerr-Newman-Taub-NUT metric with the MR parameter.
In Sec.~\ref{sec:KG_cKTN}, we describe the superradiance of a charged massive scalar in the Kerr-Newman-Taub-NUT (KNTN) spacetime.
In Sec.~\ref{sec:TME_KTN}, we investigate the superradiance of massless spinning particles in a Kerr-Taub-NUT (KTN) spacetime.
In Sec.~\ref{sec:LEDO}, we present the low-energy dynamical observables for the superradiance of massless spin particles in the KTN spacetime.
We summarize the results in the last section~\ref{sec:sum}.
In this paper, we adopt the metric signature of Minkowski spacetime as $(-1,+1,+1,+1)$. 
Throughout this paper, we work in natural and geometric units such that the reduced Planck constant $\hbar$, the speed of light $c$ in vacuum, and Newton's gravitational constant $G$ equal unity.

\section{The Kerr-Taub-NUT spacetime}
\label{sec:KTN}

\subsection{General Axial-symmetric metric}

As a generalization of the spherical, symmetric spacetime~\cite{Azreg-Ainou:2014pra}, let us consider a general axial symmetric stationary spacetime with metric
\beqa
ds^2 = - f [ dt + W d\phi ]^2 + \frac{1}{g}dr^2 + h d\Omega_{2} , \label{ds2_generic}
\eeqa
where $f,g,h$, and $w$ are functions to be determined. For the stationary metric, all of them are time-independent. 
Due to the presence of $W(r,\theta)$, the metric is rotating along the polar axis. We denote $w=W(r,0)$, which is polar angle independent. If $w=0$, it reduces to a spherical, symmetric metric. $d\Omega_{2} \equiv d\theta^2 + \sin^2\theta d\phi^2 $ is the solid angle of co-dimension $2$ spatial space.
The metric of the solutions to the Einstein equations has to be asymptotically flat at spatial infinity. Thus, the functions $f,g$, and $h$ have to satisfy the asymptotic conditions
\beqa
f(r), g(r) \overset{r\to\infty}{=} 1, \quad h(r) \overset{r\to\infty}{=} r^2.
\eeqa
Note that $w(r)$ doesn't need to be asymptotic flat at infinity.
In the NP formalism, the inverse metric of Eq.~(\ref{ds2_generic}) becomes
\beqa
g^{\mu\nu} = - 2 l^{(\mu} n^{\nu)} + 2 m^{(\mu} \bar{m}^{\nu)} ,
\eeqa
where the null tetrads are
\beqa
\begin{split}
	l^\mu & = \left( 0, 1, 0 , 0 \right), \\
	n^\mu & = \left(  \sqrt\frac{g}{f} , - \frac{g}{2},  0 , 0 \right) , \\ 
	m^\mu & = \frac{1}{\sqrt{2h}}\left( -\frac{iw}{\sin\theta} , 0, 1, \frac{i}{\sin\theta} \right) . \\
\end{split} \label{tetrad_lnm}
\eeqa
We have chosen the real tetrad $l^\mu$ as the outgoing null vector tangent to the light cone, and $n^\mu$ is the ingoing null vector. They satisfy the orthonormal conditions in Eqs.~(\ref{norm_nlm}) and (\ref{orth_nlm}). 
Firstly, let's transform the metric in Boyer-Lindquist (BL) coordinates $(t,r,\theta,\phi)$ to the outgoing or ingoing Eddington-Finkelstein (EF) coordinates $(u/v,r,\theta,\phi)$ by imposing a retarded time $u=t-r_\star$ or advanced time $v=t+r_\star$, where $r_\star$ is the tortoise coordinates and the null coordinate transformations are
\beqa
 dt = du + \frac{dr}{\sqrt{f(r)g(r)}}, \quad dt = dv - \frac{dr}{\sqrt{f(r)g(r)}} . 
\eeqa
Then the metric transforms into a generic metric in the outgoing EF coordinates as
\beqa
\begin{split}
ds^2 
& = - f (du+wd\phi)^2 - 2 \sqrt\frac{f}{g} (du + w d\phi ) dr + h d\Omega_2 \\
& = - f (du+wd\phi)(dv+wd\phi) + h d\Omega_2 ,
\end{split}  \label{ds2_fgh}
\eeqa
where $u$ and $v$ denote the retarded and advanced time, respectively. They characterize the radial coordinate of the photon at a fixed instant time $t$. The coordinate lines of constants $u$ or $v$ represent outgoing or ingoing radial null rays~\cite{Finkelstein:1958zz}. The function $r$ is a function of $u$ and $v$, defined implicitly by the relation $r_\star(r) = (v-u)/2$. When $w=0$, the metric reduces to spherical, symmetric spacetime, e.g., the Schwarzschild metric.
Secondly, by performing a complex coordinate transformation on the $(u,r)$ or $(v,r)$ plane, we have
\beqa
\begin{split}
& (u, v, r) \to  (u-ia\cos\theta, v+ia\cos\theta, r + ia \cos\theta), \\ 
\end{split} \label{ur_vr_complex}
\eeqa
where $a$ is the rotational parameter. 
The complex transformation in Eq.~(\ref{ur_vr_complex}) transforms the vectors $e^\mu_a = ( \delta_t^\mu,\delta_r^\mu,\delta_\theta^\mu, \delta_\phi^\mu ) \to ( \delta_t^\mu,\delta_r^\mu, ia\sin\theta (\delta_t^\mu  - \delta_r^\mu), \delta_\phi^\mu )$ .
Under the transformation, the functions $\{f,g,h,w\}$ are also expected to be transformed to the other functions $\{F,G,H,W\}$, respectively. $f,g,h$ are functions of radial coordinates only, while $F,G,H$ will depend on not only $r$ but also the polar angle $\theta$.
The tetrad in the EF coordinates becomes
\beqa
\begin{split}
	l^\mu & = \left( 0, 1, 0 , 0 \right), \\
	n^\mu & = \left(  \sqrt\frac{G}{F} , - \frac{G}{2},  0 , 0 \right) ,  \\ 
	m^a & = \frac{1}{ \sqrt{2H}} \left( - \frac{i W}{\sin\theta} + ia \sin\theta,  -ia \sin\theta, 1, \frac{i}{\sin\theta} \right) , \\
\end{split} \label{tetrad_lnm_NUT}
\eeqa
where we assume that $H$ is a complex function and denote $ |H| \equiv \sqrt{H\bar{H}}$ as the amplitude of $H$. The new metric in the outgoing EF coordinates is
\beqa
\begin{split}
ds^2 & = - F du^2 - 2 \sqrt\frac{F}{G}dudr \\
& + 2 \bigg[ a\sin^2\theta \bigg( F - \sqrt\frac{F}{G}\bigg) - F W \bigg] du d\phi \\
& + 2 (a \sin^2\theta - W )\sqrt\frac{F}{G} dr d\phi \\
& + |H| d\theta^2 + \bigg[ \sin^2\theta  \bigg( |H| +  2\sqrt\frac{F}{G} ( a^2\sin\theta^2 - a W )  \bigg) \\
& - F (a\sin^2\theta- W)^2 \bigg] d\phi^2 .\\
\end{split} \label{ds2_FGH}
\eeqa
For non-rotating metric (i.e., $a=0$), the metric in Eq.~(\ref{ds2_FGH}) recovers Eq.~(\ref{ds2_fgh}).

In the situation where $W=0$, the metric reduces to
\beqa
\begin{split}
ds^2 & = - F du^2 - 2 \sqrt\frac{F}{G}dudr + 2  a\sin^2\theta \bigg( F - \sqrt\frac{F}{G}\bigg)   du d\phi \\
& + 2 a \sin^2\theta \sqrt\frac{F}{G} dr d\phi \\
& + |H| d\theta^2 + \sin^2\theta  \bigg[ |H| + a^2\sin\theta^2 \bigg( 2\sqrt\frac{F}{G} - F \bigg) \bigg] d\phi^2 .\\
\end{split}
\eeqa
This recovers the metric for generating the Kerr metric without the NUT parameter.
Finally, we need to bring the rotating metric in the EF coordinates back to the BL ones by using a global coordinate transformation~\cite{Boyer:1966qh}
\beqa
du \to dt + \lambda(r)dr, \quad d\phi \to d\phi + \chi(r) dr , \label{lambda-chi}
\eeqa
where $\lambda$ and $\chi$ depend on $r$ only to ensure integrability, so that we can integrate the two equations to obtain the global coordinates $u(t,r)$ and $\phi(\phi,r)$. By requiring that the mixing components of the metric terms $g_{tr}$ and $g_{r\phi}$ are absent, we obtain
\beqa
\begin{split}
 \lambda(r) = - \frac{\Sigma + a^2 \sin^2\theta - a W }{ \Delta }  , \quad  \chi(r) = - \frac{a}{ \Delta }, \\
\end{split}
\label{chi_lambda}
\eeqa
where we have denoted metric functions
\beqa
\begin{split}
\Sigma \equiv \sqrt\frac{G}{F} |H| , \quad \Delta \equiv a^2 \sin^2\theta + G|H| . \\
\end{split}
\label{Delta-Sigma_FGH}
\eeqa
Note that when $F=G$, we have $\Sigma=|H|$ and $\Delta = a^2\sin^2\theta + G \Sigma$. 

Since $\chi$ and $\lambda$ are only radial coordinate dependent, 
\beqa
  \lambda(r) = - \frac{  k + a^2 - a w  }{ a^2 + gh }   , ~ \chi(r) = - \frac{a}{a^2+gh}  ,  ~  k = \sqrt\frac{g}{f}h , \quad
  \label{chi-lambda}
\eeqa
where $w \equiv W(\theta=\pi/2)$. We will focus on the rotation motion near the equatorial plane with $\theta=\pi/2$, then $F,G,|H|,\Sigma$ reduces to a combination of $f,g,h,k$, respectively, as
For the nonrotating metric (i.e., $a=0$), 
\beqa
 \lambda(r) =  - \frac{1}{\sqrt{fg}}  , \quad  \chi(r) =  0 ,  \quad  k = \sqrt\frac{g}{f}h ,  \label{chi-lambda_a=0}
\eeqa
and $\Delta = gh$.
The constraints imply that the combinations of $G,F,K$ as above should also be $\theta$ independent.

We obtain a stationary axis-symmetric rotating spacetime metric
\begin{widetext}
\beqa
\begin{split}
ds^2 
& = - F  dt^2 + \frac{|H|}{G|H|+a^2\sin^2\theta} dr^2 + 2 \bigg( ( a \sin^2\theta - W)  F - a \sin^2\theta\sqrt\frac{F}{G}  \bigg) dt d\phi  \\
& + |H| dr^2 +  \bigg( |H| \sin^2\theta + 2a\sin^2\theta \sqrt\frac{F}{G}(a\sin^2\theta-W)  - F (a\sin^2\theta-W)^2 \bigg) d\phi^2 . \\
\end{split}  \label{ds2_FGH_2}
\eeqa
In the NP formalism, the tetrad in the BL coordinates can be chosen as
\beqa
\begin{split}
l^\mu & = \bigg( \frac{G|H|+a \sqrt{FG}(a\sin^2\theta-W)}{F(G|H|+a^2\sin^2\theta)} , \sqrt\frac{G}{F} , 0, \frac{a}{G|H|+a^2\sin^2\theta}\sqrt\frac{G}{F} \bigg), \\ 
n^\mu & =  \bigg( \frac{G|H| - a \sqrt{FG}(a\sin^2\theta-W) + 2 a^2\sin^2\theta }{ 2(G|H|+a^2\sin^2\theta) } , - \frac{1}{2} \sqrt{FG}, 0, - \frac{a\sqrt{FG}}{2(G|H|+a^2\sin^2\theta)} \bigg), \\
m^\mu & = \bigg( - \frac{ i \csc\theta [a\sin^2\theta - \sqrt{FG}(a\sin^2\theta-W) ]  }{\sqrt{2}(G|H|+a^2\sin^2\theta)}\sqrt\frac{G|H|}{F} , - \frac{ia\sin\theta}{\sqrt{2H}}, \frac{1}{\sqrt{2H}}, \frac{i\csc\theta G\sqrt{H}}{\sqrt{2}(G|H|+a^2\sin^2\theta)} \bigg) . \\
\end{split} \label{lmn_FGH}
\eeqa
\end{widetext}
On the other hand, from Eqs.~(\ref{chi_lambda}) and (\ref{Delta-Sigma_FGH}), we can rewrite $F$ and $G$ as
\beqa
\begin{split}
& F = \frac{\Delta - a^2\sin^2\theta}{\Sigma^2} |H| = \frac{a^2\cos^2\theta + gh }{[a^2\cos^2\theta + k + a(W - w)  ]^2} |H| , \\
& G = \frac{ \Delta - a^2\sin^2\theta }{|H|} = \frac{ a^2 \cos^2\theta + g h }{|H|} . \\
\end{split}
\eeqa
In the non-rotating limit (i.e., $a \to 0$), we expect the metric functions $(F,G,H,W)$ reduces to $(f,g,h,w)$. In this case, the NP null tetrads of the axial symmetric spacetime, according to Eq.~(\ref{lmn_FGH}), reduce to
\beqa
\begin{split}
l^\mu & = \bigg( \frac{1}{f}, \sqrt\frac{g}{f} , 0, 0 \bigg), \\ 
n^\mu & =  \bigg( \frac{1}{2}, - \frac{1}{2}\sqrt{fg}, 0, 0 \bigg), \\
m^\mu & = \frac{1}{\sqrt{2h}} \bigg( - \frac{iw}{\sin\theta} , 0, 1, \frac{i}{\sin\theta} \bigg) , \\
\end{split} \label{lnm_fgh}
\eeqa
when $w(r)=0$, it recovers that in a spherically symmetric static spacetime.
Our results are applicable to the generic case, e.g., when the metric function $f \ne g$ in Eq.~(\ref{ds2_generic}).
If $F\ne G$, then $|H| \ne \Sigma$, the metric in Eq.~(\ref{ds2_FGH_2}) can be rewritten as
\beqa
\begin{split}
ds^2 & = \frac{|H|}{\Sigma} \bigg( - \frac{\Delta}{\Sigma} [ dt - (a\sin^2\theta - W ) d\phi ]^2 +  \frac{\Sigma}{\Delta} dr^2  +   \Sigma d\theta^2  \\
& + \frac{\sin^2\theta}{\Sigma} [adt - (\Sigma + a^2 \sin^2\theta - a W ) d\phi ]^2 \bigg) , \quad \\
\end{split} \label{ds2_new}
\eeqa
where $\Sigma$ and $\Delta$ in the notation of original metric functions are
\beqa
 \Sigma \equiv k + a^2 \cos^2\theta + a (W - w) , \quad \Delta \equiv a^2 + gh .\label{Delta-Sigma_ghkW}
\eeqa
In this case, $H(r,\theta)$ remains an unknown function that needs to be determined by the Einstein field equations. In particular, the arbitrary function $H=H(r,\theta)$ can be chosen in such a way that a physically acceptable rotating solution to the components $(r,\theta)$ of mixing terms of the Einstein tensor identically vanishes, i.e., $G_{r\theta}=0$.

\subsection{Properties of Taub-NUT spacetime}

\subsubsection{Taub-NUT spacetime}

The Taub-NUT (Newman-Unti-Tamburino) metric is a vacuum solution of Einstein’s equation. It is a natural generalization of the Schwarzaschild metric and takes the form of~\cite{Misner:1963fr,Misner-Taub:1969,Manko_2005}
\beqa
\begin{split}
ds^2 & = - f [ dt + 2n(\cos\theta + C)d\phi ]^2 + \frac{dr^2}{f} \\
& + (r^2 + n^2) d\Omega_2 , \\
\end{split} \qquad \label{ds2_TN}
\eeqa
where the metric function is
\beqa
f = 1 - \frac{2(M r + n^2)}{r^2+n^2} , \label{f_TN}
\eeqa
where $f(r)<0$ corresponds to the Taub universe and $f(r)>0$ corresponds to NUT space. The Taub-NUT spacetime is space-like at $r=0$ and is time-like at infinity. 
When $n=0$, it recovers the Schwarzschild metric. The metric is sourced by mass $M$ together with the NUT parameter $n$. Note that even in the limit that $M \to 0$, the metric still indicates a curved spacetime with horizons at $r_\pm = \pm n$.
When the time is periodically identified as $t \sim t + 8\pi n$, the Misner string is unobservable. By imposing Wick rotating to imaginary time $t \to - i \tau$, and $n\to -iN$, we obtain the Euclidean version of the Taub-NUT spacetime, which is asymptotically locally flat (ALF) but not asymptotically flat (AF)~\cite{Hawking:1998jf}.
The non-vanishing NUT parameter is present in the $g_{t\phi}$ component of the metric, where
$g_{t\phi} \overset{r\to \infty}{=} - 4n(\cos\theta + C)$. 
As a result, the metric is not asymptotically flat at infinity in the large distance limit (i.e., $r \to \infty$) (i.e., $r \to \infty$). 
$C$ is the MR parameter, which defines the position of the Misner string in the Taub-NUT spacetime.
The Taub-NUT metric is geodesically complete for any value of $C$, but the absence of closed time-like and null geodesics requires $|C|\le 1$~\cite{Hennigar:2019ive}.

\subsubsection{Brill spacetime}

The Brill spacetime, or charged Taub-NUT spacetime, is a natural generalization of the Reissner-Nordstr\"{o}m (RN) metric without central singularity, with the metric function $f(r)$ in a form~\cite{Brill:1964}
\beqa
f = 1 - \frac{ 2 (M r +  n^2) - Q^2 }{r^2+n^2}  . \label{f_CTN}
\eeqa
When $n \ne 0$, the Brill metric owns the property that there is no central singular in this spacetime since the metric function never diverges at $r=0$, e.g., $f(0) \ne \infty$. In fact, one can check that the Brill metric nor has curvature singularity, e.g., $R,R_{\mu\nu} R^{\mu\nu},R_{\mu\nu\rho\sigma} R^{\mu\nu\rho\sigma}$ are not divergent at $r=0$. When $n=0$, it recovers the RN metric. $Q$ denotes the charge of the spacetime. The corresponding electric gauge field is
\beqa
A = - \frac{Qr}{r^2+n^2}[ dt + 2n ( \cos\theta + C ) d\phi ] . \label{Eq:gauge-field-Taub-NUT}
\eeqa

\subsection{Kerr-Newman-Taub-NUT spacetime}

The metric of the Taub-NUT spacetime in Eq.~(\ref{ds2_TN}) belongs to a class of axial symmetric spacetime with metric functions in Eq.~(\ref{ds2_generic}) specified as
\beqa
W = 2n(\cos\theta + C ), \quad w= 2nC, \quad h = r^2 + n^2, \label{W-w-h} 
\eeqa
where $n$ is the NUT parameter and $\abs{C}\le 1$ is the MR parameter that adjusts the location of the string singularities.
By imposing the ENJA upon the Brill metric in Eqs.~(\ref{ds2_TN}) with the metric function in (\ref{f_CTN}), we obtain the exact metric of Kerr-Newman-Taub-NUT spacetime in the BL coordinates as
\beqa
\begin{split}
ds^2 & = - \frac{\Delta}{\Sigma} \big[ dt - [ a\sin^2\theta  - 2n (\cos\theta+C) ] d\phi \big]^2 + \frac{\Sigma}{\Delta}dr^2 \\
& +  \Sigma d\theta^2 + \frac{\sin^2\theta}{\Sigma} [adt - (r^2+n^2 + a^2 - 2anC) d\phi ]^2 . \\
\end{split} \label{ds2_CKTN}
\eeqa
For Taub-NUT spacetime, we have $f=g$ in Eq.~(\ref{ds2_TN}), thus $F=G$ and $\Sigma=H$, which satisfy the constraints $G_{r\theta}=0$.
The metric functions according to Eq.~(\ref{Delta-Sigma_ghkW}) are
\beqa
\begin{split}
 \Sigma & = a^2 \cos^2\theta + \sqrt\frac{g}{f} h(r) + a (W - w) \\
& = r^2 + (a \cos\theta + n)^2 , \\
 \Delta  & = a^2 + (r^2+n^2)f(r) \\
& = r^2 - 2M r + a^2 - n^2 + Q^2 , \\
\end{split} \label{Delta-Sigma}
\eeqa
where the metric function $f(r)=g(r)$ is given in Eq.~(\ref{f_CTN}), $W$, $w$, and $h$ are given in Eq.~(\ref{W-w-h}). $M$ is mass, $J=Ma$ is angular momentum, and $n$ is the NUT parameter that has an interpretation of magnetic mass or gravitomagnetic monopole (charge). The metric is a generalization of the KNTN spacetime with MR parameter $C$, which indicates the position of the Misner string.
On one hand, by setting $a=0$, the metric specializes to Brill metric as long as $Q \le n$. The metric function $ \Delta/\Sigma \equiv f(r)$ recovers Eq.~(\ref{f_CTN}). On the other hand, by setting $n=0$, It reduces to the Kerr metric in BF coordinate.
The outer and inner horizons of KNTN spacetime are
\beqa
r_\pm = M \pm \sqrt{M^2+n^2-Q^2-a^2} . \label{rpm}
\eeqa
Note that $r_\pm $ is no longer real for $a^2+Q^2>M^2+n^2$. Thus, the absence of naked singularity requires that $|a| \le \sqrt{M^2+n^2-Q^2}$.  The physical singularity at $\Sigma=0 $, i.e., $r=0$ and $a\cos\theta + n=0$ is no longer hidden behind the horizon, violating cosmic censorship. 
It's worth noticing that in Eq.(\ref{ds2_TN}), the NUT parameter $n$ presents a factor $r^2+n^2$ in front of the solid angle $d\Omega_2$ of a sphere.
The surface areas of the outer and inner horizons are
\beqa
A_\pm = 4\pi (r_\pm^2 + a^2 + n^2 - 2anC)  . \label{A_pm}
\eeqa
The entropy of the black hole is characterized by the Bekenstein-Hawking formula as
\beqa
S_\pm = \frac{A_\pm}{4}. \label{S_pm}
\eeqa
They are identified with a quarter of the areas of the horizon. $S_+=S_{BH}$ corresponds to the entropy at the event horizon.
Thus, a black hole is a thermal object with temperatures in terms of the outer and inner horizons as
\beqa
T_\pm \equiv \frac{\kappa_\pm}{2\pi} 
 = \frac{r_+-r_-}{A_\pm} = \frac{\sqrt{M^2+n^2-Q^2-a^2}}{2\pi(r_\pm^2+a^2+n^2-2anC)} , \quad \label{T_pm}
\eeqa
where $T_+$ corresponds to the Hawking temperature.
$\kappa_\pm$ is the surface gravity on the corresponding horizon of the KNTN spacetime as
\beqa
\kappa_\pm \equiv \frac{r_+-r_-}{2(r_\pm^2+a^2+n^2-2anC)}   
, \quad\label{kappa_pm}
\eeqa
where $r_\pm$ can be expressed in $(M,n,Q,a)$ by using Eq.(\ref{rpm}).
It is obvious that both Hawking temperature and surface gravity are inversely proportional to the (reduced) areas of horizons. Note that $\kappa_\pm$ carries a scaling dimension inversely proportional to the length dimension, i.e., $[L^{-1}]$. 

The outer and inner ergosurfaces, called ergosurfaces, are surfaces of infinite redshift as
\beqa
r_\pm^E = M \pm \sqrt{M^2+n^2-Q^2-a^2 \cos^2\theta}.
\eeqa
They involve not only the radial but also the polar angle for describing the rotational motion.
By using the fact that the world line of a particle has to be time-like or light-like (for massless particles such as the photon and graviton, etc.), the angular velocities of the KTN spacetime with respect to a distant observer are determined within the upper and lower critical velocities as
\begin{widetext}
\beqa
\Omega  \equiv \frac{d\phi}{dt} 
= \frac{a\sin\theta \pm \sqrt{\Delta}}{(r^2+a^2+n^2-2anC)\sin\theta \pm [ a\sin^2\theta - 2n(\cos\theta+C) ]\sqrt{\Delta}} . \label{Omega}
\eeqa
\end{widetext}
They depend on both radial and polar angle coordinates. The angular velocity $\Omega$ is vanishing at infinity radial distances, i.e., $\Omega \overset{r\to\infty}{=} 0$, except along the polar axis.
The angular velocity $\Omega$ as they approach the horizons reduces to
\beqa
\Omega_\pm  = \frac{a}{r_\pm^2+a^2+n^2-2anC} = 4\pi \frac{a}{A_\pm} , \label{Omega_pm}
\eeqa
which can be interpreted as the angular velocities of the horizons. In fact, the tidal forces become much stronger, so that even the spacetime within, including the outer and inner horizons, is rotating at angular velocities.
In this case, both upper and lower velocities reduce to the same one along the polar axis.
In this situation, $\Omega_\pm$ depends on Misner string (or wireline defect) MR parameter $C$ as
\beqa
\Omega|_{\theta=0} = \frac{1}{-2n(1+C)}, \quad \Omega|_{\theta=\pi} = \frac{1}{-2n(-1+C)}.
\eeqa
They represent the rotating velocities of the Misner string along the polar axis towards the north pole ($\theta=0$) and that towards the south pole ($\theta=\pi$), separately. Although the two Misner strings are symmetric along the polar axis, in general, they are rotating not only in the opposite direction but also at different velocities. This will lead to frame-dragging effects, which measure the difference between a particle co-rotating or counter-rotating along the rotating directions of sources.
Thus, when $C = \pm 1$, there is only one Misner string or wireline defect along the polar axis, towards either the north pole ($\theta=0$) with a finite angular velocity as
\beqa
\Omega|_{\theta=0} = - \frac{1}{4n} , \quad 
\eeqa
or towards the south pole ($\theta=\pi$) with an angular velocity
\beqa
\Omega|_{\theta=\pi} = \frac{1}{4n}.
\eeqa
For $C=1$, the south pole axis is regular; for $C=-1$, the north pole is regular. For the special case when $C=+1$ or $C=-1$, there is only a single semi-infinite singularity on the upper or lower part of the symmetry axis, namely, along the north or south pole, respectively. All other values (i.e., $C \ne \pm 1$) correspond to Taub-NUT solutions with two semi-infinite singularities.
Especially when $C=0$, the angular momentum of the middle rod vanishes since the diverging angular momentum of the two semi-infinite rods cancel each other since the two line singularities are symmetric.
In particular, when $C=0$, there is no conical singularity. In addition, the two Misner strings are rotating in opposite directions but with the same amplitude of velocities. They form a pair of counter-rotating strings with velocities as
\beqa
\Omega|_{\theta=0} = - \frac{1}{2n}, \quad \Omega|_{\theta=\pi} = \frac{1}{2n} .
\eeqa

The NUT parameter, or gravitomagnetic mass, plays the role of the source of two physical strings threading spacetime along the axis of symmetry, towards both the north pole and the south pole simultaneously.
Therefore, we can interpret the NUT parameter as a twist since the Taub-NUT spacetime with $C=0$ is also in terms of a twisting spacetime.
To be brief, when $C = \pm 1$, it corresponds to a regular south/north pole axis. The singularity presents in the north/south sphere and towards the north/south pole with a rotating frequency of $\Omega = \mp 1/4n$, respectively.
When $C=0$, there are two wire singularities going towards the south and north poles simultaneously, but with counter-rotating frequency $\Omega = \mp 1/2n$, respectively. The relative minus sign indicates that the two-line singularities are in counter-rotating directions.
The KNTN spacetime is a stationary, axially symmetric solution to the Einstein-Maxwell equation. It describes a rotating electrically charged source with the NUT parameter.
One can check that the KNTN metric satisfies the Einstein field equations as
\beqa
G_{\mu\nu} = \kappa_N T_{\mu\nu} . \label{eom_Einstein}
\eeqa
where the electromagnetic stress tensor is
\beqa
\begin{split}
T_{\mu\nu}
& =  \frac{1}{4\pi} \bigg( F_{\mu\rho} F^{~\rho}_{\nu} - \frac{1}{4}g_{\mu\nu} F^{\rho\sigma}F_{\rho\sigma}   \bigg) , \\
\end{split}
\eeqa
where the gauge field $2$-form is
\beqa
\begin{split}
F  = dA & = \frac{Q}{\Sigma^2}[ - [ r^2 - (a\cos\theta+n)^2 ] [dt \\
& - ( a \sin^2\theta - 2n ( \cos\theta + C) )  d\phi ] \wedge dr \\
& + 2 r\sin\theta ( a \cos\theta + n )[ a dt \\
& - ( r^2 + a^2 + n^2 - 2 a n C)  d\phi ] \wedge d\theta  ] .  \\
\end{split}  \label{Eq:field-strength-Kerr-Taub-NUT}
\eeqa
In the KNTN spacetime, according to Eq.~(\ref{phi_012}), we have the electromagnetic field tensor in the NP formalism as
\beqa
\begin{split}
 & \phi_0 = \phi_2 = 0 , \\ 
 & \phi_1  = \frac{Q}{2[r-i(a\cos\theta+n)]^2} .
\end{split}
\eeqa
The $1$-form electric gauge field in KNTN spacetime is
\beqa
\begin{split}
A 
= - \frac{Qr}{\Sigma}\big[ dt - [ a\sin^2\theta - 2n ( \cos\theta + C ) ] d\phi \big] . 
\end{split} \label{Eq:gauge-field-Kerr-Taub-NUT}
\eeqa
When $a=0$, it recovers that in Brill spacetime as Eq.~(\ref{Eq:gauge-field-Taub-NUT}).

The KNTN spacetime is stationary and axially symmetric.
There are two Killing vectors, $\xi_{(t)} = \partial_t$, $ \xi_{(\phi)}= \partial_\phi$, associated with the time translational and azimuthal rotational isometries of the KNTN metric, respectively. 
The surface where the Killing vector $\xi_{(t)}$ becomes a null vector is called the stationary limit surface.
The killing vector is time-like outside the surface and space-like in The ergosphere, a region between the ergosurfaces and the horizons.
An observer who follows integral curves of $\xi_{(t)}$, is a static observer with zero angular velocity relative to radial infinity.
The co-rotating Killing vectors as two null generators on the horizon are
\beqa
 \xi_\pm = \partial_t + \Omega_\pm \partial_\phi .
\eeqa
$\Omega_\pm$ are the angular velocities on the horizons as Eq.~(\ref{Omega_pm}). The surface where the Killing vector $\xi_{\pm}$ becomes a null vector is called the speed-of-light surface. The co-rotating Killing vector $\xi_{+}$ is space-like outside the surface and time-like in the region between the surface and the event horizons. 
An observer who moves along an integral curve of $\xi_+$, is a rigidly rotating observer with the same angular velocity $\Omega_+$ as that of the even horizon with respect to static observers at infinity.
As a result, the electrostatic potential $\Phi_\pm \equiv 
 - \xi_\pm^\mu A_\mu(r_\pm)$ for the Kerr-Newman-Taub-NUT spacetime measured at the north or south poles of the $2$-sphere with radius $r_\star$ with respect to the horizon is defined by
\beqa
\Phi_\pm 
= 4\pi \frac{Q r_\pm}{A_\pm} = \frac{Qr_\pm}{r_\pm^2 + a^2 + n^2 - 2an C} , \quad \label{Phi_pm}
\eeqa 
where $A_\mu(r_\pm)$ are the gauge fields in Eq.~(\ref{Eq:gauge-field-Kerr-Taub-NUT}) located at horizons $r=r_\pm$ given by Eq.(\ref{rpm}).
The electric static potential $\Phi_\pm$ represents the potential energy of a test particle with an electric charge opposite to that of the black hole on the horizons. It is not singular at horizon.
Later on, we are able to express all universal low-energy dynamical observables on superradiance in terms of the thermal dynamical variables studied above.

\section{Klein-Gordon equation in the Kerr-Newman-Taub-NUT spacetime}
\label{sec:KG_cKTN}

In this section, as a warmup, let's consider a charged massive scalar field $\Phi$ in the background of Kerr-Newman-Taub-NUT spacetime.
The scalar field satisfies the relativistic Klein-Gorden (KG) equation in curved spacetime:
\beqa
\begin{split}
( D_\mu D^\mu - \mu^2 ) \Phi = 0,\\
\end{split}
\eeqa
where $D_\mu = \nabla_\mu - iq A_\mu$ is covariant derivative. The parameters $\mu$ and $q$ are the mass and charge coupling constants of the charged massive scalar field, respectively.~\footnote{Both the parameters $q$ and $\mu$ own the dimensions of mass (i.e.,$[M]=[L^{-1}]$), since they stands for $q/\hbar$ and $\mu/\hbar$, respectively.} $A_\mu$ is the electromagnetic gauge field in the KNTN spacetime as given in Eq.~(\ref{Eq:gauge-field-Kerr-Taub-NUT}). 
In KNTU spacetime with the metric given in Eq.~(\ref{ds2_CKTN}), the KG equation can be expressed as
\begin{widetext}
\beqa
\begin{split}
 &    \frac{\partial}{\partial r} \bigg(   \Delta   \frac{\partial}{\partial r}  \Phi \bigg)  +  \frac{\omega^2(r^2+a^2+n^2-2anC)^2+m^2a^2 - 2qQr[\omega(r^2+a^2+n^2-2anC)-am]-q^2 Q^2 r^2}{\Delta} \Phi \\
 & + 2 \omega m a \bigg( 1 - \frac{(r^2+a^2+n^2-2anC)}{\Delta} \bigg) \Phi   - \mu^2 r^2 \Phi  +  \frac{1}{\sin\theta}   \frac{\partial}{\partial \theta} \bigg(   \sin\theta  \frac{\partial}{\partial \theta}  \Phi \bigg) - \frac{m^2}{\sin^2\theta} \Phi \\
 &   - \omega^2 [ a   \sin\theta - 2n (\cos\theta + C)\csc\theta ]^2 \Phi - 4 \omega m n \frac{ \cos\theta + C }{\sin^2\theta} \Phi -   \mu^2   (a\cos\theta + n)^2  \Phi = 0.
 \end{split} \label{eom_massive_KG_KTN}
\eeqa
\end{widetext}
In the static and axisymmetric spacetime background, the equations can be solved through variable separation methods. By making the variable separation assumption that the wave function decomposes into angular and radial modes as
\beqa
\Phi(t,r,\theta,\phi) = \int d\omega \sum_{lm} e^{-i\omega t} e^{im\phi}  S_{\omega lm}(\theta,\phi) R_{\omega lm}(r) ,
\eeqa
where $\omega>0$ is the wave frequency. $(l,m)$ are the spheroidal harmonic index (e.g., the orbital angular momentum) and the azimuthal harmonic index (e.g., the magnetic angular momentum) of the wave function, separately. In the spherical case, $l$ and $m$ are the eigenvalues of the usual angular momentum operator associated with the rotational and axial symmetry, respectively.
$S_{\omega lm}$ and $R_{\omega lm}$ (we have omitted the dependence on $a$, $\mu$, and $q$ for briefness) denote the angular and radial parts of the scalar wave function $\Phi$. 
We obtain two independent equations of motion: the angular equation for $S_{\omega lm}(\theta)\equiv S$ and the radial equation for $R_{\omega lm}(r)\equiv R$, separately.

\subsection{Angular equations}

The angular equations of motion can be rewritten as
\beqa
\begin{split}
& \bigg[ \frac{1}{\sin\theta}   \frac{d}{d\theta} \bigg(   \sin\theta  \frac{d}{d\theta}  \bigg)  - \bigg( \bar{s} \cot\theta + \frac{\bar{m}}{\sin\theta} - a\omega \sin\theta \bigg)^2  \\
& -   \mu^2   (a\cos\theta + n)^2   + \lambda \bigg] S = 0,  \\
\end{split} \label{eom_S_scalar_massive}
\eeqa
where we have made notations of an effective spin or twist parameter and magnetic quantum number as
\beqa
\bar{s} \equiv 2n\omega, \quad \bar{m} = m + \bar{s}C . \label{sbar_mbar}
\eeqa
To make the wave function $S$ single valued in azimuthal angle $\phi$, $\bar{m} = m + \bar{s}C$ has to be an integer, which means $m$ and $\bar{s}$ (since $C=0,\pm 1$) are either both integers or both half-integers.
We can also rewrite the angular equation in Eq.~(\ref{eom_S_scalar_massive}) as
\beqa
\begin{split}
& \bigg[ \frac{1}{\sin\theta} \frac{d}{d \theta} \bigg( \sin\theta \frac{d}{d \theta} S \bigg) - \bigg( \bar{s} \cot\theta + \frac{\bar{m}}{\sin\theta} \bigg)^2 \\
& +  a^2 (\omega^2-\mu^2)  \cos^2\theta   +  2 a ( \omega \bar{s} - n \mu^2 )\cos\theta + \bar\lambda     \bigg] S = 0 , \\
\end{split} \label{eom_S_scalar_massive_2} 
\eeqa
where we have denoted the separation constant $\lambda$ in terms of the eigenvalue $\bar\lambda$ for the massive case as
\beqa
\begin{split}
\lambda & = \bar\lambda - 2a\bar{m} \omega + a^2  \omega^2 + n^2 \mu^2  . \\
\end{split} \label{lambda-lambda_b-lambda_t_massive}
\eeqa
For the case that the position of the Misner string is symmetric, i.e., $C=0$, Eq.~(\ref{eom_S_scalar_massive_2}) is the form of the angular equation from the relativistic Klein-Gordon equations in the Kerr-Newman-dynonic black hole. The effective spin $\bar{s}$ can be identified with the product of $-qQ_m$, where $q$ is the electric charge and $Q_m$ is the magnetic or dyonic charge of the black hole~\cite{Semiz:1991kh}. In analogy to that, the Dirac quantization condition for $qQ_m\in {\mathbbm Z}/2$ implies that the electric charge $q$ is quantized given a non-vanishing magnetic charge $Q_m$. The effective spin $\bar{s}$ must be quantized as $\bar{s}\in {\mathbbm Z}/2$. For the non-rotating case (i.e., $a=0$), they reduce to the angular equations in Brill spacetime in Eqs.~(\ref{ds2_TN}) and (\ref{f_CTN}).
The Eq.~(\ref{eom_S_scalar_massive}) is a generalized angular spheroidal differential equation. The generic solutions to the angular wave equation are spin-weighted spheroidal functions, as summarized in the Appendix~\ref{app:angular}.
By making a coordinate transformation defined by
\beqa
y \equiv  \frac{1-\cos\theta}{2} \in [0,1],
\eeqa
the differential equation becomes
\begin{widetext}
\beqa
\begin{split}
& S'' + \frac{2y-1}{y(y-1)} S' - \frac{1}{y(y-1)}\bigg( \lambda - [a(1-2y)+n]^2\mu^2  + \frac{[\bar{m}+4a\omega y(y-1) + \bar{s} (1-2y)]^2}{4y(y-1)} \bigg) S = 0, \\
\end{split}
\eeqa
where the prime denotes the derivative with respect to a new coordinate $y$.
The exact solution is~\footnote{Note: $\mu$ is the mass parameter for the scalar field, and $\gamma,\delta,\epsilon,\alpha$ are parameters for the confluent Heun differential equation. The reader should not be confused with the notations for the spin coefficients in the Newman-Penrose formalism. }
\beqa
\begin{split}
S(y) & = y^{\beta_1}(y-1)^{\beta_2}e^{\beta_3 y} \big[  c_1 \text{HeunC}[q,\alpha ,\gamma ,\delta ,\epsilon ,y] + c_2 y^{1-\gamma } \text{HeunC}[q + (1-\gamma ) (\epsilon -\delta ),\alpha +(1-\gamma ) \epsilon ,2-\gamma ,\delta ,\epsilon ,y]  \big] , \\
\end{split}
\eeqa
where the exponential phases are
\beqa
\beta_1 = - (\bar{s}+\bar{m})/2 , ~ \beta_2 = - (\bar{s}-\bar{m})/2, ~ \beta_3 = -2a k, 
\eeqa
where $k \equiv \sqrt{\omega^2-\mu^2}$ is the momentum of the free massive particle, and
\beqa
\begin{split}
& \gamma = 1  - \bar{s} - \bar{m} , \quad \delta =  1  - \bar{s} + \bar{m} ,  \quad \epsilon =  - 4 a k, \\
& q = \lambda + \bar{s} + 2 a \bar{m} (\bar{m}+\bar{s}-1)+2 a \omega  (\bar{m}+\bar{s})-(a+n)^2 \mu^2  , \quad \alpha = 4 a [ k (\bar{s}-1) + \bar{s} \omega - n\mu^2 ] . \\
\end{split}  \label{gamma_delta_epsilon_q_alpha}
\eeqa
They are parameters of the confluent Heun function.
By changing the coordinates as 
\beqa
z \equiv \cos\theta = 1 - 2y \in [-1,1],
\eeqa
the analytic solution to the angular equation can be rewritten as
\beqa
\begin{split}
 S(z) 
 & =  e^{- a k (1-z) } (1+z)^{-\frac{\bar{s}-\bar{m}}{2}}  \big[ c_1 (1-z)^{ - \frac{\bar{m}+\bar{s}}{2} }   \text{HeunC}[ q,\alpha ,\gamma ,\delta ,\epsilon , {(1-z)}/{2} ] \\
 & + c_2 (1-z)^{\frac{\bar{m}+\bar{s}}{2} } \text{HeunC} [ q + (1-\gamma ) (\epsilon -\delta ),\alpha +(1-\gamma ) \epsilon , 2-\gamma ,\delta ,\epsilon , {(1-z)}/{2} ]  \big] , \\
\end{split}
\eeqa
where the coefficients are defined in Eq.~(\ref{gamma_delta_epsilon_q_alpha}).
When $n=0$ (i.e., $\bar{s}=0$ and $\bar{m}=m$), and $\mu=0$ (i.e., $k=\omega$), the coefficient reduces to be
\beqa
\begin{split}
& \gamma = 1 - m, \quad \delta = 1 + m, \quad \epsilon = - 4a\omega ,  \\
&  q =  \lambda + 2a\omega(2m-1), \quad \alpha = -4a\omega .  \\
\end{split}
\eeqa
The solution reduces to ordinary spheroidal harmonics $S_{lm}$ for massless scalars in Kerr spacetime as
\beqa
\begin{split}
S(z) & =  e^{ - c  (1-z)} \left(\frac{z+1}{z-1}\right)^{m/2} \Bigg[ c_1 2^m \text{HeunC} \left[ \lambda + 2 c (2m-1), -4c ,1-m,1+m,-4 c , \frac{1-z}{2} \right] \\
& c_2 (1-z)^m \text{HeunC}\left[ \lambda-m(m+1) - 2 c  ,- 4 c (m+1) ,1+m,1+m,-4 a \omega , \frac{1-z}{2} \right]  \Bigg] , \\
\end{split}
\eeqa
where $c\equiv a\omega$ is the spheroidal parameter. 
When $c=0$, the eigenvalue is $\lambda=l(l+1)$, and the solution reduces to the ordinary spherical harmonics $Y_{lm}$ in terms of hypergeometric functions as
\beqa
\begin{split}
S(z) & =  \bigg(\frac{1+z}{1-z}\bigg)^{m/2} \bigg[ c_1 \, _2F_1\left(-l,l+1;1-m;\frac{1-z}{2} \right) +  c_2 (1-z)^m \, _2F_1\left(m-l,l+m+1;m+1; \frac{1-z}{2} \right) \bigg], \label{Plm}
\end{split}\quad
\eeqa
\end{widetext}
This is the special case of Eq.~(\ref{Plms}) with $s=0$. We can express it as the associated Legendre polynomials in Eq.(\ref{Slm}).
For a massless scalar (i.e., $\mu=0$) in KTN spacetime, the angular equation in Eq.~(\ref{eom_S_scalar_massive}) reduces to
\beqa
\begin{split}
& \bigg[ \frac{1}{\sin\theta} \frac{d}{d \theta} \bigg( \sin\theta \frac{d}{d \theta} S \bigg) - \bigg( \bar{s} \cot\theta + \frac{\bar{m}}{\sin\theta} \bigg)^2 \\
& + [ a \omega \cos\theta + \bar{s} ]^2 - \bar{s}^2  + \bar\lambda  \bigg] S = 0 . \\
\end{split} \label{eom_S_scalar_massless} 
\eeqa
We have used the shift of the eigenvalue as
\beqa
\lambda = \bar\lambda - 2a\bar{m} \omega + a^2 \omega^2  = \tilde\lambda - 2a\bar{m}\omega . \label{lambda-lambda_b-lambda_t_massless}
\eeqa

\subsection{Radial equations}

The radial equations of motion can be rewritten as
\beqa
 \frac{1}{\Delta} \frac{d}{dr} \bigg(  \Delta   \frac{d}{dr} R \bigg) +   \bigg( \frac{ K^2 }{\Delta^2} -  \frac{\mu^2 r^2 + \lambda}{\Delta} \bigg) R = 0 ,
\label{eom_R}
\eeqa
where the metric function $\Delta$ is given in Eq.~(\ref{Delta-Sigma}), and the function $K$ for KNTN is defined as
\beqa
K = \omega (r^2+a^2+n^2-2anC) -  m a - q Q r , \label{K_Q}
\eeqa
and $\lambda$ is the separation constant for solving the angular wave function. In general, it cannot be analytically expressed only in terms of $l$ and $m$ in the presence of non-vanishing $\omega,a,n$.
We can express the $\lambda$ in terms of the eigenvalue $\bar\lambda$ as Eq.~(\ref{lambda-lambda_b-lambda_t_massive}).
%
%
In the low-frequency limit $\omega \ll 1$ and when the wavelength of the scalar field is large compared to the radius of BH curvature, i.e., $\omega M\ll 1$, we can drop both $\omega$ and $\omega^2$ terms in the second row of the bracket. For a massless scalar field ($\mu=0$) in the low-frequency limit, $ M\omega \ll 1$, the radial wave equation simplifies as
\begin{widetext}
\beqa
\begin{split}
& \frac{d}{dr} \bigg(   \Delta   \frac{d}{dr} R \bigg) +  \bigg( \frac{[(r_+^2+a^2+n^2)\omega - a\bar{m} - qQ r_+]^2}{(r-r_+)(r_+-r_-)} - \frac{[(r_-^2+a^2+n^2)\omega - a\bar{m} - qQ r_-]^2}{(r-r_-)(r_+-r_-)} + q^2 Q^2 - \lambda \bigg) R = 0.
\end{split}
\eeqa
In the statice limit (i.e., $\omega=0$, then $\bar{m}=m$) for neutral particle (i.e., $Q=0$), the radial equation simplifies as
\beqa
 \frac{d}{dr} \bigg(   \Delta   \frac{d}{dr}  R \bigg)  +  \bigg( \frac{ a^2 m^2}{\Delta} - \lambda \bigg) R  = 0 .
\eeqa

The solutions to the equations are the associated Legendre's functions with an imaginary second index as
\beqa
\begin{split}
R & = c_1 P_l^{-\frac{i a m}{\sqrt{M^2 + n^2 -a^2}}}\left(\frac{r-M}{\sqrt{M^2 + n^2 -a^2}}\right)  + c_2 Q_l^{-\frac{i a m}{\sqrt{M^2 + n^2 -a^2}}}\left(\frac{r-M}{\sqrt{M^2 + n^2 -a^2}} \right) , \\
\end{split}
\eeqa
where $a<\sqrt{M^2+n^2}$ is assumed and the corresponding eigenvalue is $\lambda=l(l+1)$ which is determined through the angular wave function.
In general, we can express the radial equation in Eq.~(\ref{eom_R}) more explicitly as
\beqa
\begin{split}
& R'' + \bigg( \frac{1}{r-r_+} + \frac{1}{r-r_-} \bigg) R'  +  \bigg(  \frac{k_0^2}{(r_+-r_-)^2} + \frac{\mu_+}{r-r_+}  + \frac{\mu_-}{r-r_-} + \frac{\nu_+^2}{(r-r_+)^2} + \frac{\nu_-^2}{(r-r_-)^2} \bigg) R  = 0 , \\
\end{split} \label{R_r-rpm}
\eeqa
where the prime is a derivative with respect to the radial coordinate $r$, etc. We have used $\Delta=(r-r_+)(r-r_-)$ with horizon radius $r_\pm$ as in Eq.~(\ref{rpm}). 
The equations have singularities at $r=r_\pm$ and $r=\infty$. The coefficients are
\beqa
\begin{split}
& k_0 = k (r_+ -r_-),  \quad \mu_\pm = - \frac{\lambda_\pm}{r_+-r_-}, \quad \nu_\pm   = \frac{\omega  \left(r_\pm^2 + a^2 + n^2 \right)- (a \bar{m}+q Q r_\pm)}{r_+-r_-} , \\ 
\end{split}   \label{ABC}
\eeqa
where $k$ is the translational momentum of a particle with rest mass $\mu$ staying at spatial infinity as
\beqa
k = \sqrt{\omega^2-\mu^2}. \label{k0}
\eeqa
To find the exact solutions, we can first make a change in the coordinate variables as
\beqa
x = \frac{r_+-r}{r_+-r_-} \in (-\infty, 0] . \label{x-r}
\eeqa
The radial equation in Eq.~(\ref{eom_R}) become of the form
\beqa
\begin{split}
& R'' + \bigg( \frac{1}{x} + \frac{1}{x-1} \bigg) R' + \bigg( k_0^2 + \frac{\lambda_+}{x} + \frac{\lambda_-}{x-1} + \frac{\nu_+^2}{x^2} + \frac{\nu_-^2}{(x-1)^2} \bigg) R = 0 , 
\end{split} \label{R_x}
\eeqa
where the prime denotes derivatives with respect to $x$, and the coefficients $A_\pm$ are
\beqa
\begin{split}
\lambda_\pm & = \pm \bigg( \lambda - k^2 r_\pm^2  - (\omega r_\pm  - q Q)^2 + \frac{q^2 Q^2}{2} + \frac{2[\omega(a^2+n^2+r_+ r_-) - a\bar{m} - q Q(r_++r_-)/2]^2}{(r_+-r_-)^2} \bigg) .\\
\end{split}   \label{lambda_pm}
\eeqa
The parameters $\nu_\pm$ in Eq.~(\ref{ABC}) can be expressed in terms of frequencies and the Hawking temperature or surface gravities in Eqs.~(\ref{T_pm}) and (\ref{kappa_pm}) as
\beqa
\begin{split}
 \nu_\pm  =  \frac{\omega-\omega_\pm}{4\pi T_\pm}  = \frac{\omega-\omega_\pm}{2\kappa_\pm} ,
\end{split} \label{nu_pm}
\eeqa
where $\omega_\pm$ are critical values of angular frequency on the outer and inner (or event and Cauchy) horizons as
\beqa
\omega_\pm = \frac{am+qQr_\pm}{r_\pm^2 + a^2 + n^2 - 2anC}  = m \Omega_\pm + q \Phi_\pm , \label{omega_pm}
\eeqa
where $\Omega_\pm$ are angular velocites of horizons in Eq.~(\ref{Omega_pm}), and $\Phi_\pm$ are electrostatice potentials of horizons in Eq.~(\ref{Phi_pm}), respectively.
The general solution to the radial equation can be expressed in terms of
\beqa
\begin{split}
R & = e^{ i k_0 x} (-x)^{i \nu_+} (1-x)^{i \nu_-} [ c_1 \text{HeunC}[q,\alpha ,\gamma ,\delta ,\epsilon ,x] + c_2 x^{1-\gamma }  \text{HeunC}[(1-\gamma ) (\epsilon -\delta )+q,\alpha +(1-\gamma ) \epsilon ,2-\gamma ,\delta ,\epsilon,x] ], \\
\end{split} \label{R_x-y_x}
\eeqa
where the coefficients are
\beqa
\begin{split}
 & \epsilon  = 2 i k_0  ,  \quad \gamma  = 1 + 2i \nu_+, \quad \delta  = 1 + 2 i \nu_- , \\
 & q = \lambda_+ + \frac{1}{2}[1+\gamma(\epsilon-\delta)] , \quad \alpha = \lambda_+ + \lambda_- + \frac{1}{2}(\gamma+\delta)\epsilon . \\ 
\end{split} \label{q-a_gamma-delta-epsilon}
\eeqa
Therefore, with Eq.~(\ref{R_x}), the general solution to the radial equation is
\beqa
\begin{split}
R(r) & = \bigg( \frac{r-r_-}{r_+-r_-} \bigg)^{i \nu_-} \Bigg[ c_1 e^{ i k (r - r_+)}  \bigg( \frac{r-r_+}{r_+-r_-} \bigg)^{i \nu_+ }  \text{HeunC}\left[ q,\alpha , 1+ 2 i \nu_+ ,\delta ,\epsilon , \frac{r_+-r}{r_+-r_-}  \right] \\
& + c_2 e^{ - i k (r - r_+)}  \bigg( \frac{r-r_+}{r_+-r_-} \bigg)^{- i \nu_+} \text{HeunC}\left[ q + 2i\nu_+(\delta-\epsilon),\alpha - 2i \nu_+ \epsilon , 1 - 2 i \nu_+ ,\delta ,\epsilon, \frac{r_+-r}{r_+-r_-} \right] \Bigg]. \\
\end{split} \label{R_r}
\eeqa
\end{widetext}

\subsubsection{Near-horizon behavior}

In the near outer horizon limit, i.e., $r \to r_+$, the radial wave function behaves as
\beqa
\begin{split}
R(r) 
& \overset{r\to r_+}{=}    c_1  e^{  ik (r-r_+) } \bigg( \frac{r-r_+}{r -r_-} \bigg)^{ i \nu_+ }   \\
&  + c_2 e^{ - ik (r-r_+) }\bigg( \frac{r-r_+}{r-r_-} \bigg)^{ - i \nu_+ } ,  \\
\end{split} \label{R_rp}
\eeqa
where $\omega_+$ is the angular frequency and $\nu_+$ is defined in Eq.~(\ref{nu_pm}).
%
%
%
For the massless case (i.e., $\mu=0$, $k=\omega$), the radial wave function can be reexpressed as
\beqa
\begin{split}
R  
& \overset{r\to r_+}{=} c_1 e^{  i (\omega-\omega_+) r_\star } + c_2 e^{ - i (\omega-\omega_+) r_\star } , \\
\end{split}
\eeqa
where $r_\star$ is the tortoise coordinate defined by
\beqa
\begin{split}
r_\star 
& = r + \frac{1}{2\kappa_+} \log\frac{|r-r_+|}{r_+-r_-} - \frac{1}{2\kappa_-} \log\frac{|r-r_-|}{r_+-r_-}  , \\
\end{split} \label{r_star-tortoise}
\eeqa
where $\kappa_\pm$ is the surface gravity of the KNTN spacetime in Eq.~(\ref{kappa_pm}). 
%
By using Eq.~(\ref{omega_pm}), we have
\beqa
\begin{split}
 r_\star \omega_+
& \overset{r\to r_+}{=}  \frac{am+qQr_+}{r_+-r-}\log\frac{r-r_+}{r-r_-} + \ldots, \\
\end{split}
\eeqa
where $\ldots$ denotes an irrelevant constant term.
The radial wave function can also be rewritten by combining the frequency and magnetic quantum number phase factor as
\beqa
\begin{split}
e^{-i\omega t} e^{im \phi} R 
& \overset{r\to r_+}{=} c_1 e^{-i\omega u} e^{i m \tilde\phi} + c_2 e^{-i\omega v } e^{im \bar\phi} , \\
\end{split}
\eeqa
where $u=t-r_\star$ and $v=t+r_\star$ are retarded and advanced time, and the corresponding retarded and advanced azimuthal angles in outgoing and ingoing EF coordinates are, respectively
\beqa
\begin{split}
& \tilde\phi = \phi - \frac{a}{r_+-r_-}\log\frac{r-r_+}{r-r_-}, \\
& \bar\phi = \phi + \frac{a}{r_+-r_-}\log\frac{r-r_+}{r-r_-}. \\
\end{split}
\eeqa
The coefficients $c_{1,2}$ are associated with the outgoing and ingoing modes, respectively.
By imposing the in-falling boundary condition (i.e., $c_1=0$), we obtain the general solution to the radial equations in Eq.~(\ref{R_r}) as
\begin{widetext}
\beqa
\begin{split}
& R(r)  =  e^{ i k (r - r_+)} \bigg( \frac{r-r_-}{r_+-r_-} \bigg)^{i \nu_-}   \bigg( \frac{r-r_+}{r_+-r_-} \bigg)^{- i \nu_+} \text{HeunC}\left[ q + 2i\nu_+(\delta-\epsilon),\alpha - 2i \nu_+ \epsilon , 1 - 2 i \nu_+ ,\delta ,\epsilon, \frac{r_+-r}{r_+-r_-} \right] , \\
\end{split} \label{R_r_2}
\eeqa
where the ingoing coefficient $c_2=\abs{R(r_+)}$ is related to the field amplitude at the horizon and is normalized to be unit due to that $\text{HeunC}[q,\alpha ,\gamma ,\delta ,\epsilon , 0] = 1 $.

On the other hand, at low frequency, the radial equation for massless scalars can be rewritten as
\beqa
\begin{split}
& x(x+1) \frac{d}{dx} \bigg( x(x+1) \frac{d}{dx} R  \bigg) + [-\bar\lambda x(x+1) + \nu_+^2 ]R = 0 , \\
\end{split}  \label{eom_R_NH}
\eeqa
where $x$ is defined in Eq.~(\ref{x-r_2}). We have set $\mu=0$ since, at the near-horizon limit, the mass $\mu$ is not relevant.
The solution to the radial equation is
\beqa
\begin{split}
R & = x^{-i\nu_+} (1+x)^{i\nu_+} [ c_1 \, _2F_1(-l,l+1;1 - 2i\nu_+;-x) + c_2 x^{2i \nu_+}  \, _2F_1(-l+2i\nu_+,l+1+2i\nu_+;1+2i \nu_+;-x) ] , \\
\end{split} 
\eeqa
\end{widetext}
where the coefficients associated with $c_{2}$ correspond to the ingoing wave for super-radiance, i.e., $\nu_+<0$. By imposing the ingoing boundary conditions on the horizon, we obtain
\beqa
R =  c_1  x^{-i\nu_+} (1+x)^{+i\nu_+} \, _2F_1(-l,l+1;1-2i\nu_+ ;-x) . \quad
\eeqa
On the other hand, at large $x$, the wave function behaves as
\beqa
\begin{split}
R(x) & \overset{x\to\infty}{=} c_1 \bigg( x^{l} \frac{ \Gamma (2 l+1)  \Gamma (-2 i \nu_+ +1)}{\Gamma (l-2 i\nu_+ +1) \Gamma (l+1)} \\
& + x^{-l-1} \frac{ \Gamma (-2 l-1)  \Gamma (-2 i \nu_+ +1)}{\Gamma (-l-2 i \nu_+ ) \Gamma (-l)} \bigg) . \\
\end{split} \label{R_infty}
\eeqa

\subsubsection{Infinite boundary behavior}

In the infinite boundary limit (i.e., $r\to \infty$) at large distance, the radial equation in the form of Eq.~(\ref{R_r-rpm}) becomes
\beqa
\begin{split}
& \frac{1}{r^2} \frac{d}{dr} \bigg( r^2 \frac{d}{dr}  R \bigg)  +  \bigg(  k^2 + \frac{\mu_+ + \mu_-}{r} \\
& + \frac{\mu_+ r_+ + \mu_- r_- + \nu_+^2 + \nu_-^2}{r^2} \bigg) R  = 0 , \\
\end{split}  \label{R_large_r}
\eeqa
where the coefficients are given in Eq.~(\ref{ABC}).
The general solution to the radial equation is
\beqa
\begin{split}
R(r) & = e^{-i k r} r^{-1 + \nu } \big[ c_1 U\left( \nu +  i \nu_k ,2 \nu ,2 i r k \right) \\
& + c_2 L_{-\nu -i \nu_k }^{ - 1 + 2 \nu }\left(2 i r k\right) \big] , \\
\end{split}
\eeqa
where $U(a,b,z)$ is the confluent hypergeometric function and $L_n^a(x)$ is the generalized Laguerre polynomial. We have made the following notations:
\beqa
\begin{split}
\nu_k & = \frac{\left(\mu_-+\mu_+\right)}{2 k} =  \frac{M(k^2+\omega^2)-qQ\omega}{k} , \\
\nu & = \frac{1}{2} [1 \pm \sqrt{1-4(\nu_-^2 + \nu_+^2 + \mu_- r_- + \mu_+ r_+)}] \\
& \overset{M\omega\ll 1}{=} \frac{1}{2}(1\pm \sqrt{1+4\bar\lambda}) =  l+1 ,   -l, \\
\end{split} \label{nu_k-nu}
\eeqa
where the coefficients $\mu_\pm,\nu_\pm$ are given in Eq.~(\ref{ABC}). $\nu_k$ is a dimensionless quantity depending on the frequency. $\bar\lambda=l(l+1)$ is the eigenvalue determined via solving the angular wave function.
If we have both $\nu_k = 0 = \nu $, the solutions recover to the spherical wave as
\beqa
R(r) = \frac{1}{r}(c_1 e^{ik r } + c_2 e^{-ik r}),
\eeqa
where $c_{1,2}$ are coefficients associated with outgoing and ingoing waves.
Motivated by the above observation, we can rewrite the radial equation in Eq.~(\ref{R_large_r}) as
\beqa
\begin{split}
& \bigg[ \frac{d^2}{dr^2}   +  \bigg(  k^2 + \frac{2k \nu_k}{r} + \frac{\nu-\nu^2}{r^2} \bigg) \bigg] (r R)  = 0 , \\
\end{split}  \label{R_far_r}
\eeqa
The generic solution to the radial wave function turns out to be
\beqa
R(r) = \frac{1}{r} [c_1 M_{-i \nu_k ,\nu -\frac{1}{2}}(2 i k r)+c_2 W_{-i \nu_k ,\nu -\frac{1}{2}}(2 i k r)] ,
\eeqa
where $W_{k,\nu}(z)$ and $M_{k,\nu}(z)$ are the Whittaker functions with indexes $k$ and $\nu$. At low frequency, we have $\nu=l+1/2$.
We can rewrite the wave function as
\beqa
\begin{split}
R(r) & =   e^{-i k r} r^{\nu -1 } [ c_1 \, _1F_1(i \nu_k +\nu ;2 \nu ;2 i k r)\\
& + c_2 U(i \nu_k +\nu ,2 \nu ,2 i k r) ] , \\
\end{split}
\eeqa
where $\, _1F_1(a,b;z)$ is Kummer confluent hypergeometric function and $U(a,b,z)$ is conventional confluent hypergeometric function.
Towards spatial infinity, the radial wave function has the asymptotic behavior as
\beqa
R(r)  \overset{r\to\infty}{=} \frac{1}{r} ( c_{5} r^{-i \nu_k } e^{-ik r} +  c_2 r^{ i \nu_k } e^{ik r}) , 
\eeqa
where $\nu_k$ plays the role of a phase factor and $c_5$ is a constant as a linear combination of $c_1$ and $c_2$. The first term corresponds to the ingoing mode, and the second term corresponds to the refraction mode at large distances.
In the low-frequency limit, i.e., $M\omega\ll 1$ and $\omega\ll 1$, the term $\nu_k\sim 0$. Then the solution reduces to
\beqa
R(r) = c_3 j_{ l }(k r) + c_4 y_{ l }(k r) , \label{R_jy}
\eeqa
where $j_\nu(z)$ and $y_\nu(z)$ are spherical Bessel functions of the first and second kinds, respectively. The solution can also be expressed in Bessel functions of the first and second kinds as
\beqa
R(r) = \sqrt\frac{\pi}{2k r}  [ c_3 J_{l+\frac{1}{2} } (k r) + c_4 Y_{l+\frac{1}{2} }(k r) ].
\eeqa
In the large radial distance, the radial wave function becomes a spherical-wave:
\beqa
\begin{split}
R(r) 
& \overset{r\to\infty}{=} \frac{1}{2 k i r } (  c_1  e^{i k r - i {\pi l}/{2} }  - c_2 e^{-i k r + i {\pi l}/{2} }  ) ,
\end{split}
\eeqa
where $c_1=c_3-ic_4$ and $c_2=c_3+ic_4$. In the large distance limit, the product $rR$ are asymptotic to plane-waves along the radial direction.
In case that $k=0$, i.e., the resonance mode $\omega=\mu$, the radial wave function reduces to
\beqa
\begin{split}
R(r) & = c_1 r^l + c_2 r^{-l-1} . \\
\end{split} \label{Rl}
\eeqa

\subsubsection{Extreme limit at low frequencies}

Note that the variable transformation in Eq.~(\ref{x-r}) applies in general to non-degenerate horizons. However, in the extreme limit of the spacetime, i.e., $r_+=r_-$, the coordinates in Eq.~(\ref{x-r}) are invalid. In this situation, we could adopt the other coordinate transformation as
\beqa
u  \equiv \frac{r-r_+}{r-r_-} = \frac{x}{x-1} \in [0,1],
\eeqa
with $r\in [r_+ ,+\infty)$. 
In the coordinates, the radial equations of motion in Eq.~(\ref{R_x}) become
\begin{widetext}
\beqa
\begin{split}
& R''(u) + \frac{1}{u}R'(u) + \bigg[ \bigg(\frac{1}{u} - \frac{1}{u-1}\bigg) (2 \nu_+^2 - \lambda_+) + \frac{\nu_+^2}{u^2} + \frac{\nu_+^2 + \nu_-^2 - \lambda_+}{(u-1)^2} + \frac{\lambda_+ + \lambda_-}{(u-1)^3} + \frac{k_0^2}{(u-1)^4} \bigg] R(u) = 0 , \\
\end{split}
\eeqa
where the parameters are given in Eq.~(\ref{ABC}). For massless particles ($\mu=0$) at low frequencies ($\omega M \ll 1$), the last two terms in the large bracket are vanishing, since 
\beqa
\begin{split}
 k_0 & = k (r_+ - r_-) = 0, \quad  
 \lambda_+ + \lambda_-  = -2(r_+-r_-)\bigg[M\bigg(\frac{k^2}{\omega^2}+1\bigg)- qQ \bigg]\omega^2  \ll 1 . \\
\end{split}\quad
\eeqa
In the low-frequency case, the radial equation turns out to be exactly solvable, and the solution is
\beqa
\begin{split}
R(u) & =   (1-u)^\beta [  c_1  u^\alpha \, _2F_1\left(\alpha +\beta -\gamma  ,\alpha +\beta +\gamma  ;2 \alpha +1;u\right)  +  c_2 u^{- \alpha } \, _2F_1\left(-\alpha +\beta -\gamma  ,-\alpha +\beta +\gamma  ;1-2 \alpha ;u\right),  \\
\end{split} ~ \label{R_u}
\eeqa
where the coefficients $(\alpha,\beta,\gamma)$ and index $\nu$ are
\beqa
\alpha =  i\nu_+, \quad \beta \equiv \frac{1}{2} - \nu , \quad \gamma = - i \nu_-  . 
\label{abg}
\eeqa
By substituting Eq.~(\ref{ABC}), we have
\beqa
\begin{split}
\nu  \equiv \sqrt{\frac{1}{4} + \lambda_+ - \nu_+^2 - \nu_-^2  } 
 \overset{\omega \ll 1}{=} \sqrt{ \frac{1}{4} + \bar\lambda} = l + \frac{1}{2}  , \\
\end{split}~
\eeqa
where the eigenvalue $\bar\lambda = l(l+1)$ is obtained from solving the angular wave functions with the regularity conditions. At low-frequency limit, i.e., $\omega M \ll 1$, $\nu=l+1/2$, and $\beta=-l$.~\footnote{Note: the other branch is $\nu = - (l+1/2) $, and $\beta = l+1$, which is dropped since it is not consistent with the large distance behavior. Another solution is $\beta=1/2+\nu$, and $\nu=-(l+1/2)$, $\beta=-l$.}
The $c_{1,2}$ are coefficients associated with ingoing and outgoing modes, respectively.
Thus, in the extremal limit $r_+ \to r_-$ (i.e., $u\to 0$), the general solution in Eq.~(\ref{R_r}) reduces to
\beqa
\begin{split}
R(r) 
   & = \bigg( \frac{r_+-r_-}{r-r_-} \bigg)^{
   \beta} \bigg[ c_1 \bigg( \frac{r-r_+}{r-r_-} \bigg)^{ \alpha }  \, _2F_1 \bigg(\alpha +\beta -\gamma ,\alpha +\beta
   +\gamma ;2 \alpha +1; \frac{r-r_+}{r-r_-} \bigg) \\
   & +  c_2 \bigg( \frac{r-r_+}{r-r_-} \bigg)^{- \alpha } \, _2F_1 \bigg(-\alpha +\beta -\gamma ,-\alpha +\beta +\gamma ;1-2 \alpha ; \frac{r-r_+}{r-r_-} \bigg) \bigg] .\\
\end{split} \label{R_r-u}
\eeqa
%
In the near-horizon limit $r \to r_+$, the equation recovers Eq.~(\ref{R_rp}) as
\beqa
R(r) \overset{r\to r_+}{=} c_1 \bigg( \frac{r-r_+}{r-r_-} \bigg)^{i\nu_+}  + c_2 \bigg( \frac{r-r_+}{r-r_-} \bigg)^{-i\nu_+} .
\eeqa
Thus, the first and second branches with coefficients $c_{1,2}$ correspond to outgoing and ingoing waves, respectively.
By imposing the in-falling boundary conditions (i.e., $c_2=0$), we are left with only the physical outgoing branch. 
In the large distance limit $r\to\infty$ (i.e.,$u\to 1$), the radial wave function has asymptotic behaviors as
\beqa
\begin{split}
R(r) 
& \overset{r\to\infty}{=} c_3 r^{-\beta} + c_4 r^{-1+\beta} = r^{-1/2} ( c_3 r^{\nu} + c_4 r^{-\nu} ) , 
\end{split}
\eeqa
where the coefficients associated with the first and second far field waves are  
\beqa
\begin{split}
& c_3 = \frac{\pi  c_1 \Gamma (2 \alpha +1) \csc (2 \pi  \beta )}{\Gamma (2 \beta ) \Gamma (\alpha -\beta -\gamma +1) \Gamma (\alpha -\beta +\gamma
   +1)}, \quad c_4 = -\frac{\pi  c_1 \Gamma (2 \alpha +1) \csc (2 \pi  \beta )}{\Gamma (2-2 \beta ) \Gamma (\alpha +\beta -\gamma ) \Gamma (\alpha +\beta +\gamma)} .  \\
\end{split}
\eeqa
The parameters $\alpha,\beta,\gamma$ are denoted in Eq.~(\ref{abg}).
Thus, we can compute the reflection coefficients for the gravitational radiation modes in the extremal limit as
\beqa
\begin{split}
{\mathrm R} \equiv \bigg| \frac{Y_{\text{out}}}{Y_{\text{in}}} \bigg|^2 \equiv \frac{|c_4|^2}{|c_3|^2} 
 &  = \bigg| \frac{\Gamma ( 1 - 2 \nu )}{\Gamma (1+2 \nu )} \frac{\Gamma \big[\frac{1}{2}+\nu + i (\nu_+-\nu_-)  \big]}{\Gamma \big[ \frac{1}{2} - \nu + i (\nu_+-\nu_-)   \big]} \frac{\Gamma \big[\frac{1}{2}+\nu + i  (\nu_+ + \nu_-)   \big]}{\Gamma \big[\frac{1}{2} - \nu + i  (\nu_+ + \nu_-)  \big]} \bigg|^2 . \\
\end{split}
\eeqa
At low frequencies $M\omega \ll 1$, $R =  c_3 r^l + c_4 r^{-l-1}$, we obtain the absorption probability or the gray body factor as
\beqa
\Gamma  \equiv 1 - \bigg| \frac{Y_{\text{out}}}{Y_{\text{in}}} \bigg|^2 =  1 - \bigg| \frac{\Gamma ( -2l )}{\Gamma (2l+2)} \frac{\Gamma \big[ l + 1 + (2 M\omega - q Q) i \big]}{\Gamma \big[ - l + (2 M\omega - q Q) i   \big]} \frac{\Gamma \big[ l + 1 + \frac{i}{2}  \big( \frac{\omega - m \Omega_+ - q \Phi_+}{2\pi T_+} + \frac{\omega - m \Omega_- - q \Phi_-}{2\pi T_-}  \big) \big]}{\Gamma \big[ -l + \frac{i}{2} \big( \frac{\omega - m \Omega_+ - q \Phi_+}{2\pi T_+}   + \frac{\omega - m \Omega_- - q \Phi_-}{2\pi T_-} \big)\big]} \bigg|^2  .
\eeqa
\end{widetext}
We have used expressions for coefficients in Eq.~(\ref{nu_pm}), the angular frequency $\omega_\pm$, and the temperatures in Eqs.~(\ref{omega_pm}) and (\ref{T_pm}).
The amplification factor is ${\mathrm A} = | {Y_{\text{out}}}/{Y_{\text{out}}} |^2 - 1 $, which is equivalent to negative absorption probability.

\subsection{Regge-Wheeler-Zerilli equations}

In a redefinition of the radial wave function as 
\beqa 
\Psi=\sqrt{r^2 + a^2+n^2-2anC}R  , \label{R_Psi}
\eeqa
we can rewrite the radial equations in Eq.~(\ref{eom_R}) in the form of the Regge-Wheeler-Zerilli (RWZ) equation as
\beqa
\bigg( \frac{\partial^2}{\partial t^2} - \frac{\partial^2}{\partial r_\star^2}  + V(r) \bigg) \Psi = 0 , \label{eom_RWZ}
\eeqa
where $V(r)$ is named Regge-Wheeler potential. $r_\star$ is a tortoise coordinate defined as
\beqa
\frac{d r_\star}{dr} = \frac{\Sigma + a^2\sin\theta - a W}{\Delta} , 
\label{r_star-tortoise-KTN}   
\eeqa
where $\Delta$ and $\Sigma$ are metric functions in Eq.~(\ref{Delta-Sigma}). By doing integration for Eq.~(\ref{r_star-tortoise-KTN}), we obtain an explicit form of tortoise coordinate.
Note that in the near outer horizon limit, $r_\star(r_+) \to -\infty$, while in the infinite boundary limit, $r_\star(+\infty) \to + \infty$. Thus, the radial region outside the event horizon $r\in[r_+,\infty)$ can be mapped into the whole real region $r_\star\in(-\infty,+\infty)$ in the tortoise coordinate. 
In this case, we can rewrite the RWZ equation to become of the form
\beqa
\bigg( \frac{d^2}{dr_\star^2} + \frac{ K^2 - \Delta (\mu^2 r^2 + \lambda) }{(r^2 + a^2+n^2-2anC)^2} - G^2 - \frac{dG}{dr_\star}    \bigg) \Psi = 0 , \quad
\eeqa
where the function $G$ is defined as
\beqa
 G = \frac{r\Delta}{(r^2+a^2+n^2-2anC)^2} .  \label{G}
\eeqa
The RWZ equation in Eq.~(\ref{eom_RWZ}) is a proper Schr\"{o}dinger form equation with the potential as
\beqa
\bigg( \frac{d^2}{dr_\star^2} + V_{\text{eff}}(r) \bigg) \Psi = 0 , \label{eom_RWZ_2}
\eeqa
with an effective potential $V_{\text{eff}}$ that is related to the Regge-Wheeler potential $V(r)$ as
\beqa
\begin{split}
V_{\text{eff}}(r) &  \equiv \omega^2 - V(r)  \\
& = ( \omega  - \omega_r )^2 - \frac{\Delta ( \mu^2  r^2  + \lambda  ) }{(r^2+a^2+n^2-2anC)^2}  \\
& - \frac{\Delta(\Delta+r \Delta')}{(r^2+a^2+n^2-2anC)^3} \\
& + \frac{3 r^2\Delta^2}{(r^2+a^2+n^2-2anC)^4} , \\ 
\end{split} \label{Veff_KTN}
\eeqa
where $\omega_r$ is a radial coordinate $r$-dependent critical frequency defined by
\beqa
\omega_r \equiv \frac{a m + q Q r}{r^2+a^2+n^2-2anC} . \label{omega_r}
\eeqa
In the near-horizon limit $r\to r_\pm$, the critical frequency recovers that on the event and Cauchy horizons as Eq.~(\ref{omega_pm}), i.e., $\omega_r(r_\pm) = \omega_\pm$. At large spatial distances $r\to \infty$, the critical frequency is vanishing, i.e., $\omega_r(\infty)=0$.
In the near-horizon and infinite limit ($r_\star \to \mp \infty$), the potential takes the form
\beqa
\begin{split}
V_{\text{eff}}(r) 
 & \overset{r_\star \to -\infty}{=} (\omega - \omega_+)^2  ,\\
 & \overset{r_\star \to +\infty}{=} \omega^2 - \mu^2 , \\
\end{split}
\eeqa
where $\omega_+$ is a critical angular frequency on the event horizon. $\kappa_+$ is the surface gravities on the event horizon, as defined in Eq.~(\ref{kappa_pm}).
In both limits, there are two linearly independent solutions to the RWZ equation in Eq.~(\ref{eom_RWZ_2}), separately. In the near outer horizon limit $r\to r_+$ (i.e., $r_\star \to -\infty$)
\beqa
\begin{split}
	\Psi
	& \overset{r_\star \to -\infty }{=} c_1 e^{i(\omega - \omega_+) r_\star } + c_2  e^{- i(\omega - \omega_+) r_\star } . \\
\end{split}
\eeqa
In the infinite large distance limit $r\to \infty$ (or $r_\star \to -\infty$), the radial wave function $\Psi_s$ has the asymptotic behavior 
\beqa
\begin{split}
\Psi & \overset{ r_\star \to + \infty }{=} c_3 e^{i k r_\star} + c_4  e^{-i k r_\star} ,  \\
\end{split}
\eeqa
where we made a notation for the momentum of a rest particle at spatial infinity as
\beqa
k \equiv \sqrt{\omega^2 - \mu^2}.
\eeqa
If $\omega>\mu$, $k>0$, the wave function $\Psi$ corresponds to that for the scattering process. If $k=0$, the wave function reduces to $c_1+c_2 r$, which corresponds to a resonance state. If $\omega<\mu$, then $k$ becomes pure imaginary, then the radial wave function becomes that of a bound state.
According to Eq.~(\ref{R_Psi}), the near-event horizon behavior of the radial wave function $R$ is the same as that of $\Psi$, but the large-distance behaviors become
\beqa
\begin{split}
&  R \overset{r \to r_+}{\propto} X_{\text{out}} e^{i(\omega - \omega_+) r_\star } + X_{\text{in}}  e^{- i(\omega - \omega_+) r_\star }, \\
&  R \overset{r \to + \infty }{\propto} Y_{\text{out}} \frac{e^{i k r_\star}}{r} + Y_{\text{in}} \frac{e^{-i k r_\star}}{r} ,  \\
\end{split} \label{R_scalar_infty}
\eeqa
where $X_{\text{out}}\propto c_1, X_{\text{in}}\propto c_2$ correspond the amplitudes of outgoing and ingoing transition, and $Y_{\text{out}}\propto c_3, Y_{\text{in}} \propto c_4$ correspond to the amplitudes of reflection and ingoing, respectively. The Wronskian condition
\beqa
i \frac{d}{dr_\star}W(R,R^\star) = 0,
\eeqa
implies the flux conservation law of reflection and transmission coefficients as
\beqa
|Y_{\text{out}}|^2 = |Y_{\text{in}}|^2 - \frac{\omega-\omega_+}{k} ( |X_{\text{in}}|^2 - |X_{\text{out}}|^2  ).
\eeqa
Thus, the amplification factor for the scattering of the massive scalar wave from curved spacetime is
\beqa
{\mathrm A} \equiv \bigg| \frac{Y_{\text{out}}}{Y_{\text{in}}} \bigg|^2 - 1   
\overset{ }{=} - \frac{\omega-\omega_+}{\sqrt{\omega^2-\mu^2}} | X_{\text{in}} |^2 .
\eeqa
where in the last equality, we have imposed the in-falling boundary condition, i.e., $X_{\text{out}}=0$, and entails the ingoing flux to be unit, i.e., $ Y_{\text{in}} = 1$.
Therefore, the amplification factor of the scattering for massive fields in a rotating spacetime is in terms of the superradiance condition as
\beqa
\mu < \omega < \omega_+ .
\eeqa
For superradiance to occur, the oscillation frequency of the perturbation, $\omega$, must be less than the critical value $\omega_+$.

\section{Teukolsky master equation in the Kerr-Taub-NUT spacetime}
\label{sec:TME_KTN}

Since the electromagnetic and gravitational perturbations of KNTN, in analogy to those of Kerr-Newman, do not decouple, in this section, we will specialize on the Kerr-Taub-NUT black hole henceforth. 
In the BF coordinates, the tetrad is related to the null tetrad in the Newman-Penrose formalism, as summarized in the Appendix~\ref{app:NP_form}
The KTN spacetime in the NP formalism can be constructed through a regular ingoing null tetrad frame that is well behaved on the past horizon ${\mathcal H}^-$ in terms of a generalized Kinnersley tetrad:
\beqa
\begin{split}
	l^\mu & = \frac{1}{\Delta} ( \Sigma + a^2 \sin^2\theta - a W , \Delta, 0, a ), \\
	n^\mu & = \frac{1}{2\Sigma} ( \Sigma + a^2 \sin^2\theta - a W , - \Delta, 0, a ), \\
	m^\mu & = \frac{1}{\sqrt{2H}} ( ia \sin\theta - i W \csc\theta , 0, 1, i \csc\theta ) . \\
\end{split} \label{general_Kinnersley}
\eeqa
The functions $\Delta$, and $\Sigma$ are given in Eq.~(\ref{Delta-Sigma}) with $Q=0$, and $H$, are
\beqa
\begin{split}
& \Delta = r^2 + a^2 - 2M r - n^2, \\
& \Sigma = r^2 + (a\cos\theta + n)^2, \\ 
& H = [r + i(a\cos\theta + n)]^2 . \\
\end{split} \quad
\eeqa
Note that $H$ is a complex function, and $\Sigma = |H| = \sqrt{H\bar{H}} $. 
Since the KTN metric belongs to Petrov type D, according to the Goldberg-Sachs theorem, the real null vectors lie in the two repeated principal null directions of the background KTN spacetime. 
$W$ includes the MR paramter $C$ as shown in Eq.~(\ref{W-w-h}). 
In the NP formalism, the $12$ independent spin coefficients (or Ricci rotation coefficients) in Eq.~(\ref{NP-SCs}) are
\beqa
\begin{split}
& \kappa = 0 , \quad \tau  = - \frac{ia\sin\theta}{\sqrt{2}} \rho\bar\rho  , \quad \epsilon = 0 , \\
& \sigma = 0 , \quad \rho  = - \frac{1}{r-i (a\cos\theta + n )} , \quad \gamma  = \mu + \frac{\rho\bar\rho}{4} \partial_r \Delta , \\
& \pi  = \frac{ia\sin\theta}{\sqrt{2}}\rho^2  , \quad \nu  = 0, \quad   \beta = - \frac{\cot\theta}{2\sqrt{2}}\bar\rho , \\
& \mu  = \frac{\rho^2 \bar\rho}{2 } \Delta  , \quad \lambda  = 0 , \quad \alpha  = \pi - \bar\beta = \pi + \frac{\cot\theta}{2\sqrt{2}} \rho  . \\
\end{split} \label{SCs_KTN}
\eeqa
Note that the NP spin coefficient $\epsilon=0$ in the generalized Kinnersly tetrad, $\kappa=0=\nu$ means geodesic, and $\sigma=0=\lambda$ means the null geodesic congruence is sheer-free. According to Eq.~(\ref{NPS_prime}), this can also be expressed as $\kappa'=\sigma'=\gamma'=0$, etc. The real and imaginary parts of $\rho \equiv \theta + i\omega$ are related to optical scalars such as expansion $\theta$ and twist $\omega$ (of the light ray congruence) along the outgoing null geodesics generated by $l_\mu$ and ingoing null geodesics generated by $n_\mu$ are, respectively, as
\beqa
\begin{split}
 \theta  = - \frac{r}{r^2+(a\cos\theta+n)^2}, ~ \omega = - \frac{a\cos\theta +n }{r^2 + (a\cos\theta +n)^2}. 
\end{split} \qquad
\eeqa
Similarly, we can define $\rho'=-\mu = \theta'+i\omega'$. 
The negative sign of the expansion means there is a trapped surface region where both ingoing and outgoing light are converging.
For the non-rotating case, i.e., $a=0$, the spacetime reduces to Taub-NUT, and the non-vanishing NP quantities are
\beqa
\begin{split}
&  \rho = - \frac{1}{r - i n} , \quad \gamma  = \frac{M-in}{2(r-in)^2}, \quad  \beta = \frac{\cot\theta}{2\sqrt{2}(r+in)} , \\
& \mu =
\frac{ (2M-r)r + n^2 }{2 (r^2+n^2)(r-in)}  ,  \quad \alpha  = - \frac{\cot\theta}{2\sqrt{2}(r-in)}  . \\
\end{split} \quad \label{SCs_TN}
\eeqa
The twist is proportional to the NUT parameter, i.e., $\omega,\omega' \propto n/r^2$ at spatial infinity along the null directions $n$ and $l$, respectively. Thus, the Taub-NUT spacetime can also be viewed as a twist spacetime, which results in a gravitomagnetic lensing effect.

In the generalized Kinnersley tetrad Eq.~(\ref{general_Kinnersley}), the five independent complex NP Weyl scalars, According to Eq.~(\ref{Psi_i}), in Kerr-Taub-NUT spacetime are
\beqa
\begin{split}
& \Psi_0 = \Psi_1 = \Psi_3 = \Psi_4 = 0 , \\
& \Psi_2 = - \frac{M-i n}{[r-i(a\cos\theta+n)]^3}   , \\ 
\end{split} \label{Psi_i}
\eeqa
where $\rho$ is the NP scalar defined in Eq.~(\ref{SCs_KTN}).
This implies that KTN spacetime belongs to Petrov type D, with a doubled pair of principal null directions or vectors, $l^\mu$ and $n^\mu$.
The components $\Psi_{0,4}$ and ($\Psi_{1,3}$) describe transverse and longitudinal degrees of freedom of gravitational waves, respectively, propagating along the null direction $l^a$ ($n^a$), while $\Psi_2$ represents the Coulomb part of the gravitational radiation field. Thus, $\Psi_{1,3}$ can be viewed as pure gauge degrees of freedom for longitudinal radiation, and the perturbations $\hat\Psi_0$ and $\hat\Psi_4$ are linearized gravitational wave perturbations for transverse radiation.
By substituting the above background invariant variables into a generic linearized perturbation equation for all spin-weighted $s$ massless fields $\hat\Phi_s$ in a Petrov type D spacetime in Eq.~(\ref{s-Teukolsky master equation}) as summarized in the Appendix~\ref{app:pert_eom_spin-s}, we obtain the generalized Teukolsky master equations for all massless spin-$s$ neutral particles ($\mu=q=0$) in KTN spacetime as below:
\begin{widetext}
\beqa
\begin{split}
&   \bigg[ [ a   \sin\theta  - 2n (\cos\theta + C )\csc\theta   ]^2- \frac{(r^2+a^2 + n^2 - 2an C)^2}{\Delta} \bigg] \frac{\partial^2}{\partial t^2}  \Phi_s \\
& + \bigg[ 2 a \bigg( 1 - \frac{r^2 + a^2 + n^2 - 2an C}{\Delta} \bigg) -4n(\cos\theta + C)\csc^2\theta \bigg]\frac{\partial^2}{\partial t \partial \phi} \Phi_s \\
& +  \bigg[ \frac{1}{\sin^2\theta} - \frac{a^2}{\Delta} \bigg]  \frac{\partial^2}{\partial \phi^2}   \Phi_s  + 2s \bigg[ \frac{a(r-M)}{\Delta} + i \frac{\cos\theta}{\sin^2\theta} \bigg]\frac{\partial}{\partial \phi} \Phi_s \\
 & + 2s \bigg[ \frac{M(r^2-a^2-n^2+2an C) + 2n(n-aC)r }{\Delta} - r - i [ a \cos\theta + 2n \csc\theta (\csc\theta + C \cot\theta) ]\bigg] \frac{\partial}{\partial t} \Phi_s \\
 & +  \Delta^{-s} \frac{\partial}{\partial r} \bigg(  \Delta^{s+1} \frac{\partial}{\partial r} \Phi_s \bigg)    +  \frac{1}{\sin\theta} \frac{\partial}{\partial \theta} \bigg( \sin\theta \frac{\partial}{\partial \theta} \Phi_s \bigg) + (s - s^2 \cot^2\theta) \Phi_s = 4\pi \Sigma T_s,
\end{split} \label{eom_Teukolsky-master_KTN}
\eeqa
where $\Delta=r^2+a^2-n^2-2Mr$ and $s$ is the ``spin weight'' of the field and $n$ is the NUT parameter. This is a generalization of the equation for a massless neutral scalar field in KTN spacetime, as Eq.~(\ref{eom_massive_KG_KTN}) with $\mu=Q=0$. 
$T$ is the source term, and $T=0$ in the vacuum case. $\Phi_s$ is a wave function of spin weight $s$. The scalar fields are represented by $s=0$, neutrinos are represented by $s=\pm 1/2$, electromagnetic fields are represented by $s=\pm 1$, and gravitational perturbations are represented by $s = \pm 2$. The $\Psi_s$ is defined as
\beqa
\begin{split}
\Phi_s 
& = \phi , ~ s = 0 , \\
& = \chi_0, ~ \text{or} ~ \rho^{-1} \chi_1, ~ s = \pm 1/2  ,  \\
& = \varphi_0, ~ \text{or} ~ \rho^{-2} \varphi_2, ~ s = \pm 1 ,  \\
& = \psi_0, ~ \text{or} ~ \rho^{-3} \psi_3, ~ s = \pm {3}/{2} ,  \\
& = \Psi_0, ~ \text{or} ~ \rho^{-4} \Psi_4, ~ s = \pm 2 , \\
\end{split}
\eeqa
where $\phi$ is the massless scalar field, $\chi_{0,1}$ are components of neutrino field, $\phi_{0,2}$ are components of electromagnetic photon, $\psi_{0,3}$ are Rarita-Schwinger vector-spinor field, and $\Psi_{0,4}$ are the Weyl components of gravitational radiation.

The equation in Eq.~(\ref{eom_Teukolsky-master_KTN}) has the property that its dependence on the angular and radial variables can be separated by decomposing the Teukolsky wave function $\Phi_s$ as
\beqa
\Phi_s = \int d\omega \sum_{lm} e^{-i\omega t} e^{im\phi} S_{\omega lm}^s(a,n,\theta) R^s_{lm}(r) , \label{psi_s} 
\eeqa
where the factor $e^{-i\omega t}$ represents stationarity, and $a,n$ indicate that KTN spacetime is axisymmetric and $\theta\in [0,\pi]$. We obtain vacuum radial and angular Teukolsky equations in KTU spacetime, respectively, as
\beqa
\begin{split}
&   \frac{1}{\sin\theta} \frac{d}{d \theta} \bigg( \sin\theta \frac{d}{d \theta} S \bigg) + \bigg(  a^2 \omega^2 - \omega^2 [ a   \sin\theta  - 2n (\cos\theta + C )\csc\theta   ]^2 -   \frac{4n (\cos\theta + C)}{\sin^2\theta}  m\omega \\
& - \frac{(m+s\cos\theta)^2}{\sin^2\theta} - 2\omega s [ a \cos\theta + 2n \csc\theta (\csc\theta + C \cot\theta) ] + s + \bar\lambda \bigg) S = 0 	, \\
& \Delta^{-s} \frac{d}{d r} \bigg(  \Delta^{s+1} \frac{d}{d r} R \bigg) + \bigg(  \frac{ [ \omega (r^2+a^2 + n^2 - 2an C) - a m ]^2}{\Delta}    \\
&  + 2i s \frac{ma(r-M)- \omega [ M(r^2-a^2-n^2+2an C) + 2n(n-aC)r + r \Delta  ]}{\Delta} + 4i s \omega r - \bar\lambda - a^2 \omega^2 + 2 a m\omega \bigg) R = 0, \\
\end{split} \label{eom_Teukolsky-master_KTN_RS}
\eeqa
\end{widetext}
where $\bar\lambda$ is the eigenvalue to be determined in solving the angular equation of motion.

\subsection{Angular equations}

The angular equation in Eq.~(\ref{eom_Teukolsky-master_KTN_RS}) can be reexpressed as
\beqa
\begin{split}
&  \bigg[ \frac{1}{\sin\theta} \frac{d}{d \theta} \bigg( \sin\theta \frac{d}{d \theta}  \bigg) - \bigg( (s+\bar{s})\cot\theta + \frac{\bar{m}}{\sin\theta} - a \omega \sin\theta \bigg)^2 \\
& + \lambda + s - 2s(\bar{s}+2a\omega \cos\theta) \bigg] S = 0 , \\
\end{split} \label{eom_S}
\eeqa
where $\lambda$ is a variable separation constant that is related to $\bar\lambda$. We have introduced new notations in terms of an effective spin $\bar{s}$ and magnetic quantum number $\bar{m}$ as defined in Eq.~(\ref{sbar_mbar}).
In this case, the product of the frequency $\omega$ together with the NUT parameter $n$, appears in the azimuthal dependence and effectively plays the role of spin, as long as $\bar{s}$ is an integer or half integer.

The presence of the Misner string MR parameter $C$ shifts the original magnetic quantum number $m$ to a new effective magnetic quantum number $\bar{m}$. For the symmetric case (i.e., $C=0$), $\bar{m}=m$.
The physical meaning of this effective spin and magnetic number will become more clear in the discussion section.
We can reorganize the equation as follows:
\beqa
\begin{split}
& \bigg[ \frac{1}{\sin\theta} \frac{d}{d \theta} \bigg( \sin\theta \frac{d}{d \theta}  \bigg) - \bigg( (s+\bar{s})\cot\theta + \frac{\bar{m}}{\sin\theta} \bigg)^2 \\
& + [ a \omega \cos\theta - s + \bar{s} ]^2 - s^2 - \bar{s}^2 + s + \bar\lambda  \bigg] S = 0 , \\
\end{split} \label{eom_S_2}
\eeqa
where $\bar\lambda = \lambda + 2a\bar{m}\omega - a^2\omega^2 $, according to Eq.~(\ref{lambda-lambda_b-lambda_t_massive}). The solution to the equation is a generalization of the spin-weighted spheroidal harmonic function with $n \ne 0$.
When $s=0$, the equation recovers the massless version of Eq.~(\ref{eom_S_scalar_massive_2}) with $\mu=0$.
When $a=0$, the equation becomes
\beqa
\begin{split}
& \bigg[ \frac{1}{\sin\theta} \frac{d}{d \theta} \bigg( \sin\theta \frac{d}{d \theta}  \bigg) - \bigg( (s+\bar{s})\cot\theta + \frac{\bar{m}}{\sin\theta} \bigg)^2 \\
& - 2 s\bar{s} + s + \bar\lambda  \bigg] S = 0 . \\
\end{split} \label{eom_S_3}
\eeqa
The solution to the differential equation is spin-weighted spherical harmonics, $Y^{s+\bar{s}}_{l\bar{m}}$ with eigenvalue as
\beqa
\bar\lambda  
 = l(l+1) - s(s+1) - \bar{s}^2 . \label{lambda_b_l-s}
\eeqa
When $\bar{s}=0$ (i.e., $n=0$), it recovers the ordinary spin-weighted spherical harmonics as defined in Eq.~(\ref{Ys_omega_lm}).
In general, the eigenvalue, which corresponds to the spin-weighted spheroidal wave function, can only be solved numerically. We can analytically expand it around $\bar\lambda$ in $c=a\omega$, e.g., $\bar\lambda+O(a\omega)$. 

The spin-weighted spherical harmonics can be analytically evaluated. The are in terms of the generalized Legendre functions, e.g., Jacobi polynomials, and their explicit form within $0\le s\le 2$ is listed in the Appendix~\ref{App_JP}.
When $n=0$ (i.e., $\bar{s}=0$ and $\bar{m}=m$), the equation reduces to that for massless spin-weighted particles in Kerr spacetime~\cite{Teukolsky:1972my}.
\beqa
\begin{split}
& \bigg[ \frac{1}{\sin\theta} \frac{d}{d \theta} \bigg( \sin\theta \frac{d}{d \theta}  \bigg) - \bigg( s\cot\theta + \frac{m}{\sin\theta} \bigg)^2 \\
& + (a \omega \cos\theta - s )^2 - s^2 + s + \bar\lambda  \bigg] S = 0 . \\
\end{split} \label{eom_S_4}
\eeqa
The solution to the equation is spin-weighted spheroidal harmonics $Z^s_{lm}(a\omega,\cos\theta)$, as shown in the Appendix~\ref{app:angular}.

For a scalar particle, i.e., $s=0$, the equation becomes that for a massless scalar in Kerr-NUT spacetime, as shown in Eq.~(\ref{eom_S_scalar_massless}) with $a \ne 0 $.
For a scalar field ($s=0$) with $a=0$, i.e., in Schwarzschild spacetime, the angular equation in Eq.~(\ref{eom_S_3}) becomes
\beqa
\bigg[ \frac{1}{\sin\theta}   \frac{d}{d\theta} \bigg(   \sin\theta  \frac{d}{d\theta}  \bigg)  - \bigg( \frac{  \bar{m} }{\sin\theta} + \bar{s} \cot\theta \bigg)^2  + \bar\lambda \bigg] S   =0 , \quad  \label{eom_S_5}
\eeqa
where an angular function can be redefined as an effective spin-weighted spherical harmonic function.
The angular eigenvalue of the generalized spin-weighted spherical harmonic wave function is
\beqa
\bar\lambda  = l(l+1) - \bar{s}^2 . 
\eeqa
It recovers the Casimir equation of Killing vectors, that satisfy the Lie algebra of $SU(2)$. 
%
%
By making identifications in analogy, $(e,g) \sim (\omega, - 2n)$, the frequency/energy $\omega$ plays the role of the electric charge $e$, and the NUT parameter $n$ plays the role of magnetic charge $g$ in a charged gauge theory.

The effective spin $\bar{s}=-eg$ is identified with the product of the electric and the monopole charges, which should be an integer due to the Dirac quantization condition in quantum mechanics. In analogy to that, the quantization of the electric charge $e$ is due to the presence of a magnetic monopole charge $g$. Similarly, the quantization of the energy $\hbar\omega$ is due to the presence of a gravitomagnetic monopole charge $n$.

\subsection{Radial Teukolsky equations}

The radial Teukolsky equation in the BF coordinate system in Eq.~(\ref{eom_Teukolsky-master_KTN_RS}) can be re-expressed as
\beqa
&& \Delta^{-s} \frac{d}{d r} \bigg(  \Delta^{s+1} \frac{d}{d r} R^s \bigg) \nn \\
&& + \bigg( \frac{K^2 - is K \Delta'}{\Delta} + 4is\omega r - \lambda  \bigg) R^s = 0 , 
\label{eom_Rs}
\eeqa
where the function $K$ is
\beqa
K = \omega (r^2+a^2 + n^2 - 2an C) - a m. \label{K}
\eeqa
To be more explicit, the radial Teukolsky equation is
\begin{widetext}
\beqa
\begin{split}
& \bigg( \Delta \frac{d^2}{dr^2} + (s+1) \Delta' \frac{d}{dr} + \frac{ \omega^2(r^2+a^2+n^2-2anC)^2 - 4am\omega (M r + n^2 - an C) + a^2 m^2  }{\Delta} \\
&  + 2i s \frac{ma(r-M)- \omega [ M(r^2-a^2-n^2+2an C) + 2n(n-aC)r ]}{\Delta} + 2i s \omega r - \bar\lambda - a^2 \omega^2 \bigg) R^s = 0 . \\
\end{split}
\eeqa
\end{widetext}
In the derivation of the above equation, we have expressed $K = \omega (2M r + 2 n^2 - 2an C + \Delta) - am$.

\subsection{Regge-Wheeler-Zerilli equations}

\begin{figure}[htb!]
  \includegraphics[width=8cm]{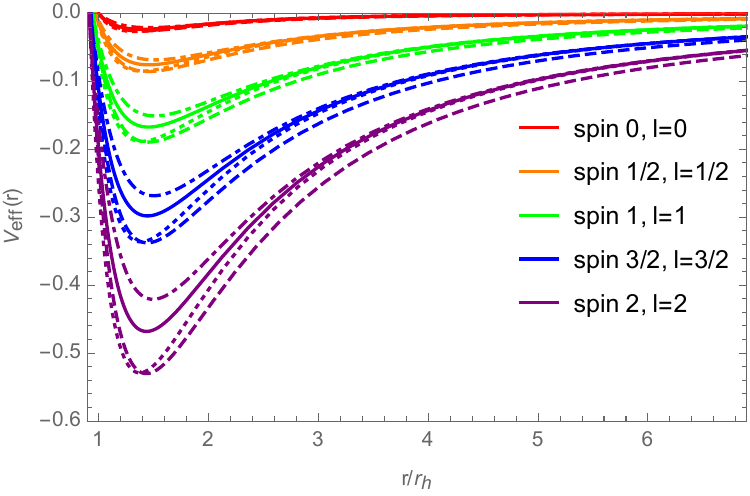}
  \caption{The effective potential for particles with spin.   \label{graph1} }
\end{figure}

The transformation to the tortoise coordinates is the same as in Eq.~(\ref{r_star-tortoise-KTN}), but the redefinition of the radial wave function in Eq.~(\ref{R_Psi}) should be generalized as
\beqa 
\Psi = \Delta^{s/2} \sqrt{r^2 + a^2+n^2-2anC} R  . \label{R_Psi_s}
\eeqa
Comparing to the scalar case, the wave function $\Psi$ contains an additional spin-weighted factor, which controls its near-horizon and large distance behaviors.
Then we can rewrite the radial Teukolsky equations for $R^s(r)$ in Eq.~(\ref{eom_Rs}) in the form of a spin-weighted Regge-Wheeler-Zerilli (RWZ) equation as Eq.~(\ref{eom_RWZ}),
\beqa
&& \bigg( \frac{d^2}{dr_\star^2} + \frac{K^2 - i  s \Delta' K  +  ( 4is\omega r  - \lambda) \Delta }{(r^2 + a^2+n^2-2anC)^2} - G^2 - \frac{dG}{dr_\star}  \bigg) \Psi = 0 , \nn\\
\label{RWZ_s}
\eeqa
where the function $K$ is defined in Eq.~(\ref{K}), and the function $G$ is defined as
\beqa
 G = \frac{s(r-M)}{r^2+a^2+n^2-2anC} + \frac{r\Delta}{(r^2+a^2+n^2-2anC)^2} . \quad \label{G}
\eeqa
The RWZ equation can be rewritten in a Schr\"{o}dinger form equation in Eq.~(\ref{eom_RWZ_2}), and the corresponding effective potential in Eq.~(\ref{Veff_KTN}) is generalized to be a spin-weighted one as
\beqa
\begin{split}
V_{\text{eff}}(r) & = ( \omega - \omega_r )^2 + \frac{ - i s \Delta'   }{r^2+a^2+n^2-2anC}(\omega-\omega_r) \\
& + \frac{ \Delta ( 4is\omega r  - \lambda) - s(\Delta + s \Delta'^2/4) }{(r^2+a^2+n^2-2anC)^2}\\
& - \frac{\Delta(\Delta+ r \Delta')}{(r^2+a^2+n^2-2anC)^3} \\
& + \frac{3 r^2\Delta^2}{(r^2+a^2+n^2-2anC)^4}  ,\\
\end{split} \label{Veff}
\eeqa
where $\omega_r$ is a frequency depending on the radial coordinate $r$ as defined in Eq.~(\ref{omega_r}) but with $Q=0$ as
\beqa
\omega_r \equiv \frac{a m }{r^2+a^2+n^2-2anC} .
\eeqa
The effective potential in Kerr-Taub-NUT spacetime, as defined in Eq.~(\ref{Veff}), is given in Fig.~\ref{graph1}. 
$r_+$ is the event horizon radius. We have chosen input parameters $\lambda=\bar\lambda - 2am \omega$, $\bar\lambda=l(l+1)$, $M=1$, $n=0.5$, $a=0.5$, $C=0$ and set $\omega=m=0$ so that $V(r)$ is a real function. 

From up to down, the solid line corresponds to that of gravitational waves (purple), Rarita-Schwinger waves (blue), electromagnetic waves (green), neutrino waves (orange), and scalar waves (red), respectively. The dashed lines correspond to reference ones in the Kerr spacetime, separately.

In order to facilitate comparison, in the same figure, we have plotted the effective potential with different MR parameters: $C=+1$ (dotted lines) and $C=-1$ (dot-dashed lines), separately.

The shape change of the potential can be understood since the $C$ parameter only affects the potential through the metric function $\Sigma$ appearing in the denominator. However, in the far region ($r\to\infty$), the MR parameter becomes irrelevant. The $C$ parameter is relevant in the angular part through the effective magnetic angular momentum in Eq.~(\ref{sbar_mbar}).

In the near-horizon and infinite limit ($r_\star \to \mp \infty$), the effective potential takes the form
\beqa
\begin{split}
V_{\text{eff}}(r) 
 & \overset{r_\star \to -\infty}{=}  [ (\omega - \omega_+) - i s \kappa_+ ]^2  \\
 & \overset{r_\star \to +\infty}{=} \omega(\omega + 2i s/r), \\
\end{split}
\eeqa
where $\omega_+$ is the critical value of angular frequency at the event horizon in Eq.~(\ref{omega_pm}). $\kappa_+$ is the surface gravities on the event horizon, as defined in Eq.~(\ref{kappa_pm}). According to the RWZ equation in Eq.~(\ref{eom_RWZ_2}), there are two linearly independent solutions. In the near outer horizon limit $r\to r_+$ (i.e., $r_\star \to -\infty$), we have
\beqa
\begin{split}
	\Psi^s 
	& \overset{r_\star \to -\infty}{=} c_1 e^{s \kappa_+ r_\star } e^{i(\omega - \omega_+) r_\star } + c_2 e^{- s \kappa_+ r_\star } e^{- i(\omega - \omega_+) r_\star } \\
	& \overset{r \to r_+ }{=} c_1 \Delta^{s /2} e^{i(\omega - \omega_+) r_\star } + c_2 \Delta^{-s /2} e^{- i(\omega - \omega_+) r_\star } , \\
\end{split}
\eeqa
where in the second row, we have used tortoise coordinate in Eq.~(\ref{r_star-tortoise}) with the near event horizon behavior as 
\beqa
r_\star \overset{r\to r_+}{=} r_+ + \frac{1}{2\kappa_+} \log\frac{r-r_+}{r_+-r_-}.
\eeqa
The mode associated with the coefficients $c_{1,2}$ corresponds to the outgoing and ingoing waves in the near horizons, respectively.
In the infinite large distance limit $r\to \infty$ (or $r_\star \to -\infty$), the radial wave function $\Psi_s$ has the asymptotic behavior
\beqa
\begin{split}
\Psi^s & \overset{ r_\star \to + \infty }{=} r_\star e^{-i \omega r_\star} \big[ c_3 U\left( -s + 1,2,2 i  \omega r_\star \right) \\
& + c_4 \, _1F_1\left( -s+1;2;2 i \omega r_\star \right) \big] ,
\end{split}
\eeqa
where $_1F_1(a;b;z)$ is the Kummer confluent hypergeometric function and $U(a,b,z)$ is the confluent hypergeometric function.
In the second row, we have used the tortoise coordinate in Eq.~(\ref{r_star-tortoise}) with the lager distance asymptotic behavior as
\beqa
r_\star \overset{r\to +\infty}{=} r + 2M\log\frac{r}{r_+-r_-}.
\eeqa
where the first term is a linearly increasing radial coordinate $r$, while the second term is a logarithmic increasing term.
In this case, the asymptotic behavior of the radial wave function becomes
\beqa
\Psi^s =  \overset{ r \to + \infty }{=} d_1 r^{-s} e^{i\omega r} + d_2 r^{s} e^{-i\omega r}  ,
\eeqa
where $d_1$ is a constant involving both $c_3$ and $c_4$, and $d_2$ is one proportional to $c_4$.
The mode associated with the coefficients $d_{1,2}$ corresponds to the outgoing and ingoing waves at large distances, respectively.
According to Eq.~(\ref{R_Psi_s}), we have
\beqa
\begin{split}
R^s  & \overset{r\to r_+}{=} \Delta^{-s/2} \Psi^s , \quad R^s   \overset{r\to\infty}{=} r^{-1-s} \Psi^s . \\
\end{split}
\eeqa
The near-horizon behavior of the radial wave function can be reexpressed as
\beqa
\begin{split}
	R^s & =  c_1 \Delta^{-s } e^{- i(\omega - \omega_+) r } (r-r_+)^{-i \nu_+} \\
	& + c_2  e^{i(\omega - \omega_+) r } (r-r_+)^{i \nu_+} \\
	& \overset{r \to r_+}{=} c_1 (r-r_+)^{-s - i \nu_+} + c_2 (r-r_+)^{i\nu_+}.
\end{split} \label{Rs_IR}
\eeqa
It has a divergent phase for $r=r_+$. The exponential index $ \nu_+ $ is defined in Eq.~(\ref{nu_pm}), where $T_+$ is the Hawking temperature and $\kappa_+$ is the surface gravity.
Therefore, the near-event horizon behavior and the large-distance behavior of the radial wave function $R$ are, respectively,
\beqa
\begin{split}
	R^s & \overset{r \to r_+ }{=} X_{\text{out}}^s  e^{i(\omega - \omega_+) r_\star } + X_{\text{in}}^s \Delta^{-s } e^{- i(\omega - \omega_+) r_\star }  ,  \\
	R^s & \overset{r \to + \infty }{=} Y_{\text{out}}^s   r^{-2s-1} e^{i\omega r} + Y_{\text{in}}^s r^{-1} e^{-i\omega r} ,  \\
\end{split} \label{Rs_scalar_infty}
\eeqa
where $X_{\text{out}}^s \propto c_1, X_{\text{in}}^s \propto c_2$ corresponds to the amplitudes of outgoing and ingoing (transition) waves at the near event horizon region, and $Y_{\text{out}}^s \propto d_1, Y_{\text{in}}^s \propto d_2$ corresponds to the amplitudes of outgoing (reflection) and ingoing waves at spatial infinity, respectively. This is a generalization of the scalar case in Eq.~(\ref{R_scalar_infty}).

The near-horizon behavior of the radial wave function at $r=r_+$ is $R \sim e^{ i (\omega-\omega_+) r_\star} $ and $R \sim e^{-i (\omega-\omega_+) r_\star }\Delta^{-s}$ (ingoing). $r_\star(r_+)\to -\infty$ is the tortoise coordinate. The factor $\Delta^{-s}$ is not a physical singularity, but due to the choice of the generalized Kinnersley tetrad in Eq.~(\ref{general_Kinnersley}), which is singular at horizons.
Thus, the asymptotic solution to the radial wave function $R$ at spatial infinity $r=\infty$ is $R \sim r^{-1-2s} e^{ i\omega r}$ (outgoing waves) and $R \sim r^{-1} e^{- i \omega r }$ (ingoing waves), respectively. They correspond to
\begin{widetext}
\beqa
\begin{split}
& \phi, \chi_1, \varphi_2, \psi_3, \Psi_4 \sim r^{-1}e^{i\omega r}, \quad \chi_0 \sim r^{-2} e^{i\omega r} , \quad \varphi_0 \sim r^{-3} e^{i\omega r} , \quad \psi_0 \sim r^{-4} e^{i\omega r} , \quad \Psi_0 \sim r^{-5} e^{i\omega r} , \quad (\text{outgoing}), \\
& \phi, \chi_0, \varphi_0, \psi_0, \Psi_0  \sim r^{-1} e^{-i\omega r}, \quad \chi_1 \sim r^{-2} e^{-i\omega r} , \quad \varphi_2 \sim r^{-3} e^{-i\omega r} , \quad \psi_3 \sim r^{-4} e^{-i\omega r} , \quad \Psi_4 \sim r^{-5} e^{-i\omega r} , \quad (\text{ingoing}) , \\
\end{split}
\eeqa
where we adopt the notations $\Phi^s=\{\phi,\chi,\varphi,\psi,\Psi\}$ to denote scalar, neutrino, Maxwell, Rarita-Schwinger, and gravitational fields, respectively.
We identify $\{ \phi,\chi_1,\varphi_2,\psi_3,\Psi_4 \}$ ($s\le 0$) as the outgoing radiative part of the fields, and identify $\{\psi,\chi_0,\varphi_0,\psi_0,\Psi_0\}$ as the ingoing radiative part of those fields, since they all decay as $1/r$.
All other Newman-Penrose components of the fields decay more rapidly than $1/r$. Note that for a massless scalar field, both the outgoing and ingoing waves decay as $r^{-1}$. This is obvious by observing Eq.~(\ref{R_scalar_infty}).
In general, the complex tetrad components of gravitational waves at null infinity behave as
\beqa
\begin{split}
& \chi_{i=0,1} \sim r^{-2+i} e^{i\omega r} ,\quad \varphi_{i=0,1,2} \sim r^{-3+i} e^{i\omega r} ,\quad \psi_{i=0,1,2,3} \sim r^{-4+i} e^{i\omega r} ,\quad \Psi_{i=0,1,2,3,4} \sim r^{-5+i} e^{i\omega r} \quad (\text{outgoing}), \\
& \chi_{i=0,1} \sim r^{-1-i} e^{-i\omega r} ,\quad \varphi_{i=0,1,2} \sim r^{-1-i} e^{-i\omega r} ,\quad \psi_{i=0,1,2,3} \sim r^{-1-i} e^{-i\omega r} ,\quad \Psi_{i=0,1,2,3,4} \sim r^{-1-i} e^{-i\omega r} \quad (\text{ingoing}). \\
\end{split}
\eeqa
\end{widetext}
For massless scalar fields, outgoing and ingoing waves behave as $r^{-1}e^{\pm i\omega r}$. In summary, for an arbitrary spin-weighted particle, we have
\beqa
\begin{split}
& \Phi^{s}_{n=0,1,\ldots 2s} = Y^s_{\text{out}} r^{n-2s-1} e^{i\omega r_\star} \quad (\text{outgoing}), \\
& \Phi^{s}_{n=0,1,\ldots 2s} = Y^s_{\text{in}} r^{n-1} e^{-i\omega r_\star} \quad (\text{ingoing}) , \\
\end{split}
\eeqa
where $Y^s_{\text{out}}$ and $Y^s_{\text{in}}$ are associated amplitudes. The ratio of powers $r^{|2s|}$ is due to the ``peeling off'' property~\cite{Penrose:1962ij}.

\section{Low-energy dynamics of superradiance}
\label{sec:LEDO}

\subsection{Radial equations at low frequencies}

We can also rewrite the radial wave function in Eq.~(\ref{eom_Rs}) as
\begin{widetext}
\beqa
\begin{split}
& \bigg[ x(1+x) \frac{d^2}{dx^2}  + (1+s)(1+2x) \frac{d}{dx} + \bigg( - \lambda + 4is[ (r_+-r_-)x + r_+ ]\omega + \frac{[ \omega( (r_+-r_-)x + r_+  )^2 + a^2 + n^2 -2anC) - am  ]^2  }{(r_+-r_-)^2x(1+x)} \\
& - \frac{ 2i s [ (r_+-r_-)x + r_+ - M ] [\omega( (r_+-r_-)x + r_+  )^2 + a^2 + n^2 -2anC) - am ]}{(r_+-r_-)^2x(1+x)}  \bigg) \bigg] R_s = 0 , \\
\end{split} \label{eom_Rs_NH}
\eeqa
\end{widetext}
where the primes are with respect to a variable $x$ as
\beqa
x \equiv \frac{r-r_+}{r_+-r_-} .  \label{x-r_2} 
\eeqa
Similarly, we can express the angular parameter $a$ as
\beqa
a = \frac{(r_+-r_-)\Omega_+}{4\pi T_+} = \frac{ A \Omega_+}{4\pi } ,
\eeqa
where $\Omega_+$ is the angular velocity of the event horizon as defined in Eq.~(\ref{Omega_pm}), $T_+$ is the Hawking temperature in Eq.~(\ref{T_pm}), and $A_+$ is the area of the event horizon in Eq.~(\ref{A_pm}).
The equation becomes:
\beqa
\begin{split}
& x^2(1+x)^2R_s'' + (1+s)x(1+x)(1+2x)R_s' \\
& + \big[ -x(1+x) \lambda +  ( \nu_+ + k_0 x^2 + 2 r_+ \omega x  )^2 \\
& + is [ - (1+2x)\nu_+ + k_0 x^2(3+2x) + 2 r_+ \omega x  ]  \big] R_s = 0 , \\
\end{split}
\eeqa
where $\nu_+$ is defined in Eq.~(\ref{nu_pm}). $k_0$ is a dimensionless variable introduced in Eq.~(\ref{ABC}) with $k=\omega$, which can also be expressed in terms of physical observables as
\beqa
k_0 =   \omega T_\pm A_\pm ,
\label{omega_k0}
\eeqa
where $A_\pm$ is the area of horizon defined in Eq.(\ref{A_pm}).
In the approximation that the Compton wavelength of the particle is much larger than the gravitational size of the black hole, i.e., $M\omega \ll 1$ and the slowly rotating limit $a\omega\ll 1$, we have
\beqa
\begin{split}
& x^2(1+x)^2R_s'' + (1+s)x(1+x)(1+2x)R_s' \\
& + [ -  x(1+x) \lambda  + \nu_+^2 - i s\nu_+(2x+1) \\
& + (2\nu_++3is)k_0 x^2 + 2is k_0 x^3+ k_0^2 x^4 ] R_s = 0 , \\
\end{split} \label{eom_Rs_x}
\eeqa
with $k_0 \ll 1$. The integral constant is defined through Eq.~(\ref{lambda-lambda_b-lambda_t_massive}) as
\beqa
\lambda  = \bar\lambda - 2a\bar{m} \omega + a^2  \omega^2  .
\eeqa
In the $a\omega \ll 1$, the angular eigenvalue becomes 
\beqa
\begin{split}
\lambda & = \bar\lambda + a^2 \omega^2 - 2am\omega 
 = (l-s)(l+s+1) ,
\end{split}
\eeqa
where $l-s$ is a non-negative integer.

\subsubsection{Near-region solution}

In the near region limit, i.e., $k_0 x\ll 1$, the equation becomes
\beqa
\begin{split}
& x^2(1+x)^2R_s'' + (1+s)x(1+x)(1+2x)R_s' \\
& + [ - \bar\lambda x(1+x) + \nu_+^2 - i s\nu_+(2x+1)  ] R_s = 0 . \\
\end{split}
\eeqa
In the $a\omega \ll 1$, the eigenvalue of angular equation is given by Eq.~(\ref{lambda_b_l-s}), i.e., $\bar\lambda = (l-s)(l+s+1) -\bar{s}^2 \ge 0 $. $l-s$ is a non-negative integer.
The most general solution to the equation is a spin-weighted radial wave function in the form of
\beqa
\begin{split}
R^s & = x^{\alpha-s/2} (1+x)^{\beta-s/2} [ c_1 \, _2F_1(a,b;c;-x) \\
& + c_2 x^{1-c} \, _2F_1(a-c+1,b-c+1;2-c;-x) ]  ,
\end{split}
\eeqa
where the coefficients are
\beqa
\begin{split}
& \alpha = - \frac{s}{2} - i \nu_+ , \quad \beta = - \frac{s}{2} + i\nu_+, \\
& a = - l - s, \quad b = l - s + 1, \quad c = 1 - s - 2i\nu_+. \\
\end{split}
\eeqa
The mode associated with $c_{2}$ corresponds to the ingoing wave for super-radiance, i.e., $\nu_+<0$. 
In the near-horizon limit, we have
\beqa
R^s \overset{x\to 0}{=} c_1 x^{-s - i\nu_+} + c_2 x^{i\nu_+} .
\eeqa
It recovers the near-horizon behavior in Eq.~(\ref{Rs_IR}) in the radial coordinate.
The causality entails that we have to choose the ingoing boundary condition so that at the horizon, the wave is always ingoing, i.e., there is no outgoing mode at the horizon.
By imposing the ingoing boundary conditions on the horizon, we obtain
\beqa
\begin{split}
R_s & =  c_1  x^{-s-i\nu_+} (1+x)^{-s+i\nu_+} \\
& \, _2F_1(-l-s,l-s+1;1-s-2i\nu_+ ;-x) . \\
\end{split}
\eeqa
In the infinity limit of $x$, the radial wave function behaves as
\beqa
\begin{split}
R_s & \overset{x\to\infty}{=} c_1 \bigg( x^{l-s} \frac{ \Gamma (2 l+1)  \Gamma (-s-2 i \nu_+ +1)}{\Gamma (l-2 i\nu_+ +1) \Gamma (l-s+1)} \\
& + x^{-l-s-1} \frac{ \Gamma (-2 l-1)  \Gamma (-s-2 i \nu_+ +1)}{\Gamma (-l-2 i \nu_+ ) \Gamma (-l-s)} \bigg) . \\
\end{split} \label{Rs_infty}
\eeqa

\subsubsection{Far region solution}

At a large distance for $x \gg 1 + |\nu| $, the first three terms in the second row of Eq.~(\ref{eom_Rs_x}) can be dropped, and the differential equation is approximated by
\beqa
\begin{split}
& R_s''(x) + \frac{2(1+s)}{x} R_s'(x) \\
& + \bigg( - \frac{\bar\lambda}{x^2} + \frac{(2\nu_+ + 3is)k_0}{x^2} + \frac{2is k_0}{x} + k_0^2 \bigg) R_s(x) = 0 . \\
\end{split}
\eeqa
Given the eigenvalue in Eq.~(\ref{lambda_b_l-s}), in the low-frequency limit, i.e., $k_0\ll 1$, the solution simplifies as
\beqa
\begin{split}
R_s(x) & = e^{-i k x} x^{l-s} [ c_3 U(l-s+1,2 l+2,2 i k x) \\
& +c_4 L_{-l+s-1}^{2 l+1}(2 i k x) ], \\
\end{split} \label{Rs_far}
\eeqa
where $U(a,b,c)$ is the confluent hypergeometric function, and $L_n(x)$ is the Laguerrel polynomial.
In the far region, the equation can also be rewritten as
\beqa
\begin{split}
& \bigg( \frac{d^2}{dx^2} - [ \bar\lambda + s(s+1) \\
& - 2 k_0 \nu_+ - i k_0 s(3+2x) - k_0^2 x^2 ] \bigg) \Psi_s = 0 ,
\end{split}
\eeqa
where $\Psi_s=x^{1+s}R_s$.
The exact solution in the far region is
\beqa
R_s = x^{-s-1} [ c_5 W_{s,l+1/2}(2 i k_0 x) + c_6 M_{s,l+1/2}(2 i k_0 x) ] . \quad
\eeqa
In the near-horizon region, the solution is behaviors as
\beqa
R_s(x) \overset{x\ll 1}{=} d_1^s x^{l-s} + d_2^s x^{-1-l-s},
\eeqa
where
\begin{widetext}
\beqa
\begin{split}
& d_1^s  = \frac{c_3 \Gamma (-2 l-1)}{\Gamma (-l-s)}+\frac{c_4 \Gamma (l+s+1)}{\Gamma (2 l+2) \Gamma (s-l)}, \quad 
 d_2^s = -\frac{i c_3 4^{-l} e^{-i \pi  l} l k_0^{-2 l-1} \Gamma (2l)}{\Gamma (l-s+1)}  . \\
\end{split}
\eeqa
On the other hand, by matching to Eq.~(\ref{Rs_infty}), the coefficients $d_{1,2}$ are determined as
\beqa
\begin{split}
	& d_1^s = c_1 \frac{ \Gamma (2 l+1)  \Gamma (-s-2 i \nu_+ +1)}{\Gamma (l-2 i\nu_+ +1) \Gamma (l-s+1)} , \quad 
	 d_2^s = c_1 \frac{ \Gamma (-2 l-1)  \Gamma (-s-2 i \nu_+ +1)}{\Gamma (-l-2 i \nu_+ ) \Gamma (-l-s)} .
\end{split}
\eeqa
From which, we can determine $c_{3,4}$ as
\beqa
\begin{split}
c_3^s & = \frac{i c_1 4^l e^{i \pi  l} k^{2 l+1} \Gamma (-2 l-1) \Gamma (l-s+1) \Gamma (-s-2 i \nu_+ +1)}{l \Gamma (2 l) \Gamma (-l-2 i \nu_+ ) \Gamma (-l-s)}, \\
c_4^s & = c_1
   \Gamma (-s-2 i \nu_+ +1)  \bigg( \frac{\Gamma (2 l+1) \Gamma (2 l+2) \Gamma (s-l)}{\Gamma (l-2 i \nu_+ +1) \Gamma (l-s+1) \Gamma (l+s+1)}  -\frac{i 2^{2 l+1} e^{i \pi  l} (2
   l+1) k_0^{2 l+1} \Gamma (-2 l-1)^2 \sin (\pi  (l+s))  }{\Gamma (-l-2 i \nu_+ ) \Gamma (-l-s) \sin (\pi  (l-s)) }  \bigg) . \\
\end{split} \quad
\eeqa
%
In the infinite limit, we have
\beqa
\begin{split}
	R_s 
	& \overset{x\to \infty}{=} 
	Y^s_{\text{in}} e^{-i \omega(r-r_+)} r^{-1} + Y^s_{\text{out}} e^{i \omega(r-r_+) } r^{-2s-1} , \\
\end{split}
\eeqa
where the ingoing and outgoing wave amplitudes at infinity are
\beqa
\begin{split}
	& Y^s_{\text{in}} = \frac{i^{1-l-s}}{(2k_0)^{1+l+s}} (2k_0)^{2s} \bigg( \frac{e^{il\pi} c_4 }{\Gamma(-l+s) } - c_3 e^{i\pi s}  \bigg) \frac{k_0}{\omega}  , \quad Y^s_{\text{out}} = -    \frac{i^{1-l-s}}{(2k_0)^{1+l+s}} \frac{c_4 \Gamma(l+s+1)}{\Gamma(l-s+1)\Gamma(s-l)}   \bigg( \frac{k_0}{\omega} \bigg)^{2s+1}  . 
\end{split} \label{Ys}
\eeqa
The factor $k_0/\omega$ comes from the coordinate transformation, since according to Eq.~(\ref{x-r_2}) and (\ref{omega_k0}), we have
\beqa
x = \frac{\omega}{k_0} (r-r_+). \label{x-omega}
\eeqa
Thus, we obtain the reflection coefficient:
\beqa
\begin{split}
\frac{Y^s_{\text{out}}}{Y^s_{\text{in}}} & = - \frac{e^{-il\pi}}{(2\omega)^{2s}}  \frac{\Gamma(l+s+1)}{\Gamma(l-s+1) + \pi [ \cot[\pi(l-s)] - i ] (c_3^{s}/c_4^{s}) }  , \\
\end{split} \quad
\eeqa
where the ratio of the constants is
\beqa
\begin{split}
& \frac{c_3^{(s)}}{c_4^{(s)}}  = -\frac{\Gamma(l-s+1)}{\pi}  \frac{\sin{(2\pi l)}}{\cos{(2\pi s)} - i  \frac{e^{-il\pi}}{(2k_0)^{2l+1}} \frac{\Gamma(l+1)\Gamma(2l+2)\Gamma(-l-s)^2\Gamma(-l-2i\nu_+)}{\Gamma(-2l-1)\Gamma(-2l)\Gamma(l-s+1)^2 \Gamma(l+1-2i\nu_+)}  }  . \\
\end{split}
\eeqa
From this, we immediately obtain the radial wave function $R_{-s}$ by solving the radial equation with $s$ replaced by $-s$ to get the asymptotic form as
\beqa
\begin{split}
R^{-s} 
& = Z^s_{\text{in}} e^{-i \omega(r-r_+)} r^{-1} + Z^s_{\text{out}} e^{i \omega(r-r_+)} r^{2s-1} , 
\end{split}
\eeqa
where the superscript index $s$ denotes its spin dependence with the relevant sign and
\beqa
\begin{split}
& Z^s_{\text{in}} \equiv Y^{-s}_{\text{in}} = (2k_0)^{-2s} \bigg( \frac{e^{il\pi} d_2^{-s} }{\Gamma(-l-s) } - d_1^{-s} e^{-i\pi s}  \bigg) \frac{k_0}{\omega}, \quad Z^s_{\text{out}} \equiv Y^{-s}_{\text{out}} = -    \frac{\Gamma(l-s+1)}{\Gamma(l+s+1)\Gamma(-s-l)} d_2^{-s}  \bigg( \frac{k_0}{\omega} \bigg)^{-2s+1} .
\end{split} \quad \label{Zs}
\eeqa
\end{widetext}

\subsection{low-energy dynamical observables}

\subsubsection{Amplification factor}

The amplification factor is defined as
\beqa
{\mathrm A}^s \equiv {\mathrm R}^s - 1 
 = - \Gamma^s , \label{A_R_Gamma}
\eeqa
where ${\mathrm R}^s \equiv |{\mathcal R}^s|^2  $ is the reflection coefficient and $\Gamma^s \equiv |{\mathcal T}^s|^2$ is the transition coefficient. They stand for the reflection probability and transmission probability, respectively.
The reflection coefficient ${\mathrm R}^s$ is defined as
\beqa
{\mathrm R}^s  = \left| \frac{Y^s_{\text{out}}}{Y^s_{\text{in}}} \frac{Z^s_{\text{out}}}{Z^s_{\text{in}}}  \right| = 1 - \Gamma^s ,
\eeqa
where $Y^s$ and $Z^s$ are given in Eqs.~(\ref{Ys}) and (\ref{Zs}). Thus, we obtain $\Gamma^s  = 1- {\mathrm R}^s = - {\mathrm A}^s$. It is equivalent to the absorption probability, also in terms of the graybody factor~\footnote{When ${\mathrm R}^s<1$, the scattering waves are not amplified but absorbed. $\Gamma^s>0$ represents absorption probability.}.
Note that at low frequencies, we can make perturbative expansions as
\begin{widetext} 
\beqa
\begin{split}
 \frac{Y^s_{\text{out}}}{Y^s_{\text{in}}} \frac{Z^s_{\text{out}}}{Z^s_{\text{in}}} & = e^{-2il\pi} \bigg[ 1 -  e^{i\pi(s-1/2)} \bigg( 3+\frac{\sin{\pi(l-3s)}}{\sin{\pi(l+s)}} \bigg) \cos{\pi(l-s)}   \frac{\Gamma (-2 l-1) \Gamma (-2 l)  \Gamma (l-s+1)^2}{\Gamma (2 l+1) \Gamma (2 l+2) \Gamma (-l-s)^2} \\
 & \times \frac{ \Gamma (l + 1 -2 i \nu_+ ) }{ \Gamma (-l-2 i \nu_+ ) }  (2k_0)^{2l+1}  + O (k_0^{4l+2})\bigg] ,  \\
\end{split} \qquad
\eeqa
where we have imposed the constraint that $2s\in {\mathcal Z}$. Thus, $\sin[\pi(l+s)] = \sin{[\pi (l-s)]}$ etc., and $l-s>0$.

\subsubsection{Gray body factor}

The absorption probability, or gray body factor, is
\beqa
\begin{split}
\Gamma^s 
& = \text{Re}\bigg[ 4 e^{-i\pi(l-s)} \cos{\pi(l-s)} \frac{\sin^2{[\pi(l+s)]}}{\sin^2(2\pi l)}   \frac{ \Gamma (l-s+1)^2 \Gamma (l+s+1)^2}{\Gamma (2 l+1)^2 \Gamma (2 l+2)^2 } \frac{\Gamma (l + 1 -2 i \nu_+ )}{\Gamma (-l-2 i \nu_+ )} (2ik_0)^{2l+1} \bigg] . \\
\end{split}
\eeqa
Only the real part in the bracket is relevant, since the reflection coefficient is a real number.
Assuming that $2s,2l\in {\mathbb Z}$, and $l-s \in {\mathbb Z}^+$, then we can write the absorption probability as
\beqa
\begin{split}
\Gamma^s 
& = \text{Re}\bigg[ \bigg( \frac{ (l-s)! (l+s)! }{ (2l)! (2l+1)! } \bigg)^2 \frac{ \Gamma (l + 1 -2 i \nu_+ ) }{\Gamma (-l-2 i \nu_+ )} (2ik_0)^{2l+1}  \bigg] , \\
& \\
\end{split} \qquad
\eeqa
where the real index $\nu_+$ is defined in Eq.~(\ref{nu_pm}). We can express it in terms of observables as
\beqa
\begin{split}
\nu_+ &  = \frac{\omega-m\Omega_+}{2  \kappa_+}  = \frac{\omega-m \Omega_+}{4\pi T_+} , \\
\end{split}
\eeqa
where $\kappa_+$ is the surface gravity on the event horizon in Eq.~(\ref{kappa_pm}) and $T_+$ is the Hawking temperature in Eq.~(\ref{T_pm}). $\Omega_+$ is the angular velocity on the event horizon defined through Eq.~(\ref{omega_pm}).
When $a=n=0$, $\nu_+=2M\omega$ it recovers the Schwarzschild case. 
By replacing the dimensionless variable $k_0$ in Eq.~(\ref{omega_k0}), we obtain the absorption probability as low frequency as
\beqa
\begin{split}
	\Gamma^s_{lm} & = \bigg( 2^l l! \frac{ (l-s)! (l+s)! }{ (2l)! (2l+1)! } \bigg)^2  \prod_{n=1}^l \bigg[ 1 + \bigg( \frac{1}{n} \frac{\omega-m\Omega_+}{2 \pi T_+} \bigg)^2 \bigg] \frac{\omega-m\Omega_+}{\pi T_+} (A_+ T_+\omega)^{2l+1} , \quad 2l \in \text{even}, \\
	& = \bigg( 2^l l! \frac{ (l-s)! (l+s)! }{ (2l)! (2l+1)! } \bigg)^2  \prod_{n=1}^{l+1/2} \bigg[ 1 + \bigg( \frac{1}{n-1/2} \frac{\omega-m\Omega_+}{2 \pi T_+}  \bigg)^2 \bigg]  \frac{2}{\pi} (A_+ T_+\omega)^{2l+1}, \quad 2l \in \text{odd} , \\
\end{split} \label{Gamma_b-f}
\eeqa
\end{widetext}
where we have to take into account that $2l$ can be an even or odd integer, which corresponds to $l$ being an integer or half integer integer, separately. We have added the subscript $(l,m)$ to indicate that $\Gamma^s$ depends on the angular orbital and magnetic quantum number.
On the right side of the formula, we have replaced the dimensionless variable $k_0$ by observables by substituting Eq.~(\ref{omega_k0}).
The common factor in front of the absorption probability formula for both bosons and fermions is 
\beqa
2^l l! \frac{ (l-s)! (l+s)! }{ (2l)! (2l+1)! }  \overset{l=s}{=} \frac{2^s s!}{(2s+1)!} .
\eeqa
The maximal value is achieved at $l=s$ mode, which is the dominant contribution to the absorption probability. Since $l \ge s$, it decays rapidly when $l \gg s$. 
The first row corresponds to the bosons, and the second line corresponds to fermions. It is clear the absorption probability for bosons is negative, which means the bosonic fields (particles or waves) are amplified rather than absorbed at low frequency. However, the absorption probability for femions is always positive, which implies there is no superradiance for fermions at low frequencies.
We can also reexpress the gray body factor in a form
\begin{widetext}
\beqa
\begin{split}
	\Gamma^s_{lm} 
	& = \bigg( 2^l  \frac{ (l-s)! (l+s)! }{ (2l)! (2l+1)! } \bigg)^2 \abs{ \Gamma \left(l + 1  +\frac{i (\omega -m \Omega_+)}{2 \pi  T_+}\right) }^2  \frac{2}{\pi } (A_+ T_+\omega)^{2l+1} \sinh\bigg(\frac{\omega-m\Omega_+}{2T_+}\bigg) , \quad 2l \in \text{even}, \\
	& = \bigg( 2^l  \frac{ (l-s)! (l+s)! }{ (2l)! (2l+1)! } \bigg)^2 \abs{ \Gamma \left( l+1+ \frac{i (\omega -m \Omega_+)}{2 \pi T_+}\right) }^2  \frac{2}{\pi} (A_+ T_+\omega)^{2l+1} \cosh\bigg(\frac{\omega-m\Omega_+}{2T_+}\bigg) , \quad 2l \in \text{odd}.\\
\end{split} \label{Gamma_b-f_2}
\eeqa
The $\sinh{x}$ and $\cosh{x}$ are odd and even functions, respectively. For bosons, $\Gamma^s_{lm}$ reverses the sign when $\omega$ crosses the resonance point $m\Omega_+$. For fermions, $\Gamma^s_{lm}$ is always positive. Thus, there is no superradiance for fermions.
To be concrete, let's consider the scalar field ($s=0$), neutrino field ($s=1/2$), electromagnetic wave ($s=1$), Rarita-Schwinger field (vector spinor) ($s=3/2$), and gravitational wave ($s=2$), separately. The low-frequency absorption probability for these fields is
\beqa
\begin{split}
	 \Gamma^0_{0m} &  = \frac{A_+}{\pi} (\omega - m \Omega_+) \omega   , \\
	 \Gamma^{1/2}_{1/2 m} &  = \bigg[ (M^2-a^2+n^2) + \bigg( \frac{A_+}{2\pi} \omega - 2 am \bigg)^2 \bigg] \omega^2 , \\
	 \Gamma^1_{1 m} & = \frac{4A_+}{9\pi} \bigg[ M^2+n^2 - a^2 + \bigg( \frac{A_+}{4\pi} \omega -  am \bigg)^2 \bigg] (\omega-m\Omega_+)   \omega^{3} , \\
	 \Gamma^{3/2}_{3/2 m} &  = \frac{1}{8} \bigg[ M^2+n^2-a^2 + \bigg( \frac{A_+\omega}{ 2\pi} - 2 am \bigg)^2 \bigg] \bigg[ M^2+n^2-a^2 + \bigg( \frac{A_+\omega}{6 \pi} - \frac{2am}{3}\bigg)^2 \bigg] \omega^4 , \\
	 \Gamma^2_{2 m} & = \frac{16}{225} \frac{A_+}{\pi} \bigg[ M^2 + n^2 - a^2 + \bigg( \frac{A_+\omega}{4 \pi} - am \bigg)^2 \bigg] \bigg[ M^2 + n^2 - a^2 + \bigg( \frac{A_+\omega}{8 \pi } - \frac{1}{2}am \bigg)^2 \bigg]  (\omega-m\Omega_+) \omega^5 , \\
	\end{split}
\eeqa
\end{widetext}
where $A_+$ is the area of the outer horizon as defined in Eq.~(\ref{A_pm}). It contributes to the BH entropy of the KTN black hole, according to Eq.~(\ref{S_pm}).
We have also used that
\beqa
 A_+ \Omega_+  = 4\pi a , \quad A_+ T_+ 
 = 2 \sqrt{M^2 + n^2 - a^2} .
\eeqa
By keeping the lowest order terms in $\omega$, except for the $\omega-m\Omega_+$ factor for bosons, which guarantees that the superradiant condition $\omega < a\omega_+$ satisfies, we obtain the absorption probability at ultra low frequency as
\beqa
\begin{split}
	 \Gamma^0_{0m} &  = \frac{A_+}{\pi} (\omega - m \Omega_+) \omega  , \\
	 \Gamma^{1/2}_{1/2 m}  & \overset{\omega \ll 1}{=} [ M^2 + n^2 + (4m^2-1)a^2 ] \omega^2 , \\
	 \Gamma^1_{1 m} 	 & \overset{\omega\ll 1}{=} \frac{4}{9}  [ M^2+n^2 + (m^2-1) a^2 ]    \omega^{2} \Gamma^0_{0m} ,  \\
	 \Gamma^{3/2}_{3/2 m} 	& \overset{\omega\ll 1 }{=} \frac{1}{8}  [ M^2+n^2 + (4m^2/9-1) a^2  ] \omega^2 \Gamma^{1/2}_{1/2 m} , \\
	 \Gamma^2_{2 m} ,
	 & \overset{\omega \ll 1}{=} \frac{4}{25}  [ M^2 + n^2 + (m^2/4-1) a^2 ]   \omega^{2}  \Gamma^1_{1 m} . \\
\end{split}
\eeqa
To compare the relative amplitudes of the absorption probability, we have expressed that of higher spin fields with that of lower spin fields.
The absorption probability of the Rarita-Schwinger wave (spin-$3/2$) is expressed in terms of that of the neutrino wave (spin-$1/2$). The absorption probability of the gravitational wave (spin-$2$) is expressed in terms of that of an electromagnetic wave (spin-$1$), which is expressed in terms of that of a scalar wave (spin-$0$).
We also obtain a new result for the massless Rarita-Schwinger field. In contrast to the case of the bosons, i.e., integer spin perturbation fields, the lack of the factor $(\omega-m\Omega_+)$ in the gray body factor indicates that fermions, i.e., half-integer spin perturbation fields, do not exhibit superradiance.

\subsubsection{Absorption cross section}

With the analytic absorption probabilities at low frequencies for the various angular modes for spin-$s$ weighted massless boson and fermion fields, including scalar, neutrino, electromagnetic, Rarita-Schwinger, and gravitational waves in the KTN black hole, we can obtain the low-frequency, i.e., $M\omega\ll 1$ absorption cross section for a massless particle of spin $s$ averaged over all orientations of the spacetime~\footnote{
If $\Gamma^s$ is independent of $m$, then we have
\beqa
\sigma^s = \frac{\pi}{\omega^2} \sum_{lm}\Gamma^s_{lm} = \frac{\pi}{\omega^2} \sum_{l=s}^\infty (2l+1) \Gamma^s_l .
\eeqa
To obtain the total absorption cross section, we can sum over all of the spin states by multiplying a factor of $g_s=2s+1$, which accounts for the spin degrees of freedom. For each spin-weighted species, e.g., the neutrino, one also needs to take flavor into account.
}
\beqa
\sigma^s  = \sum_{l=s}^\infty \sigma^s_l ,
\eeqa
where the $l$-mode of the partial wave cross section of the spin-$s$ particle is
\beqa
\sigma^s_{l} =  \sum_{m=-l}^l \sigma^s_{lm} = \frac{\pi}{\omega^2} \sum_{m=-l}^l \Gamma^s_{lm} .
\eeqa
The dominant $l=s$ mode contributes to the absorption cross section as
\begin{widetext}
\beqa
\begin{split}
\sigma^0_0(\omega) & = A_+ , \\
\sigma^{1/2}_{1/2}(\omega) & = 2\pi (M^2 + n^2) + \frac{A_+^2}{2\pi}\omega^2, \\
\sigma^1_1(\omega) & = \frac{4}{3}A_+ [ (M^2 + n^2 + a^2) + 
\frac{A_+^2 }{16\pi^2} \omega^2   ] \omega^2 , \\
\sigma^{3/2}_{3/2}(\omega) & = \frac{\pi}{18} \bigg( (M^2+n^2) [9(M^2+n^2)+32a^2] +\frac{5A_+^2}{2\pi^2}(M^2+n^2+2a^2)\omega^2 + \frac{A_+^4}{16\pi^4}\omega^4 \bigg) \omega^2 , \\
\sigma^2_2(\omega) & = \frac{8}{45}A_+ \bigg( [2(M^2+n^2)^2 + 11(M^2+n^2)a^2 + 4a^4] + \frac{5A_+^2}{32\pi^2}(M^2+n^2+3a^2) \omega^2 + \frac{A_+^4}{512\pi^4}\omega^4 \bigg) \omega^4 . \\
\end{split}
\eeqa
\end{widetext}
It's worth noticing that at low frequency, the absorption cross section for the bosons is proportional to the area of the black hole $A_+$ in Eq.~(\ref{A_p}). It can be expressed more explicitly as
\beqa
A_+ = 8\pi [  M^2 + n^2 - an C + M \sqrt{M^2-a^2+n^2}  ]  . \label{A_p}
\eeqa
It's clear that the cross section depends not only on the NUT parameter but also on the MR parameter $C$ which describes the position of the defect along the polar axis.
In the Schwarzschild limit, i.e., $a=n=0$, the dominant absorption cross section for $s=l=0$ reduces to the correct result for Schwarzschild black holes at low frequency as
\beqa
\sigma^0_0 =  4 \pi (2M)^2.
\eeqa
The cross sections at low frequencies $M\omega \ll 1$ are smaller than those at high frequencies $M\omega\gg 1$.~\footnote{At high frequencies $M\omega \gg 1$, the (angle-averaged) cross section for each species of particle approaches the geometric optical limit as
\beqa
\sigma^0(\omega) \overset{\omega \gg 1}{=} \frac{27}{4} \pi [ 2M^2 + 2M\sqrt{M^2+n^2-a^2} + n^2 - a^2  ]  .
\eeqa
The high-frequency absorption cross section is independent of MR parameter $C$. In the Schwarzschild limit, i.e., $a=n=0$, the high-frequency cross section is reduced to $\sigma^0_0 = 27\pi M^2$~\cite{Sanchez:1976fcl,Sanchez:1977si}.}
While for fermions, the absorption cross section is independent of the horizon area or entropy of the black hole. As frequencies go to zero, the absorption cross section for both electromagnetic waves and gravitational waves is zero, but that for massless scalar particles and neutrino particles remains at finite values. The cross sections go to zero as the frequency goes to the second power for photons as well as Rarita-Schwinger fermions, and to the fourth power for gravitons.

\subsubsection{Emission rate of Hawking radiation}

By combining the low-frequency absorption probability ${\mathrm A}^s = -\Gamma^s$ with the thermal factor $n_s(\omega,\omega_+,T_+)$ for a black hole with finite temperature, we can obtain the expected particle number (of the spin-$s$ species) with
\beqa
\langle N^s_{lm} \rangle = \Gamma^s_{lm} n_s  ,
\eeqa
where $n_s$ is the thermal statistical distribution of the spin-$s$ particle, which can be expressed in a unified form as
\beqa
n_s 
 =  \frac{1}{e^{(\omega-\omega_+)/T_+} - (-1)^{2s} } .
\eeqa 
$T_+$ is the Hawking temperature in Eq.~(\ref{T_pm}), and $\omega_+$ is the critical angular frequency at the event horizon in Eq.~(\ref{omega_pm}), which is related to the horizon angular velocity $\Omega_+$ and horizon electrostatic potential $\Phi_+$.
In the low-frequency limi $\omega\ll 1$,
\beqa
\begin{split}
& n_b(\omega) = \frac{1}{e^{(\omega-\omega_+)/T_+}-1} \overset{\omega \ll 1}{\approx} \frac{T_+}{\omega} - \frac{1}{2} + O(\omega) \ \\
& n_f(\omega) = \frac{1}{e^{(\omega-\omega_+)/T_+}+1} \overset{\omega \ll 1}{\approx} \frac{1}{2} + O(\omega) . \\
\end{split}
\eeqa
The minus and plus signs in the statical distribution for bosons ($b$) and fermions ($f$) are used, separately. For massive particles ($\omega \sim m \gg 1$), the statistical distribution is the Boltzmann one.
The averaged emission rate per frequency interval $d\omega$ is related to the expected particle numbers through
\beqa
\frac{dN^s_{lm}}{dt} = \vev{N^s_{lm}} \frac{d\omega}{2\pi} .
\eeqa
where $2\pi$ is a phase space factor.
The emission rate is 
\beqa
\frac{dN^s_{lm}}{dtd\omega} = \frac{1}{2\pi} \vev{N^s_{lm}} = \frac{\Gamma^s_{lm}}{2\pi} n_s , \label{dN_dt-domega}
\eeqa
where $\Gamma^s_{lm}$ is the gray body factor, and at low frequencies it is given in Eq.~(\ref{Gamma_b-f}).
For slow rotating (i.e., $a\ll 1$) or non-rotating black hole (i.e.,$a=0$), $\Omega_+\approx 0$, $A_+ \Omega_+\approx 0$ and $A_+ T_+ \approx 2\sqrt{M^2+n^2}$, the Hawking temperature is $T_+ = [4\pi(M+\sqrt{M^2+n^2})]^{-1} \ll 1 $. Thus, the emission rate at low frequencies for both bosons and fermions is given by
\beqa
&& \frac{dN^s_{lm}}{dtd\omega}  = \frac{1}{2\pi^2} \bigg( 2^l l! \frac{ (l-s)! (l+s)! }{ (2l)! (2l+1)! } \bigg)^2  ( 2 \sqrt{M^2+n^2}\omega)^{2l+1}, \nn \\
\label{dN_dt-domega}
\eeqa
where we have approximated as below
\beqa
\begin{split}
& \prod_{n=1}^l \bigg[ 1 + \bigg( \frac{1}{n} \frac{\omega}{2 \pi T_+} \bigg)^2 \bigg]  \overset{\omega \ll 1}{\approx} 1 + O(\omega^2) , \\
& \prod_{n=1}^{l+1/2} \bigg[ 1 + \bigg( \frac{1}{n-1/2} \frac{\omega}{2 \pi T_+}  \bigg)^2 \bigg] \overset{\omega \ll 1}{\approx} 1 + O(\omega^2). \\
\end{split}
\eeqa
This means at low frequencies, the emission rate goes as $\omega^{2l+1}$, thus the power $dN^s_{lm}/dt $ goes as $\omega^{2l+2}$. Consequently, the lower spin weighted particle (thus lower $l$ allowed since $l\ge s$) was emitted faster from a non-rotating black hole, thus dominating the low-frequency power drain from the black hole.

\subsubsection{Power spectrum}

The scattering of spin-$s$ particles will carry off energy $\epsilon = \hbar\omega$ and angular momentum $j=m$ about the axis of the black hole.
According to the averaged emission rate, the mass and angular momentum of the black hole decrease at the rate given by the total energy and torque emitted as
\beqa
\frac{d}{dt} \bpm
M \\
J \\
\epm = - \sum_{l,m} \int  \frac{dN^s_{lm}}{dt} \bpm
\omega \\
m \\
\epm ,
\eeqa
where $l,m$ are the angular momentum and its azimuthal magnetic component, and $\omega$ is the energy of the massless particle.
For the Hawking radiation of rotating spacetime, the emission rates of energy and angular momentum are~\footnote{
For KNTN, $\omega_+ = m \Omega_+ + e \Phi_+$, where $\Phi_+$ is the electric static potential on the event horizon. If the electrostatic potential is large enough, the dominant component of the emission becomes the superradiant discharge process. In this case, the upper bound of the integral over frequency should be replaced by $e\Phi_+$.}
\beqa
\begin{split}
	\frac{d}{dt} \bpm
M \\
J \\
\epm & = - \frac{1}{2\pi}\sum_{l,m} \int_0^\infty d\omega \Gamma^s_{lm} n_s 
\bpm
\omega \\
m \\
\epm , 
\end{split} \label{dM_dJ}
\eeqa
where $\Gamma^s$ is the emission rate in Eq.~(\ref{dN_dt-domega}). It is related to the amplification factor ${\mathrm A}^s$ and reflection coefficient ${\mathrm R}^s $ at the infinite boundary through Eq.~(\ref{A_R_Gamma}). It can be obtained by solving the radial Teukolsky equation.
$\omega_+ = m \Omega_+ $ is the critical angular frequency in Eq.~(\ref{omega_pm}). $\Omega_+$ is the angular velocity on the event horizon. $T_+$ is the Hawking temperature in Eq.~(\ref{T_pm}).
\begin{figure}[h!]
  \includegraphics[width=8cm]{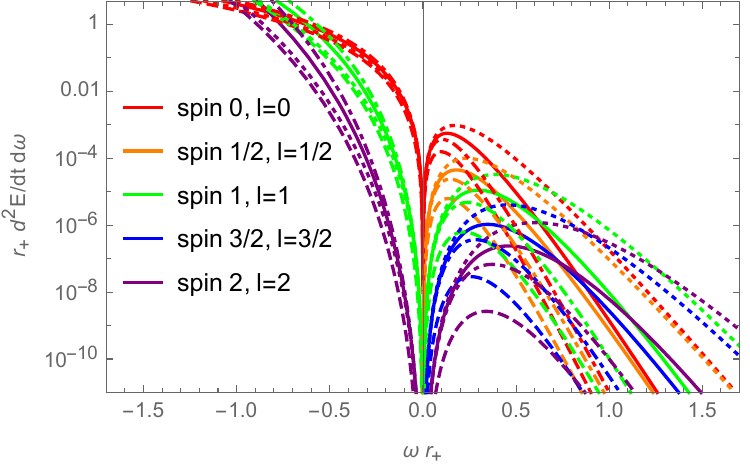}
  \caption{Energy power spectrum for particles with spin
     \label{graph2} }
\end{figure}
In the extreme limit, the Hawking temperature $T_+$ vanishes, and Hawking radiation becomes purely superradiant, i.e., ${\mathrm A}^s>0$, that is, when $\omega < \omega_+$ and $T_+ \to 0$. In this case, the emission rates are
\beqa
\begin{split}
	\frac{d}{dt} \bpm
M \\
J \\
\epm & = (-1)^{2s} \frac{1}{2\pi}\sum_{l,m} \int_0^\infty d\omega  ({\mathrm R}^s-1) \bpm
\omega \\
m \\
\epm .
\end{split}
\eeqa
There is a minus sign or a plus sign corresponding to the boson and fermion cases, separately. This indicates that the superradiant scattering of bosons leads to a decrease in the total mass and angular momentum of the KTN spacetime. 
Once we obtain the absorption probability (or the graybody factor) $\Gamma^s$, we can calculate the angular dependent power spectrum of the Hawking radiation for each massless spin-$s$ particle as
\beqa
\frac{d^2 E^s_{lm}}{ dt  d\Omega } = \frac{1}{2\pi} \Gamma^s_{lm} n_s |S^s_{lm}(c,\theta,\phi)|^2 \omega d\omega , 
\eeqa
where $d\Omega=d\phi d\cos\theta$ is the infinitesimal solid angle, $\omega_+ = m \Omega_+$ is the critical angular frequency, $\Omega_+$ is the angular velocity at the horizon. $\Gamma^s_{lm}$ is the graybody factor, which is identical to the absorption probability of the incoming wave of the corresponding modes carrying angular orbital and magnetic quantum numbers. $S^s_{lm}$ is a spin-weighted spheroidal function.
By integrating over the angular variables $(\theta,\phi)$, we obtain the angle-averaged energy power spectrum as
\beqa
\frac{d^2 E^s_{lm}}{dt d\omega} = \frac{1}{2\pi} \Gamma^s_{lm} n_s \omega. \label{dE_dt-domega}
\eeqa
At low frequencies, the power spectrum behaves as $\propto \omega^{2l+2}$.
The energy power spectrum for massless scalar, neutrino, photon, Rarita-Schwinger field, and gravitons is shown in Fig.~\ref{graph2}.
We have chosen input parameters $M=1$, $n=0.5$, and $a=0.9$ (rapidly rotating), and set $m=0$ so that $d^2J/dtd\omega=0$.
As has been discussed before, the MR parameter becomes irrelevant in the far region. The effect of $C$ is only incorporated through the definition of the Hawking temperature $T_+$ in Eq.~(\ref{T_pm}), as well as the angular velocity in Eq.~(\ref{Omega_pm}).

From bottom to top, it corresponds to the gravitational wave (purple), the Rarita-Schwinger wave (blue), the electromagnetic wave (green), the neutrino wave (orange), and the scalar wave (red), respectively. The dashed line corresponds to reference ones in the Kerr black hole, separately.
In order to facilitate comparison, in the same figure, we have also plotted the energy power spectrum with different MR parameters: $C=0$ (solid lines), $C=+1$ (dotted lines), and $C=-1$ (dot-dashed lines), separately.

The figure shows that the energy power spectrum with the $C=-1$ case is larger than that with the $C=0$ and $C=+1$ cases.
For three cases, the energy power spectrum for each massless spin-$s$ wave in the KTN spacetime is enhanced than that in the Kerr black hole.

In particular, they have different behaviors for $\omega>0$ (co-rotating) and $\omega<0$ (counter-rotating) regions. It is clear that the superradiance condition is always fulfilled in the counter-rotating region. Only scalar, electromagnetic, and gravitational waves are amplified.

The figure also verifies that there is no superradiance for fermions, i.e., half-integer spin perturbation fields do not exhibit superradiance.
Intuitively, the superraidnce means boson fields in a rotating spacetime will gain extra energy, given that the fields are counter-rotating with respect to the rotation direction of the spacetime.

In the case that $S^s_{lm}(c,\theta,\phi)$ are frequency independent, i.e., $c=0$, the spin-weighted spheroidal harmonics $S^s_{lm}$ reduce to spin-weighted spherical harmonics $Y^s_{lm}$. By integrating over the angular variable $\phi$, we obtain the angular spectrum as
\beqa
\frac{d^3 E^s_{lm}}{dt d\cos\theta d\omega} = \frac{1}{2\pi} \Gamma_{lm}^s n_s  I^s_{lm}(\cos\theta) \omega ,
\eeqa
where $\theta\in[0,\pi]$ is the polar angle, and $\theta=0,\pi$ stands for forward and backward scattering, respectively. $s = 0, 1/2, 1, 3/2, 2$, $l = 0, 1/2, 3/2, 1, 2$, and $m = -l, \ldots, l$. $I^s_{lm}(\cos\theta)$ are polar angle distribution functions, after integrating over azimuthal angle as
\beqa
I^s_{lm}(\cos\theta) = \int_0^{2\pi} d\phi |Y^s_{lm}(\cos\theta,\phi)|^2 . \label{Islm}
\eeqa 
It stands for angular (spin, orbital, and azimuthal) dependent probability of wave function with quantum numbers $(s,l,m)$ for $s = 0, 1/2, 1, 3/2, 2$, $l = 0, 1/2, 3/2, 1, 2$, and $m = -l, \ldots, l$.
With the explicit form of the spin-weighted spherical harmonics, as listed in the Appendix~\ref{app:Islm}, we can exactly integrate out the integral $I^s_{lm}$.
For scalar wave ($l=s=0$), $Y^s_{0m}$ is ordinary scalar spherical harmonics, we have
\beqa
I^0_{00} = \frac{1}{2}.
\eeqa
For neutrino wave ($l=s=1/2$), $Y^{s}_{lm}$ is spinor-valued spherical harmonics, we have
\beqa
\begin{split}
& I^{\pm 1/2}_{1/2,\mp 1/2} = \cos^2\frac{\theta}{2}, \quad I^{\pm 1/2}_{1/2,\pm 1/2} = \sin^2\frac{\theta}{2} . 
\end{split}
\eeqa
For electromagnetic wave ($l=s=1$), $Y^{s}_{lm}$ is vector spherical harmonics, we have
\beqa
\begin{split}
& I^{-1}_{1,-1} = I^{1}_{1,1} = \frac{3}{2}\sin^4\frac{\theta}{2}, \quad I^{-1}_{1,1} = I^{1}_{1,-1} = \frac{3}{2}\cos^4\frac{\theta}{2}, \\
& I^{0}_{1,0} = \frac{3}{2}\cos^2\theta, \quad I^{-1}_{1,0} = I^{0}_{1,-1} = I^{0}_{1,1} = I^{1}_{1,0} = \frac{3}{4}\sin^2\theta. \\
\end{split}
\eeqa
For Rarita-Schwinger field ($l=s=3/2$), $Y^{s}_{lm}$ is spinor-valued vector spherical harmonics, we have
\beqa
\begin{split}
 I^{-3/2}_{3/2,-3/2} & = I^{3/2}_{3/2,3/2} = 2 \sin ^6\frac{\theta }{2}, \\
  I^{-3/2}_{3/2,3/2} & = I^{3/2}_{3/2,-1/2} = 2 \cos ^6\frac{\theta }{2} ; \\
 I^{-3/2}_{3/2,-1/2} & = I^{3/2}_{3/2,1/2} = I^{1/2}_{3/2,3/2} = I^{-1/2}_{3/2,-3/2} \\
 & = \frac{3}{2} \sin ^2\frac{\theta}{2} \sin ^2\theta, \\
 I^{-3/2}_{3/2,1/2} & = I^{3/2}_{3/2,-1/2} = I^{3/2}_{3/2,-3/2} = I^{-1/2}_{3/2,3/2} = I^{1/2}_{3/2,-3/2} \\
 & = \frac{3}{2}  \cos^2\frac{\theta}{2} \sin^2\theta, \\
 I^{-1/2}_{3/2,-1/2} & = I^{1/2}_{3/2,1/2} = \frac{1}{2} \sin ^2\frac{\theta }{2} (3 \cos \theta+1)^2, \\
 I^{-1/2}_{3/2,1/2} & = I^{1/2}_{3/2,-1/2} = \frac{1}{2} \cos^2\frac{\theta }{2} (3 \cos \theta-1)^2. \\
\end{split}
\eeqa
For gravitational wave ($l=s=2$), $Y^{s}_{lm}$ is tensor spherical harmonics, we have
\beqa
\begin{split}
&	I^{-2}_{2,-2} = I^{2}_{2,2} = \frac{5}{2} \sin ^8\frac{\theta }{2}, \\
&   I^{-2}_{2,2} = I^{2}_{2,-2} = \frac{5}{2} \cos ^8\frac{\theta }{2} ; \\
&   I^{-2}_{2,-1} = I^{2}_{2,1} = I^{-1}_{2,-2} = I^{1}_{2,2} = \frac{5}{2} \sin^4\frac{\theta }{2} \sin^2\theta, \\
&   I^{-2}_{2,0} = I^{2}_{2,0} = I^{0}_{2,-2} = I^{0}_{2,2} = \frac{15}{16}\sin^4\theta, \\
&   I^{-2}_{2,1} = I^{2}_{2,-1} = I^{-1}_{2,2} = I^{1}_{2,-2} = \frac{5}{2}  \cos^4\frac{\theta}{2} \sin ^2\theta , \\
&   I^{-1}_{2,-1} = I^{1}_{2,1} = \frac{5}{2} \sin^4\frac{\theta }{2} (2 \cos \theta+1)^2, \\
&   I^{-1}_{2,0} = I^{1}_{2,0} = I^{0}_{2,-1} = I^{0}_{2,1} = \frac{15}{16} \sin ^2(2\theta ), \\
&   I^{-1}_{2,1} = I^{1}_{2,-1} = \frac{5}{2} \cos ^4\frac{\theta }{2} (2 \cos \theta-1)^2, \\
&   I^{0}_{2,0} = \frac{5}{32} (3\cos (2 \theta )+1)^2 . \\
\end{split}
\eeqa
All of the angular distribution functions $I^s_{lm}(\cos\theta)$ satisfy the normalization condition:
\beqa
\int_{0}^{\pi} \sin\theta d\theta I^s_{lm}(\cos\theta) = 1.
\eeqa
The angular distribution $I^s_{lm}(\cos\theta)$ stands for the angular dependence of the scattering probability of spin-$s$ waves.

\section{Summary}
\label{sec:sum}

In this paper, we generalized the ENJA from the spherical, symmetric spacetime metric to an axial, symmetric metric.
The method is generic and applicable to a general axial symmetric metric, such as the Taub-NUT and Brill metric.
By applying the methods, we obtained the metric functions of the Kerr-Newman-Taub-NUT (KTN) metric from the Brill metric.
The null tetrad of the metric in both EF and BF coordinate systems is expressed in the NP formalism.
Black hole (BH) thermal dynamical variables, including entropy/area, temperature/surface gravity, angular velocities/frequencies, and electrostatic potential, are investigated.
The NUT parameter and the MR parameter determine the amplitude and rotating directions of the wireline defect.
We study the superradiance of (massive or massless) particles in a (charged or neutral) Kerr-Taub-NUT spacetime.

We first studied the superradiance of a charged massive scalar field in KNTN spacetime. Its dynamics are described by the relativistic Klein-Gordon equation. We decompose the wave function into angular and radial parts through variable separation methods, and the angular and radial equations turn out to be exactly solvable.
We obtain exact solutions to both angular and radial equations in KTN spacetime. The corresponding solutions can be expressed in terms of Heun confluent functions, so that it is possible to carry out an accurate calculation. The eigenvalues are determined by solving angular equations.
By studying the behaviors of radial wave functions in the extremal limit, we can calculate the reflection coefficients and the absorption probability (or gray body factor) at low frequencies. The radial wave function is reformulated in the form of the RWZ equations, and we analyze the near-horizon and infinite boundary or large distance behaviors. We give the superradiance amplification factor of charged massive scalar fields in KNTN spacetime and illustrate the superradiance condition for the massive scalar field.

Then we studied the superradiance of neutral massless particles with spin in KTN spacetime in the NP formalism. It includes the spherical, symmetric metric as a special case. We gave a generalized Kinnersley tetrad that is regular on the past horizon.
We calculated the complex Weyl scalars, which contain information about background spacetime. One of them is related to optical scalars, such as expansion, twist, shear, etc. The NUT parameter has an interpretation as the twist of the geodesic congruence of null rays in a symmetric axial rotational spacetime.
We derived generalized Teukolsky master equations for all spin particles in KTN spacetime. Our approach is applicable to more generic axial-symmetric stationary spacetime.
The perturbation field equations govern the superradiant scattering of spin particles in KTN spacetime. The spin particles include scalar waves, neutrinos, electromagnetic waves, the Rarita-Schwinger field, and gravitational radiation.
The equation is separated into two parts with angular and radial coordinate dependence, separately.
The angular equations result in generalized spin-weighted spheroidal harmonics. To describe the relation among the NUT parameter, frequency, and spin, we make a new notation named ``effective spin''. The quantization of the effective spin recovers Misner's quantization condition in general relativity, in analogy to Dirac's quantization condition in electromagnetics.
We also reformulated the radial equation into the form of a RWZ equation with an effective potential for all spin particles. We analyze the near-horizon and infinite boundary asymptotic behavior of independent radiating components of spinning particles.
We studied the near-horizon and infinite boundary behaviors of the radial wave function.

In the end, we provided analytical expressions of universal low-energy dynamical observables, such as amplification factor, absorption probability (or gray body factor), emission rate, power spectrum, cross section, etc.
We verified that there is no superradiance for the massless fermion perturbation field.
We showed that the energy power spectrum for each massless spin particles in the KTN spacetime is larger than that in the Kerr black hole.
We obtained an exact result on the low-frequency absorption cross section for all spin massless fields, including scalar, neutrino, electromagnetic, Rarita-Schwinger, and gravitational waves.
It is found that for all spin particles in the KTN spacetime, the superradiant absorption cross sections are enhanced compared to those in the Kerr spacetime.

\begin{acknowledgments}
We appreciate Sunggeun Lee's early collaboration and valuable discussion of the work.
We would like to thank Yun Soo Myung, Hongsu Kim, and Dong-Han Yeom for their valuable discussions.
The Basic Science Research Program supports B.-H.~Lee (NRF-2020R1F1A1075472) and W.~Lee (NRF-2022R1I1A1A01067336) through the National Research Foundation of Korea, funded by the Ministry of Education.
The National Research Foundation of Korea supports Y.-H.~Qi (Grant No. 2020R1A6A1A03047877) through the Center for Quantum Spacetime (CQUeST) research center to carry on fundamental research in the field of fundamental physics.
He would like to thank Peng-Ming Zhang (National Natural Science Foundation of China under Grant No. 11975320) for the hospitality during his visit to the School of Physics and Astronomy, Sun Yat-sen University, Zhuhai.

\end{acknowledgments}

\appendix
%
%
%
%
%
%
%
%
%
%
%

\section{Angular wave functions}
\label{app:angular}

\subsection{Spin-weighted spheroidal harmonics}

By making a variable transformation, we can rewrite the angular equations in Eq.~(\ref{eom_S_2}) as
\beqa
\begin{split}
& \frac{d}{dz}\bigg( (1-z^2) \frac{d}{dz} S \bigg) + \bigg( c^2 z^2 + 2c(\bar{s}-s)z \\
& + \bar\lambda + s - 2 s \bar{s} + 2 a \omega \bar{s}  C - \frac{[\bar{m}+(s+\bar{s})z]^2}{1-z^2} \bigg) S =0 , \\
\end{split}
\eeqa
where $z=\cos\theta$ and $\theta$ is the polar angle. $\bar\lambda = \lambda - c^2 + 2 c m $ with $c = a\omega$, according to Eq.~(\ref{lambda-lambda_b-lambda_t_massive}).
When $n=0$ (i.e., $\bar{s}=0$ and $\bar{m}=m$), the angular equation recovers that for spin-weighted spheroidal harmonic wave functions as
\beqa
\begin{split}
& \frac{d}{dz}\bigg( (1-z^2) \frac{d}{dz} S \bigg) + \bigg( (c z)^2 - 2sc z \\
& + \bar\lambda + s - \frac{(m+sz)^2}{1-z^2} \bigg) S =0 , \\
\end{split} \label{eom_angular_spheroidal_s}
\eeqa
where $\bar\lambda$ is the angular separation constant, and $m$ is the azimuthal separation constant.
The equation is essentially a two-parameter eigenvalue equation with boundary conditions that $S$ is regular for $z \in [-1, 1]$. The two parameters are $\lambda(m,s)$ and $c$.
The generic solution to the angular equation is the spin-weighted spheroidal harmonic function as
\beqa
\begin{split}
Z_{lm}^s(c,\theta,\phi) 
= \sqrt{\frac{2l+1}{4\pi}\frac{(l-m)!}{(l+m)!}}  \text{\textit{PS}}^s_{lm}(c,\cos\theta) e^{ im\phi} 
,  
\end{split} \quad\label{Zs_omega_lm}
\eeqa
where $c \equiv a\omega$ indicating the frequency appears in the angular dependence.
The spheroidal wave function $S^s_{lm}(c,z)$ is oblate if $c = c_R$ is real, or prolate if $c = i c_I$ is pure-imaginary. In general, if $c \ne 0$, we would expect that the generalized spin-weighted spheroidal functions can be perturbatively expanded. At leading order, it is the spin-weighted spheroidal harmonics with corresponding eigenvalues and obtain corrections from $ a \omega $.
The spin-weighted spheroidal wave functions satisfy the orthonormal condition as
\beqa
\int_0^{2\pi} \int_0^\pi |S^{s+\bar{s}}_{lm}|^2 \sin\theta d\theta d\phi = 1.
\eeqa
When $c=0$, the (spin-weighted) spheroidal function reduces to (spin-weighted) spherical harmonics, and the eigenvalue reduces to the orbital angular momentum eigenvalue $l(l+1)$.
The spheroidal functions can be calculated by using recurrence relations.

When $s = 0$, Eq.~(\ref{eom_angular_spheroidal_s}) recovers the spheroidal equation as 
\beqa
\begin{split}
\frac{d}{dz}\bigg( (1-z^2) \frac{d}{dz} S \bigg) + \bigg( (c z)^2  + \bar\lambda  - \frac{m^2}{1-z^2} \bigg) S =0 . \\
\end{split} \quad \label{eom_angular_spheroidal}
\eeqa
The equation has regular singularities at $z = \pm 1$ with exponents $\pm m/2$ and an irregular singularity of rank $1$ at $z = \infty$ if $c \ne 0$. If $c^2>0$ (i.e., $c$ is real), the equation is called the oblate spheroidal equation. If $c^2<0$ (i.e., $c$ is pure imaginary), the equation is called a prolate spheroidal equation. When $c=0$ and $\abs{z}<1$, the equation reduces to the Legendre differential equation in Eq.~(\ref{Plm}).
If $|z|<1$, the equation in Eq.~(\ref{eom_angular_spheroidal}) is an angular spheroidal equation.
If $|z|>1$, Eq.~(\ref{eom_angular_spheroidal}) is a radial spheroidal equation. The solution can be expressed in terms of the radial spheroidal wavefunctions.
When $s=0$, $Z^s_{lm}(c,\theta,\phi)$ in Eq.~(\ref{Zs_omega_lm}) reduces to the ordinary spheroidal harmonic function $Z_{lm}(c,\theta,\phi)$.
\beqa
\begin{split}
Z_{lm}(c,\theta,\phi) 
= \sqrt{\frac{2l+1}{4\pi}\frac{(l-m)!}{(l+m)!}} \text{\textit{PS}}_{lm}(c,\cos\theta) e^{ im\phi}  
.
\end{split} \quad \label{Z_omega_lm}
\eeqa
The angular wave function can be expressed in terms of two linearly independent functions as
\beqa
S(z) = c_1 \text{\textit{PS}}_{lm}(ic ,z)+c_2 \text{\textit{QS}}_{lm}(ic ,z) ,
\eeqa
where $\text{\textit{PS}}_{lm}(c,z)$ and $\text{\textit{QS}}_{lm}(c,z)$ are angular spheroidal functions of the first and second kind, respectively. When $c \ne 0$, the eigenvalues $\lambda(c)$ can be solved numerically.
The angular spheroidal function can be perturbatively expanded with respect to the spheroidal parameter as
\begin{widetext}
\beqa
\begin{split}
\text{\textit{PS}}_{lm}(c,z) & = P_l^m(x) - c^2 \left(\frac{(-m+l+1) (-m+l+2) P_{l+2}^m(x)}{2 (2 l+1) (2 l+3)^2}-\frac{(m+l-1) (m+l) P_{l-2}^m(x)}{2 (2 l-1)^2 (2 l+1)}\right) + O(c^4) , \\
\text{\textit{QS}}_{lm}(c,z) & = Q_l^m(x) - c^2 \left(\frac{(-m+l+1) (-m+l+2) Q_{l+2}^m(x)}{2 (2 l+1) (2 l+3)^2}-\frac{(m+l-1) (m+l) Q_{l-2}^m(x)}{2 (2l-1)^2 (2 l+1)}\right)+ O(c^4) . \\
\end{split}
\eeqa
\end{widetext}
The leading order expansions are ordinary associated Legendre functions. The next-leading order expansion involves associated Legendre functions with $l \pm 2$.

\subsection{Spin-weighted spherical harmonics}

In the $c\to 0$, the spin-weighted spheroidal harmonics $Z^s_{lm}(c,\theta,\phi)$ in Eq.~(\ref{Zs_omega_lm}), reduce to
\beqa
Z^s_{lm}(c, \theta, \phi) = Y^s_{lm}(\theta, \phi) + O(c) , \label{S_Y}
\eeqa
where $c = a \omega$ is the oblateness parameter. The leading order (when $c=0$) recovers the spin-weighted spherical harmonics as
\beqa
\begin{split}
 Y^s_{lm}(\theta,\phi)  & = \sqrt{\frac{2l+1}{4\pi}\frac{(l-m)!}{(l+m)!}}  P^s_{lm}(\cos\theta) e^{ im\phi}   ,  \\
\end{split} \label{Ys_omega_lm}
\eeqa
where $P^s_{lm}$ is called the spin-weighted associated Legendre function. When $s=0$, $P^s_{lm}$ reduces to the ordinary associated Legendre polynomials $P_l^m(z)$. 
The explicit form of the spin-weighted spherical harmonics $Y^s_{lm}$ is related to an orbital angular momentum $l$-weighted representation of the rotation group $SO(3)$. It is the rotator function $D^l_{-sm}(\theta,\phi,\psi)$ and determined to be~\cite{Goldberg1967}
\begin{widetext}
\beqa
\begin{split}
Y^s_{lm}(\theta,\phi) 
& = (-1)^m\sqrt\frac{2l+1}{4\pi}D^l_{-sm}(\theta,\phi,0) \\
& = (-1)^m e^{im\phi} \sqrt{\frac{2l+1}{4\pi}\frac{(l+m)!(l-m)!}{(l+s)!(l-s)!}} \bigg(\sin\frac{\theta}{2}\bigg)^{2l} \sum_{j} \binom{l-s}{j}\binom{l+s}{j+s-m}(-1)^{l-j-s}\bigg( \cot\frac{\theta}{2} \bigg)^{2j+s-m},\\
\end{split}
\eeqa
\end{widetext}
where $\theta,\phi,\psi$ are the Euler angles, the summation over integer $j$ is understood in the regions $j \in [0,l-s]$ and $j+s-m \in [0,l+s]$.
A factor $(-1)^m$ is inserted as the Condon-Shortley phase, so that when $s=0$, it reduces to the standard spherical harmonics $Y^0_{lm}(\theta,\phi)=Y_{lm}(\theta,\phi)$.
The spin-weighted spherical harmonics have the property as
\beqa
\begin{split}
& Y^{-s}_{lm} = (-1)^{m+s} Y^{s\star}_{l-m} . \\
\end{split} \label{Y_star}
\eeqa
They also satisfy the orthogonality relation as well as that for spherical harmonics as
\beqa
\begin{split}
 \sum_{lm} Y^s_{lm}(\theta',\phi')Y^s_{lm}(\theta,\phi) & = \delta(\cos\theta-\cos\theta')\delta(\phi-\phi') , \\
 \int Y^{s\star}_{lm} Y^s_{l'm'} d\Omega & = \delta_{ll'} \delta_{mm'} . \\
\end{split} \label{Yslm_orth}
\eeqa

\subsubsection{Spin-weighted associated Legendre equations}

In the spherical case, i.e., $c = 0$ ($\omega = 0$ or $a = 0$), the angular equation recovers that for spin-$s$ weighted spherical harmonic wave functions as
\beqa
\begin{split}
\frac{d}{dz}\bigg( (1-z^2) \frac{d}{dz} S \bigg) + \bigg( \bar\lambda + s - \frac{(m+sz)^2}{1-z^2} \bigg) S =0 , \\
\end{split} \quad \label{ode_spin-Plm}
\eeqa
where $z=\cos\theta$.
When the eigenvalue is $\bar\lambda = l(l+1) - s(s+1)$, where the first and second terms correspond to the total orbital/spin angular momentum, the solution to the angular equation is spin weighted associated Legendre polynomials as
\begin{widetext}
\beqa
\begin{split}
P_{lm}^s(z) & = 	(z+1)^{\frac{m-s}{2}} \bigg[ c_1  (1-z)^{-\frac{1}{2} (m+s)}  \, _2F_1\left(-l-s,l-s+1;-m-s+1;\frac{1-z}{2}\right)  \\
& + c_2 (1-z)^{\frac{m+s}{2}}  \,_2F_1\left(m-l,l+m+1;m+s+1;\frac{1-z}{2}\right) \bigg] .
\end{split} \label{Plms}
\eeqa
\end{widetext}
We can determine the coefficients $c_{1,2}$ by using the orthogonality condition in Eq.~(\ref{Yslm_orth}).
When $s=0$, the eigenvalue is $\lambda=l(l+1)$, and the equation reduces to a Legendre differential equation as
\beqa
\frac{d}{dz}\bigg( (1-z^2) \frac{d}{dz} S \bigg) + \bigg( \bar\lambda  - \frac{m^2}{1-z^2} \bigg) S =0 , \label{Plm}
\eeqa
and the solution becomes 
\beqa
S_{lm}(z) = c_1 P_{l}^m(z)+c_2 Q_{l}^m(z) , \label{Slm}
\eeqa
where $P_l^m(z)$ is the associated Legendre polynomial and $Q_l^m(z)$ is the associated Legendre's function of the second kind. The first one can be defined through the hypergeometric function as~\cite{Petrich:1988zz}
\beqa
\begin{split}
P_l^{m}(z) & = \frac{  1}{\Gamma(1-m)}  \bigg( \frac{1 + z}{1 - z} \bigg)^{m/2} \\
& \times \, _2F_1\left(-l,l+1;1 - m;{(1-z)}/{2}\right) , \\
\end{split}
\eeqa
where $\Gamma(z)$ is the Gamma function. The second one is a linear combination of $P_l^m$ and $P_l^{-m}$. For regularity consideration, we may drop the second term.

\subsubsection{Jacobi polynomials}
\label{App_JP}

The spin-weighted spherical harmonics are related to and can be expressed as a sum over Jacobi polynomials. This can be viewed since the spin-weighted differential equations in Eq.~(\ref{ode_spin-Plm}) can be re-expressed as
\beqa
\begin{split}
& \frac{d}{dz}\bigg( (1-z^2) \frac{d}{dz}  P^s_{lm}(z)  \bigg) \\
& + \bigg( l(l+1)  -\frac{(m-s)^2}{2 (z+1)} - \frac{(m+s)^2}{2 (1-z)} \bigg) P^s_{lm}(z)  = 0 .  \\
\end{split} \label{ode_spin-Plm-2}
\eeqa
The differential equation has regular singular points at $\pm 1,\infty$. The solutions to the equations above are orthonormal spin-weighted associated Legendre functions in terms of the Jacobi polynomials as
\beqa
P^s_{lm}(z)  = N^s_{lm}(1-z)^{a/2}(1+z)^{b/2} P_n^{(a,b)}(z) ,
\eeqa
where $N^s_{lm}$ is the normalization constant as~\cite{Ripley:2020xby}
\beqa
N^s_{lm} = (-1)^{\text{max}(m,-s)}\sqrt{\frac{2n+a+b+1}{2^{a+b+1}} \frac{n! (n+ a+b)! }{ (n+a)! (n+b)! }  } . \qquad
\eeqa 
The explicit form of the Jacobi polynomials is given by the Rodrigues formula:
\beqa
\begin{split}
 P_n^{(a,b)}(z) & = \frac{(-1)^n}{2^n n!} (1-z)^{-a} (1+z)^{-b} \\
& \times \partial_z^n [ (1-z)^{a + n}(1+z)^{b+n}   ]. \\
\end{split}\qquad
\eeqa
The variables $a$, $b$, and $n$ are all positive integers determined to be
\beqa
\begin{split}
& a = \abs{m+s}, \quad b = \abs{m-s},  \\
& n = l - \bar{l} , ~ -l -1 - \bar{l} ,  \quad \bar{l} \equiv \frac{a+b}{2} . \\
\end{split}
\eeqa
Since for the Jacobi polynomials, we need $n$ to be a positive integer, we have dropped the branches of the solutions with $n=-l-1-\bar{l}$, where $\bar{l} = \pm s, \pm m$, with $\pm$ associated with positive and negative $s$ or $m$, respectively. 

It can be determined by utilizing the orthonormal relations of the Jacobi polynomials as
\beqa
\begin{split}
& \int_{-1}^{+!} dz (1-z)^a (1+x)^b P_m^{(a,b)}(z) P_n^{(a,b)}(z) \\
& = \frac{2^{a+b+1}}{2n+a+b+1}\frac{\Gamma(n+a+1)\Gamma(n+b+1)}{n!\Gamma(n+a+b+1)} \delta_{mn} , \\
\end{split}
\eeqa
so that 
\beqa
\int_{-1}^{+1} dz P^s_{lm}(z) P^s_{l'm}(z) = \delta_{ll'}.
\eeqa
The spin-weighted spherical harmonics are a generalization of the traditional spherical harmonics. It can be expressed in terms of the classical Jacobi polynomials with different parameters in full form as
\beqa
\begin{split}
Y^s_{lm}(\theta,\phi) & = N^s_{lm} \bigg(\sin\frac{\theta}{2}\bigg)^{|m+s|} \bigg(\cos\frac{\theta}{2}\bigg)^{|m-s|} \\
& \times P^{|m+s|,|m-s|}_{l-\bar{l}}(\cos\theta) e^{im\phi} \frac{1}{\sqrt{2\pi}}, \\
\end{split}
\eeqa
where the function $P_n^{(a,b)}(z)$ denotes a Jacobi polynomial of degree $n=l-\bar{l}$. $1/\sqrt{2\pi}$ is the normalization constant associated with $e^{im\phi}$. $N^s_{lm}$ is the normalization factor determined through using the orthonormal relations of the Jacobi polynomials as
\beqa
N^s_{lm} = (-1)^{\text{max}(m,-s)} \sqrt\frac{(2l+1)(l+\bar{l})!(l-\bar{l})!}{2(l+\bar{l}')!(l-\bar{l}')!}, 
\eeqa 
with $\bar{l} = (a+b)/2=\text{max}(|m|,|s|)$ and $\bar{l}' = \text{min}(|m|,|s|)$.
For $s = 0$, $l$, and $m = 0$, the function becomes
\beqa
Y^0_{l0}(\theta,\phi) = \sqrt\frac{2l+1}{4\pi} P_l(\cos\theta) . 
\eeqa

\subsubsection{Explicit forms of higher-spin spherical harmonics}
\label{app:Islm}
The scalar ($l=s=0$) spherical harmonic is
\beqa
Y^0_{00} = \frac{1}{\sqrt{4\pi}}.
\eeqa
The spinor ($l=s=1/2$) spherical harmonics are
\beqa
\begin{split}
&  Y^{-\frac{1}{2}}_{\frac{1}{2},-\frac{1}{2}} = \frac{i}{\sqrt{4\pi}}\sqrt{1-\cos\theta} e^{-i\frac{\phi}{2}}, \\
&  Y^{-\frac{1}{2}}_{\frac{1}{2},\frac{1}{2}} = \frac{i}{\sqrt{4\pi}}\sqrt{1+\cos\theta} e^{i\frac{\phi}{2}}, \\
&  Y^{\frac{1}{2}}_{\frac{1}{2},-\frac{1}{2}} = -\frac{i}{\sqrt{4\pi}}\sqrt{1+\cos\theta} e^{-i\frac{\phi}{2}}, \\
&  Y^{\frac{1}{2}}_{\frac{1}{2},\frac{1}{2}} = \frac{i}{\sqrt{4\pi}}\sqrt{1-\cos\theta} e^{i\frac{\phi}{2}} . \\
\end{split}
\eeqa
The vector ($l=s=1$) spherical harmonics are
\beqa
\begin{split}
& Y^{-1}_{11} = - \frac{1}{2}\sqrt\frac{3}{4\pi}(1+\cos\theta)e^{i\phi} , \\
& Y^{0}_{1-1} = \frac{1}{\sqrt{2}}\sqrt\frac{3}{4\pi}\sin\theta e^{-i\phi}, \\
& Y^{0}_{10} = \sqrt\frac{3}{4\pi} \cos\theta, \\
& Y^{0}_{11} = - \frac{1}{\sqrt{2}} \sqrt\frac{3}{4\pi}\sin\theta e^{i\phi} , \\
& Y^{\pm 1}_{1-1} = - \frac{1}{2}\sqrt\frac{3}{4\pi}(1\pm\cos\theta)e^{-i\phi}, \\
& Y^{\pm 1}_{10} = \pm \frac{1}{\sqrt{2}}\sqrt\frac{3}{4\pi} \sin\theta, \\
& Y^{1}_{11} = - \frac{1}{2}\sqrt\frac{3}{4\pi}(1-\cos\theta)e^{i\phi} . \\
\end{split}
\eeqa
Note the relations due to property in Eq.~(\ref{Y_star}).
The spinor-vector ($l=s=3/2$) spherical harmonics are
\beqa
\begin{split}
& Y^{-\frac{3}{2}}_{\frac{3}{2},-\frac{3}{2}} = - i \sqrt\frac{1}{8\pi} (1-\cos\theta)^{3/2} e^{-i\frac{3\phi}{2}}, \\
& Y^{-\frac{3}{2}}_{\frac{3}{2},-\frac{1}{2}} = - i \sqrt\frac{3}{8\pi} \sin\theta \sqrt{1-\cos\theta} e^{-i\frac{\phi}{2}}, \\
& Y^{-\frac{3}{2}}_{\frac{3}{2},\frac{1}{2}} = - i \sqrt\frac{3}{8\pi} \sin\theta \sqrt{1+\cos\theta} e^{i\frac{\phi}{2}}, \\
& Y^{-\frac{3}{2}}_{\frac{3}{2},\frac{3}{2}} = - i \sqrt\frac{1}{8\pi} (1+\cos\theta)^{3/2} e^{i\frac{3\phi}{2}}, \\
& Y^{-\frac{1}{2}}_{\frac{3}{2},-\frac{3}{2}} = i \sqrt\frac{3}{8\pi} \sin\theta \sqrt{1-\cos\theta} e^{-i\frac{3\phi}{2}}, \\
& Y^{-\frac{1}{2}}_{\frac{3}{2},-\frac{1}{2}} = i \sqrt\frac{1}{8\pi} (1+3\cos\theta) \sqrt{1-\cos\theta} e^{-i\frac{\phi}{2}}, \\
& Y^{-\frac{1}{2}}_{\frac{3}{2},\frac{1}{2}} = -  i \sqrt\frac{1}{8\pi} (1-3\cos\theta) \sqrt{1+\cos\theta} e^{i\frac{\phi}{2}}, \\
& Y^{-\frac{1}{2}}_{\frac{3}{2},\frac{3}{2}} = - i \sqrt\frac{3}{8\pi} \sin\theta \sqrt{1+\cos\theta} e^{i\frac{3\phi}{2}}, \\
& Y^{\frac{1}{2}}_{\frac{3}{2},-\frac{3}{2}} =  - i \sqrt\frac{3}{8\pi} \sin\theta \sqrt{1+\cos\theta} e^{-i\frac{3\phi}{2}}, \\
& Y^{\frac{1}{2}}_{\frac{3}{2},-\frac{1}{2}} = i \sqrt\frac{1}{8\pi} (1-3\cos\theta) \sqrt{1+\cos\theta} e^{-i\frac{\phi}{2}}, \\
& Y^{\frac{1}{2}}_{\frac{3}{2},\frac{1}{2}} =  i \sqrt\frac{1}{8\pi} (1+3\cos\theta) \sqrt{1-\cos\theta} e^{i\frac{\phi}{2}}, \\
& Y^{\frac{1}{2}}_{\frac{3}{2},\frac{3}{2}} = -  i \sqrt\frac{3}{8\pi} \sin\theta \sqrt{1-\cos\theta} e^{i\frac{3\phi}{2}}, \\
& Y^{\frac{3}{2}}_{\frac{3}{2},-\frac{3}{2}} =  i \sqrt\frac{1}{8\pi} (1+\cos\theta)^{3/2} e^{-i\frac{3\phi}{2}}, \\
& Y^{\frac{3}{2}}_{\frac{3}{2},-\frac{1}{2}} =  - i \sqrt\frac{3}{8\pi} \sin\theta \sqrt{1+\cos\theta} e^{-i\frac{\phi}{2}}, \\
& Y^{\frac{3}{2}}_{\frac{3}{2},\frac{1}{2}} =  i \sqrt\frac{3}{8\pi} \sin\theta \sqrt{1-\cos\theta} e^{i\frac{\phi}{2}}, \\
& Y^{\frac{3}{2}}_{\frac{3}{2},\frac{3}{2}} =  - i \sqrt\frac{1}{8\pi} (1-\cos\theta)^{3/2} e^{i\frac{3\phi}{2}} .  \\
\end{split}
\eeqa
The explicit forms of gravitational wave ($s=2$) spherical harmonics are
\beqa
\begin{split}
	& Y^{-2}_{2-2} = \sqrt\frac{5}{64\pi} (1-\cos\theta)^2e^{-2i\phi}, \\
	& Y^{-2}_{2-1} = \sqrt\frac{5}{16\pi} (1-\cos\theta)\sin\theta e^{-i\phi}, \\
	& Y^{-2}_{20} = \sqrt\frac{15}{32\pi} \sin^2\theta, \\
	& Y^{-2}_{21} = \sqrt\frac{5}{16\pi}  (1+\cos\theta)\sin\theta e^{i\phi}, \\
	& Y^{-2}_{22} = \sqrt\frac{5}{64\pi}  (1+\cos\theta)^2 e^{2i\phi}, \\
	& Y^{-1}_{2-2} = -\sqrt\frac{5}{16\pi} (1-\cos\theta) \sin\theta e^{-2i\phi}, \\
	& Y^{-1}_{2-1} =  -\sqrt\frac{5}{16\pi} (1+\cos\theta-2\cos^2\theta)\sin\theta e^{-i\phi}, \\
	& Y^{-1}_{20} = -\sqrt\frac{15}{8\pi}\sin\theta \cos\theta, \\
	& Y^{-1}_{21} = \sqrt\frac{5}{16\pi} (1-\cos\theta-2\cos^2\theta) e^{i\phi}, \\
	& Y^{-1}_{22} = \sqrt\frac{5}{16\pi} (1+\cos\theta)\sin\theta e^{2i\phi}, \\
	& Y^{0}_{2-2} = \frac{1}{2} \sqrt\frac{15}{8\pi} \sin^2\theta e^{-2i\phi}, \\
	& Y^{0}_{2-1} =  \sqrt\frac{15}{8\pi} \sin\theta \cos\theta e^{-i\phi}, \\
	& Y^{0}_{20} = -\sqrt\frac{5}{16\pi} (1-3\cos^2\theta), \\
	& Y^{0}_{21} = - \sqrt\frac{15}{8\pi} \cos\theta \sin\theta e^{i\phi}, \\
	& Y^{0}_{22} = \frac{1}{2} \sqrt\frac{15}{8\pi} \sin^2\theta e^{2i\phi}, \\
	& Y^{1}_{2-2} = -\sqrt\frac{5}{16\pi} (1+\cos\theta) \sin\theta e^{-2i\phi}, \\
	& Y^{1}_{2-1} =  \sqrt\frac{5}{16\pi} (1-\cos\theta-2\cos^2\theta)\sin\theta e^{-i\phi}, \\
	& Y^{1}_{20} = \sqrt\frac{15}{8\pi}\sin\theta \cos\theta, \\
	& Y^{1}_{21} = - \sqrt\frac{5}{16\pi} (1+\cos\theta-2\cos^2\theta) e^{i\phi}, \\
	& Y^{1}_{22} = \sqrt\frac{5}{16\pi} (1-\cos\theta)\sin\theta e^{2i\phi}, \\
	& Y^{2}_{2-2} = \sqrt\frac{5}{64\pi} (1+\cos\theta)^2 e^{-2i\phi}, \\
	& Y^{2}_{2-1} =  - \sqrt\frac{5}{16\pi} (1+\cos\theta)\sin\theta e^{-i\phi}, \\
	& Y^{2}_{20} = \sqrt\frac{15}{32\pi} \sin^2\theta, \\
	& Y^{2}_{21} = - \sqrt\frac{5}{16\pi} (1-\cos\theta)\sin\theta e^{i\phi}, \\
	& Y^{2}_{22} = \sqrt\frac{5}{64\pi} (1-\cos\theta)^2 e^{2i\phi} .  \\
\end{split}
\eeqa
They are all normalized in such a way that
\beqa
\int d\Omega Y^{s \star}_{lm} Y^s_{l'm'} = \delta_{ll'} \delta_{mm'}.
\eeqa

\section{The Newman-Penrose Formalism}
\label{app:NP_form}

\subsection{Complex null tetrad}


The Newman-Penrose (NP) formalism is a null-tetrad formulation of gravity theory with complex tetrads~\cite{Newman:1961qr}. It is an approach to describing spacetime with the congruences of null geodesics. 
We denote the NP null tetrad by a complex null tetrad as
\beqa
e^\mu_a = (l^\mu, n^\mu, m^\mu, \bar{m}^\mu),
\eeqa
where the index $a=1,2,3,4$ labels the tetrad vectors. $l^\mu = e_1^\mu$, $n^\mu = e_2^\mu$, and $m^\mu = e_3^\mu$, etc. In abbreviation, $(l,n,m,\bar{m})$ can be expressed in terms of the basis vectors $(e^1,e^2,e^3,e^4)$. It consists of two real null vectors $l$, $n$, and one complex null vector $m$, where $\bar{m}^\mu = m^{\mu \star}$. The real null vectors $\{l^\mu, n^\mu \}$ span a time-like $2$-plane in the tangent space at each point of spacetime, while the complex vectors $\{m^\mu,\bar{m}^\mu\}$ span the orthogonal spacelike $2$-plane. In the $(-1,+1,+1,+1)$ signature of Minkowski spacetime, the null tetrad of world-vectors is normalized as~\cite{Frolov1979}
\beqa
\begin{split}
 n_\mu l^\mu = - 1, \quad m_\mu \bar{m}^\mu =1 . \\
\end{split} \label{norm_nlm}
\eeqa
They are all null vectors with respect to the metric $g_{\mu\nu}$ and also orthogonal to each other as
\beqa
\begin{split}
& n_\mu n^\mu = l_\mu l^\mu = m_\mu m^\mu = \bar{m}_\mu \bar{m}^\mu = 0 , \\
& n_\mu m^\mu = n_\mu \bar{m}^\mu = l_\mu m^\mu = l_\mu \bar{m}^\mu = 0 . 
\end{split} \label{orth_nlm}
\eeqa
$l^\mu$ and $n^\mu$ are real null orthonormal vectors (i.e., $l^\mu=\bar{l}^\mu$, $n^\mu=\bar{n}^\mu$), while $(m^\mu, \bar{m}^\mu)$ is a pair of complex conjugated orthonormal vectors.

The spacetime metric can be written as
\beqa
\begin{split}
g_{\mu\nu}  \equiv \eta_{ab}  e_{\mu}^a e_\nu^b = - 2 l_{(\mu} n_{\nu)} + 2 m_{(\mu} \bar{m}_{\nu)} , \\
\end{split}
\eeqa
where $\eta_{ab}$ is the flat-space metric in tangent spacetime and its inverse $\eta^{ab}$, lower or raise the tetrad indices as
\beqa
\eta_{ab} = \eta^{ab} = \bpm
0 & -1 & 0 & 0 \\
-1 & 0 & 0 & 0 \\
0 & 0 & 0 & 1 \\
0 & 0 & 1 & 0 \\
\epm . \label{eta_ab}
\eeqa
The coframe $e^a$ and frame $e_a$ are related by $e^a=\eta^{ab}e_b$. The non-vanishing components of the symmetric metric are $\eta^{12} = -\eta^{34} = -1$. Thus, the dual basis of the $1$-form, i.e., the coframe null tetrad, is $e_\mu^a = (-n_\mu, -l_\mu, \bar{m}_\mu, m_\mu)$.
Thus, the covariant derivative can be decomposed into a linear combination of directional derivative operators as
\beqa
\begin{split}
\nabla_\mu 
= - n_\mu D - l_\mu D'  + \bar{m}_\mu \delta + m_\mu \delta' , \\
\end{split} \label{covD}
\eeqa
where $(D,D',\delta, \delta')$ are tangent differential operators along the tetrad direction:
\beqa
\begin{split}
& D = l^\mu \nabla_\mu = \bar{D}, \quad 
 D' = n^\mu \nabla_\mu = \bar{D}', \\
& \delta = m^\mu \nabla_\mu = \bar\delta', \quad
 \delta' = \bar{m}^\mu \nabla_\mu  = \bar\delta. \\
\end{split} \label{D-delta}
\eeqa
The last equality implies that $D$ and $D'$ are real operators, while $\delta$ and $\delta'$ are complex conjugate operators of each other.

\subsection{Spin coefficients}

The connection coefficients are defined as
\beqa
\nabla_c e_b^\mu = \gamma^a_{~bc} e_a^\mu,
\eeqa
where $\nabla_c = e_c^\nu \nabla_\nu$. Thus, the coefficients can be expressed in the complex Ricci rotation coefficients defined by
\beqa
\gamma_{abc}  = e_{b\mu;\nu} e_a^{\mu} e_c^{\nu } , \label{gamma_e}
\eeqa
where the semicolon represents the covariant derivative. The various Ricci rotation coefficients are called spin coefficients. Assuming the non-metricity is vanishing, we have
\beqa
\gamma_{abc} + \gamma_{bac} = 0 .
\eeqa
Thus, the Ricci rotation coefficients are antisymmetric in the first two indices. Therefore, the Ricci rotation coefficients $\gamma_{abc}$ have $24$ real components. The connection can be expressed in terms of $12$ complex NP spin coefficients denoted by specific symbols as below:
\beqa
\begin{split}
& \kappa \equiv \gamma_{131} , \quad \tau \equiv  \gamma_{132}, \quad \epsilon \equiv  (\gamma_{121} -  \gamma_{341})/2, \\
& \sigma \equiv \gamma_{133}, \quad \rho \equiv  \gamma_{134}, \quad \gamma \equiv  (\gamma_{122} - \gamma_{342})/2,  \\ 
& \pi \equiv \gamma_{421}, \quad \nu \equiv \gamma_{422}, \quad \beta \equiv ( \gamma_{123} - \gamma_{343})/2, \\ 
& \mu \equiv \gamma_{423}, \quad \lambda \equiv \gamma_{424}, \quad \alpha \equiv  (\gamma_{124} - \gamma_{344})/2 . \\
\end{split} \label{SCs}
\eeqa
Note that the definitions have opposite signs compared to those in the flat Minkowski metric signature $(+1,-1,-1,-1)$.
The complex conjugates of these spin coefficients are obtained through the replacement of $3 \leftrightarrow 4$ in $\gamma_{abc}$. These NP spin coefficients are defined in tensor notation as
\beqa
\begin{split}
& \kappa  = - m^\mu D l_\mu, \quad \tau = - m^\mu D' l_{\mu}, \\
& \epsilon = -( n^\mu D l_{\mu} + m^\mu D \bar{m}_{\mu}  )/2 , \\
& \sigma  = - m^\mu \delta l_{\mu}, \quad \rho  = - m^\mu \delta' l_{\mu}, \\
& \gamma = - (  n^\mu  D' l_{\mu } + m^\mu D' \bar{m}_{\mu}   )/2 , \\
& \pi = \bar{m}^\mu D n_{\mu} , \quad \nu  = \bar{m}^\mu D' n_\mu , \\
& \beta = -  ( n^\mu \delta l_{\mu} +  m^\mu \delta  \bar{m}_{\mu} )/2 , \\
&  \mu = \bar{m}^\mu \delta  n_{\mu} , \quad \lambda  = \bar{m}^\mu \delta' n_{\mu} , \\
& \alpha = - ( n^\mu \delta'  l_{\mu} + m^\mu \delta' \bar{m}_{\mu} )/2 , \\
\end{split} \label{NP-SCs}
\eeqa
where $(D,D',\delta,\delta')$ are four locally defined covariant derivative operators in Eq.~(\ref{D-delta}). These differential operators satisfy some general identities, as shown below. When $\kappa=\sigma=\nu=\lambda=0$ holds, the null vector $l^\nu$ is tangent to affine parametrized geodesics, i.e., $Dl_\nu=0$, and the geodesics are shear-free, i.e., $\sigma=0$. According to the Goldberg-Sachs theorem, the identities are valid for any Petrov-type $D$ metrics as
\beqa
\begin{split}
& [D-(p+1)\epsilon + \bar\epsilon + q\rho -\bar\rho](\delta - p\beta + q\tau) \\
& - [\delta -(p+1)\beta - \bar\alpha + \bar\pi + q\tau](D-p\epsilon+q\rho) = 0 , \\
& [D'+(p+1)\gamma-\bar\gamma - q\mu + \bar\mu ](\delta'+p\alpha-q\pi) \\
& -[\delta'+(p+1)\alpha+\bar\beta-\bar\tau-q\pi](D'+p\gamma-q\mu) = 0 , \\
\end{split} \label{ID_delta-D}
\eeqa
for a pair of arbitrary constants $(p,q)$. We have used the second and complex conjugate of the third commutation relations in Eq.~(\ref{com_delta-D}) and adopted the notations as~\cite{Penrose:1985bww}
\beqa
\begin{split}
& \sigma' = -\lambda, \quad \tau' = -\pi, \quad \kappa' = -\nu, \\
& \rho' = -\mu, \quad \beta' = -\alpha , \quad \epsilon' = - \gamma , \\
& \lambda' = -\sigma, \quad \pi' = -\tau, \quad \nu' =- \kappa, \\
&  \mu' = - \rho, \quad \alpha' = - \beta , \quad \gamma' = - \epsilon . \\
\end{split} \label{NPS_prime}
\eeqa
The second equality in Eq.~(\ref{ID_delta-D}) can be obtained from the first one. By considering the fact that the full set of NP equations is invariant under the interchange $l\leftrightarrow n$ and $m\leftrightarrow \bar{m}$, which is equivalent to
\beqa
\begin{split}
& D \leftrightarrow D', \quad \delta \leftrightarrow \delta', \quad \kappa \leftrightarrow - \nu, \quad \rho \leftrightarrow - \mu, \\
& \sigma \leftrightarrow - \lambda, \quad \tau \leftrightarrow - \pi , \quad \epsilon \leftrightarrow - \gamma, \quad \beta \leftrightarrow - \alpha . \\
\end{split} \label{sym_exc}
\eeqa
For a spherically symmetric static metric,
\beqa
\begin{split}
& \tau=\pi=\epsilon = \gamma = 0 , \quad \bar\rho=\rho , \quad \bar\beta = -\alpha , \\
& \pi' = \tau' = \gamma' = \epsilon' = 0 , \quad \bar\mu' = \mu', \quad \bar\alpha' = - \beta'.
\end{split} \label{NPS_SSS}
\eeqa
The identities are reduced to
\beqa
\begin{split}
& (D+q\rho-\rho)(\delta + p\alpha) = (\delta + p\alpha)(D+q\rho). \\
& (D'- q\mu + \mu)(\delta'+p\alpha) = (\delta' + p\alpha)(D'-q\mu) . \\
\end{split}
\eeqa

\subsection{Newman-Penrose scalar components of the Weyl, Ricci, and Maxwell tensors}
In four dimensions, the Riemann curvature tensor has $20$ independent real components. It was split into three irreducible parts: the Weyl tensor $C_{\mu\nu\rho\sigma}$, the symmetric Ricci tensor $R_{\mu\nu}-g_{\mu\nu}R/4$, and the Ricci scalar $R$. The Weyl tensor and the symmetric Ricci tensor admit $10$ real components each. The Weyl tensor $C_{\mu\nu\rho\sigma}$ is defined as
\beqa
\begin{split}
C_{\mu\nu\rho\sigma} & = R_{\mu\nu\rho\sigma} - (g_{\mu[\rho}R_{\sigma]\nu} - g_{\nu[\rho} R_{\sigma]\mu}   ) \\
 & +  g_{\mu[\rho}g_{\sigma]\nu} R/3 . \\
\end{split} 
\label{Weyl-Riemann}
\eeqa
It reduces to the Riemann tensor when the metric is Ricci flat (i.e., $R_{\mu\nu}=0$). The Weyl tensor contains ten independent real, or, equivalently, five independent complex components. They are in terms of physical gravitoelectric and gravitomagnetic field components measured by an observer at rest. The five complex components of the Weyl curvature tensor are named NP Weyl-scalars as
\beqa
\begin{split}
& \Psi_0  =  C_{lmlm}  , \quad \Psi_1  =  C_{lmln}  ,  \\
& \Psi_2  =  (C_{lnln}+C_{ln\bar{m}m})/2  ,  \\  
& \Psi_3  =  C_{ln\bar{m}n}  ,   \quad  \Psi_4  =  C_{\bar{m}n\bar{m}n} , \\
\end{split} \label{Psi_01234}
\eeqa
where we have expressed the quantities in terms of the null tetrad, e.g., $C_{lmlm} \equiv C_{\mu\nu\rho\sigma}l^\mu m^\nu l^\rho m^\sigma $, etc. The five complex NP Weyl scalars are independent of spacetime coordinates but depend on the choice of null tetrad.
The Ricci tensor has nine independent components. They can be expressed with the null-tetrad components of the traceless part of the Ricci tensor $ \hat{R}_{\mu\nu} = R_{\mu\nu} - g_{\mu\nu}R/4 $ and the orthogonal conditions in Eq.~(\ref{orth_nlm}) as
\beqa
\begin{split}
& \Phi_{00} = R_{ll}/2  , \quad \Phi_{01} = R_{lm}/2  , \quad \Phi_{02} = R_{mm}/2, \\ 
& \Phi_{10} = R_{l\bar{m}}/2 , ~ \Phi_{11} = (R_{ln}+R_{m\bar{m}})/4 , ~\Phi_{12} = R_{mn}/2  , \\
& \Phi_{20} = R_{\bar{m}\bar{m}}/2  , \quad \Phi_{21} = R_{\bar{m}n}/2  , \quad \Phi_{22} = R_{nn}/2 , \\
\end{split} \quad \label{Rij}
\eeqa
where $R_{ll}=R_{\mu\nu}l^\mu l^\nu$, etc. Note that $\Phi_{10}=\bar\Phi_{01}$, $\Phi_{20}=\bar\Phi_{02}$, and $\Phi_{21}=\bar\Phi_{12}$. The traceless Ricci tensor $\hat{R}_{\mu\nu}$ is equal to the stress energy tensor $T_{\mu\nu}$ through the Einstein equation:
\beqa
\hat{R}_{\mu\nu} = \kappa_N T_{\mu\nu} .
\eeqa
Thus, all the Ricci tensors in Eq.~(\ref{Rij}) can be replaced with corresponding stress energy tensors, e.g., $R_{ll} = \kappa_N T_{ll}$ with $T_{ll} = T_{\mu\nu}l^\mu l^\nu$, etc.
The last independent component is the Ricci scalar, which can be expressed as
\beqa
R = - 2 (R_{ln} - R_{m\bar{m}}) = 24 \Lambda,
\eeqa
and $\Lambda$ is the cosmological constant.
In the presence of an electromagnetic gauge field, the stress energy tensor can be expressed in three complex NP Maxwell scalars:
\beqa
\begin{split}
 \phi_0 =  F_{lm}  , \quad \phi_1 = (F_{ln} + F_{\bar{m}m})/2 , \quad \phi_2 =  F_{\bar{m}n}  , 
\end{split} \quad \label{phi_012}
\eeqa
where $F_{lm} = F_{\mu\nu}l^\mu m^\nu$, etc. $F_{\mu\nu}=2A_{[\nu;\mu]}$ is the antisymmetric Maxwell tensor, and $A_\mu$ is the electromagnetic potential.

\subsection{Newman-Penrose equations}

\subsubsection{Commutation relations}

The differential operators $D_a=(D,D',\delta,\delta')$ with $a=1,2,3,4$ in Eq.~(\ref{D-delta}) satisfy the commutation relations as
\beqa
[D_a, D_b] = -T^c_{~ab}D_c,
\eeqa
where $T^c_{~ab} = 2 \Gamma^c_{~[ab]}$ is torsion tensor.
To be more explicit, the independent commutation relations between the null bases are
\beqa
\begin{split}
	& [D',D] =  (\gamma+ \bar\gamma)D + (\epsilon + \bar\epsilon)D' - (\bar\tau + \pi)\delta - (\tau+\bar\pi)\delta' , \\
	& [\delta, D] = (\bar\alpha + \beta - \bar\pi) D + \kappa D' - (\epsilon-\bar\epsilon + \bar\rho)\delta - \sigma \delta' , \\
	& [\delta, D'] = - \bar\nu D + (\tau - \bar\alpha - \beta)D' + (\mu-\gamma+\bar\gamma)\delta + \bar\lambda \delta' ,  \\
	& [\delta', \delta] = (\bar\mu - \mu )D + (\bar\rho-\rho)D' + (\alpha-\bar\beta)\delta + ( \beta - \bar\alpha )\delta' . \\
\end{split} \label{com_delta-D}
\eeqa

\subsubsection{Covariant derivative of a null tetrad}

The orthogonality and normalization conditions of the null basis vectors satisfy the following relations:
\beqa
\begin{split}
	& D l = (\epsilon+\bar\epsilon)l - \kappa \bar{m} - \bar\kappa m , \\
	& D n = - (\epsilon+\bar\epsilon)n - \pi m -\bar\pi \bar{m}, \\
	& D m = - (\epsilon-\bar\epsilon)m + \bar\pi l - \kappa n, \\
	& D' l = (\gamma+\bar\gamma)l - \bar\tau m - \tau \bar{m}, \\ 
	& D' n = -(\gamma+\bar\gamma)n - \nu m - \bar\nu \bar{m}, \\ 
	& D' m =  (\gamma-\bar\gamma)m + \bar\nu l - \tau n , \\
	& \delta l = (\bar\alpha + \beta)l - \bar\rho m - \sigma \bar{m}, \\
	& \delta' l = (\alpha + \bar\beta)l - \rho \bar{m} - \bar\sigma m, \\
	& \delta  n = - (\bar\alpha + \beta)n - \mu m - \bar\lambda \bar{m}, \\
	& \delta' n = - (\alpha + \bar\beta)n - \bar\mu \bar{m} - \lambda m, \\
	& \delta  m = - (\beta - \bar\alpha)m + \bar\lambda l - \sigma n , \\
	& \delta' m =  - (\alpha - \bar\beta) m + \bar\mu l - \rho n . \\
\end{split}
\eeqa

\subsubsection{Radial equations}

The NP equations are equivalent to the Ricci identities of the Riemann curvature tensor as
\beqa
\begin{split}
	D\rho - \delta' \kappa & = \rho^2 + (\epsilon + \bar\epsilon)\rho + \sigma \bar\sigma  + \kappa (\pi - 3\alpha - \bar\beta ) - \bar\kappa \tau \\
	& + \Phi_{00} , \\
	D\sigma - \delta \kappa & = \sigma(\rho+\bar\rho + 3\epsilon - \bar\epsilon) - \kappa(\bar\alpha + 3 \beta - \bar\pi + \tau) \\
	& +  \Psi_0 , \\
	D\tau - D' \kappa & = \rho( \bar\pi + \tau ) + \sigma ( \pi + \bar\tau ) + \tau (\epsilon - \bar\epsilon) - \kappa (3 \gamma + \bar\gamma) \\
	& + \Phi_{01} + \Psi_1 , \\
	D\alpha - \delta' \epsilon & =  \rho(\alpha + \pi) + \bar\sigma \beta - (2\alpha + \bar\beta - \pi)\epsilon + \alpha \bar\epsilon \\
	& - \kappa \lambda - \gamma \bar\kappa +  \Phi_{10} , \\
	D\beta - \delta \epsilon & = \sigma (\alpha + \pi) - \epsilon (\bar\alpha - \bar\pi) + \beta (\bar\rho - \bar\epsilon)  \\
	& - \kappa (\gamma + \mu ) + \Psi_1 , \\
	D\gamma - D' \epsilon & =  \alpha(\bar\pi + \tau) + \beta\bar\tau + \pi(\beta+\tau)   - (\gamma+\bar\gamma) \epsilon \\
	& - \gamma (\epsilon+\bar\epsilon) - \kappa \nu  + \Phi_{11} - \Lambda + \Psi_2 , \\
	D\lambda - \delta' \pi & =  \mu\bar\sigma + \lambda (\rho+ \bar\epsilon) + \pi(\alpha -\bar\beta +\pi) - \bar\kappa \nu - 3 \epsilon \lambda \\
	& + \Phi_{20}, \\
	D\mu - \delta \pi & = \lambda\sigma + \mu ( \bar\rho - \epsilon - \bar\epsilon ) + \pi (\beta+\bar\pi) - \bar\alpha \pi - \kappa \nu \\
	& + 2\Lambda + \Psi_2 ,  \\
	D\nu - D'\pi & = \mu(\pi + \bar\tau) +  \lambda(\bar\pi + \tau) + (\gamma - \bar\gamma) \pi - (3\epsilon + \bar\epsilon)\nu \\
	& + \Phi_{21} + \Psi_3 .  \\ 
\end{split}
\eeqa

Since the generators are null geodesics living in a null hypersurface, $\kappa=0$ and $\rho=\bar\rho$. The convergence $\rho$ and shear $\sigma$ obey the radial equations:
\beqa
\begin{split}
	D\rho & = \rho^2 + \sigma \bar\sigma + (\epsilon + \bar\epsilon)\rho + \Phi_{00}, \\
	D\sigma & =  ( 2\rho + 3\epsilon-\bar\epsilon)\sigma + \Psi_0  ,\\
	D\tau & = \tau \rho +\bar\tau \sigma + \Psi_1, \\
	D\alpha & = \alpha \rho + \beta \bar\sigma, \\
	D\beta & = \beta \rho + \alpha \sigma + \Psi_1, \\
	D\gamma & = \tau \alpha + \bar\tau \beta + \Psi_2 - \Lambda, \\
	D\lambda & = \lambda \rho + \mu \bar\sigma, \\
	D\mu & = \mu \rho + \lambda \sigma + \Psi_2 + 2 \Lambda, \\
	D\nu & = \tau \lambda + \bar\tau \mu + \Psi_3,  \\
\end{split}
\eeqa
where $D\equiv d/dt$ with $t$ is the affine parameter along the generators such that $l^\mu=dx^\mu/dt$. The quantities $\theta$, $\omega$, and $\sigma$ are named optical scalars. They represent the expansion, vorticity, and shear of the null ray, respectively. The first two are related to the real and imaginary parts of $\rho$ as
\beqa
\theta = - \text{Re}\rho, \quad  \omega = \text{Im}\rho.
\eeqa
The NP equations of $\rho$ and $\sigma$ can also be expressed explicitly as
\beqa
\begin{split}
& D\theta = \omega^2 - \theta^2  - \sigma\bar\sigma + \Phi_{00}, \\
& D\omega = - 2\theta \omega, \\
& D\sigma = - 2\theta \sigma - \Psi_0 . \\
\end{split}
\eeqa
These are nothing but expand-, vorticity-, and shear-propagation equations, since the Ricci identities (i.e., the NP equations) yield Raychaudhuri's equations.

\subsubsection{Hypersurface equations}

The hypersurface equations are
\beqa
\begin{split}
	D' \lambda - \delta \nu & = \nu(3\alpha + \bar\beta + \pi - \bar\tau ) - \lambda(3\gamma-\bar\gamma + \mu + \bar\mu ) \\
	& - \Psi_4, \\
	\delta \rho - \delta' \sigma & = (\rho-\bar\rho)\tau + (\bar\beta-3\alpha)\sigma + (\bar\alpha + \beta)\rho \\
	& + (\mu - \bar\mu)\kappa + \Phi_{01} - \Psi_1,  \\
	\delta \alpha - \delta' \beta & =    ( \rho - \bar\rho ) \gamma + \mu \rho - \lambda \sigma + \epsilon(\mu-\bar\mu) + \alpha \bar\alpha \\
	& - 2\alpha \beta + \beta \bar\beta + \Phi_{11} + \Lambda - \Psi_2 , \\
	\delta \lambda - \delta' \mu & = (\rho-\bar\rho)\nu + (\mu-\bar\mu)\pi + (\alpha+\bar\beta)\mu \\
	& + (\bar\alpha -3\beta) \lambda + \Phi_{21} - \Psi_3 , \\ 
	D'\mu - \delta \nu & = (\bar\alpha + 3\beta - \tau)\nu + \bar\nu \pi - \mu(\mu+\gamma + \bar\gamma) \\
	& - \lambda\bar\lambda - \Phi_{22} , \\
	D'\beta - \delta \gamma & =  \nu \sigma + \epsilon \bar\nu - \tau (\mu+\gamma) + \beta ( 2\gamma-\bar\gamma - \mu) \\
	& -\alpha \bar\lambda + \bar\alpha \gamma - \Phi_{12} , \\
	D' \sigma - \delta \tau & =  - \rho \bar\lambda + \sigma (3\gamma-\bar\gamma-\mu) + \tau(\bar\alpha -\beta - \tau) \\
	& + \kappa \bar\nu - \Phi_{02} , \\
	D' \rho - \delta' \tau & =  \rho(\gamma+\bar\gamma-\bar\mu) - \lambda \sigma -\tau (\alpha-\bar\beta + \bar\tau ) \\
	& + \kappa \nu - 2 \Lambda - \Psi_2, \\
	D' \alpha - \delta' \gamma & = \nu (\epsilon+\rho) - \lambda(\beta+\tau) + \gamma(\bar\beta - \bar\tau) \\
	& + \alpha(\bar\gamma - \bar\mu) - \Psi_3 . \\
\end{split}
\eeqa

\subsubsection{Bianchi identities}

The second Bianchi identity of the curvature tensor is
\beqa
\nabla_{[a} R_{bc]df} 
= 0 . \quad
\eeqa
We can express it in the null tetrad. By replacing the curvature in terms of the null tetrad, in total, there are $11$ independent Bianchi identities.
In NP formalism, the Bianchi identities are~\cite{Stewart1991}
\beqa
\begin{split}
	& D \Psi_1 - \delta' \Psi_0 - D \Phi_{01} + \delta \Phi_{00} \\
	& = (\pi-4\alpha) \Psi_0 + 2 (2\rho+\epsilon ) \Psi_1 - 3\kappa \Psi_2 \\
	& + (2 \bar\alpha + 2\beta - \bar\pi ) \Phi_{00} - 2 (\epsilon + \bar\rho) \Phi_{01} \\
	&- 2\sigma \Phi_{10}  + 2 \kappa \Phi_{11} + \bar\kappa \Phi_{02} , \\
	& D' \Psi_1 - \delta \Psi_2 - D' \Phi_{01} + \delta' \Phi_{02} - 2\delta \Lambda \\
	& = \nu \Psi_0 + 2(\gamma-\mu )\Psi_1 - 3\tau \Psi_2 + 2 \sigma \Psi_3 \\
	& - \bar\nu \Phi_{00} + 2(\bar\mu - \gamma)\Phi_{01} + (2\alpha - 2\bar\beta + \bar\tau)\Phi_{02} \\
	& + 2\tau \Phi_{11} - 2\rho \Phi_{12} ,  \\
	& D' \Psi_0 - \delta \Psi_1 + D \Phi_{02} - \delta \Phi_{01} \\
	& = (4\gamma-\mu) \Psi_0 - 2 (2\tau+\beta ) \Psi_1 + 3 \sigma \Psi_2 \\
	& - \bar\lambda \Phi_{00}  - 2 (\beta - \bar\pi) \Phi_{01}  + 2 \sigma \Phi_{11} \\
	& + (2\epsilon-2\bar\epsilon+\bar\rho) \Phi_{02} - 2 \kappa \Phi_{12} , \\
	& D' \Psi_1 - \delta \Psi_2 - \delta \Phi_{11} + D \Phi_{12} + \delta \Lambda \\
	& = \nu \Psi_0 + 2(\gamma-\mu)\Psi_1 - 3\tau \Psi_2 + 2 \sigma \Psi_3 \\
	& - \mu \Phi_{01} + \pi \Phi_{02} - \bar\lambda \Phi_{10} + 2\pi \Phi_{11} \\
	& + (\bar\rho - 2 \bar\epsilon)\Phi_{12} + \sigma \Phi_{21} - \kappa \Phi_{22} . \\
\end{split} \quad \label{BIs}
\eeqa
Combining the last equation with the second one, it can be re-expressed as
\beqa
\begin{split}
& D\Phi_{12} - \delta \Phi_{11} + D' \Phi_{01} - \delta' \Phi_{02} + 3 \delta \Lambda \\
& = \bar\nu \Phi_{00} + (2\gamma-\mu - 2\bar\mu)\Phi_{01} \\
& + (\pi-\tau - 2\alpha + 2 \bar\beta) \Phi_{02} - \bar\lambda \Phi_{10}  + 2 (\bar\pi -\tau)\Phi_{11}  \\
& + (2\rho+\bar\rho-2\bar\epsilon) \Phi_{12} + \sigma \Phi_{21} - \kappa \Phi_{22} ,
\end{split}
\eeqa
and
\beqa
\begin{split}
	& D \Psi_2 - \delta' \Psi_1  + \delta \Phi_{10} - D \Phi_{11} - D \Lambda \\
	& = -\lambda \Psi_0 + 2 (\pi - \alpha ) \Psi_1 + 3\rho \Psi_2 - 2\kappa \Psi_3  + \mu \Phi_{00} \\
	& - \pi \Phi_{01} + (2\bar\alpha - \bar\pi) \Phi_{10}  - 2\bar\rho \Phi_{11} + \bar\kappa \Phi_{12} - \sigma \Phi_{20} + \kappa \Phi_{21}, \\
	& D' \Psi_2 - \delta \Psi_3 + D \Phi_{22} - \delta \Phi_{21} + 2 D' \Lambda \\
	& = 2\nu \Psi_1 - 3\mu \Psi_2 + 2(\beta-\tau)\Psi_3 + \sigma \Psi_4 - 2\mu \Phi_{11} + 2 \pi \Phi_{12} \\
	& - \bar\lambda \Phi_{20} + 2(\beta+\bar\pi)\Phi_{21} + (\bar\rho - 2\epsilon - 2\bar\epsilon) \Phi_{22} , \\
	& D\Psi_2 - \delta' \Psi_1 + D' \Phi_{00} - \delta' \Phi_{01} + 2 D \Lambda \\
	& = - \lambda \Psi_0 + 2(\pi-\alpha) \Psi_1 + 3 \rho \Psi_2 - 2 \kappa \Psi_3 \\
	& + (2\gamma + 2\bar\gamma - \bar\mu)\Phi_{00} - 2 (\alpha +\bar\tau)\Phi_{01} - 2\tau \Phi_{10} \\
	& + 2 \rho \Phi_{11} + \bar\sigma \Phi_{02} , \\ 
	& D'\Psi_2 - \delta \Psi_3 - D' \Phi_{11} + \delta' \Phi_{12} - D' \Lambda \\
	& =  2\nu \Psi_1 - 3\mu \Psi_2 + 2(\beta-\tau)\Psi_3 + \sigma \Psi_4 - \nu \Phi_{01} - \bar\nu \Phi_{10} \\
	& + 2 \bar\mu \Phi_{11} + \lambda \Phi_{02} + (\bar\tau - 2\bar\beta)\Phi_{12} + \tau \Phi_{21} - \rho \Phi_{22} . \\
\end{split}
\eeqa
Assuming the third and second equations are independent, the first and fourth equations can be re-expressed as
\beqa
\begin{split}
	& D \Phi_{11} - \delta \Phi_{10} + D' \Phi_{00} - \delta' \Phi_{01} + 3 D\Lambda \\
	& = (2\gamma + 2\bar\gamma-\mu - \bar\mu)\Phi_{00} + (\pi - 2\alpha - 2\bar\tau)\Phi_{01} + \bar\sigma \Phi_{02} \\
	& + (\bar\pi - 2\bar\alpha - 2\tau)\Phi_{10} + 2(\rho+\bar\rho)\Phi_{11} - \bar\kappa \Phi_{12} \\
	& + \sigma \Phi_{20} - \kappa \Phi_{21} ,  \\
	& D \Phi_{22} - \delta \Phi_{21} + D' \Phi_{11} - \delta' \Phi_{12} + 3 D' \Lambda \\
	& = \nu \Phi_{01} - \lambda \Phi_{02} + \bar\nu \Phi_{10} - 2(\mu+\bar\mu)\Phi_{11} \\
	& + (2\bar\beta + 2\pi - \bar\tau)\Phi_{12} - \bar\lambda \Phi_{20} \\
	& + (2\beta+2\bar\pi - \tau) \Phi_{21} + (\rho + \bar\rho - 2\epsilon - 2\bar\epsilon)\Phi_{22} . \\
\end{split}
\eeqa

Another two independent equations are
\beqa
\begin{split}
	& D\Psi_4 - \delta' \Psi_3 + D' \Psi_{20} - \delta' \Phi_{21} \\
	& = - 3\lambda \Psi_2 + 2(\alpha+2\pi) \Psi_3 + (\rho-4\epsilon)\Psi_4 + 2\nu \Phi_{10} - 2\lambda \Phi_{11} \\
	& - (2\gamma - \bar\gamma + \bar\mu) \Phi_{20} + 2(\alpha-\bar\tau)\Phi_{21} + \bar\sigma \Phi_{22} . \\
	& D' \Psi_3 - \delta \Psi_4 - D' \Phi_{21} + \delta' \Phi_{22} \\
	& =  3\nu \Psi_2 - 2(\gamma+2\mu) \Psi_3 + (4\beta-\tau)\Psi_4 - 2\nu \Phi_{11} + 2\lambda \Phi_{12} \\
	& - \bar\nu \Phi_{20} + 2(\gamma+\bar\mu) \Phi_{21} + (\bar\tau  - 2\bar\beta - 2\alpha )\Phi_{22} . \\
\end{split}
\eeqa

The last independent equation is
\beqa
\begin{split}
	& 3(D\Psi_3- \delta' \Psi_2) + \delta \Phi_{20} - D\Phi_{21} + 2(D' \Phi_{10}-\delta' \Phi_{11} ) \\
	& = - 6 \lambda \Psi_1 + 9 \pi \Psi_2 + 6 (\rho-\epsilon)\Psi_3 - 3\kappa \Psi_4 \\
	& + 2\nu \Phi_{00} - 2\lambda \Phi_{01} + 2(\mu - \bar\mu + 2\bar\gamma )\Phi_{10} \\
	& - 2(\pi+ 2\bar\tau)\Phi_{11} + 2\bar\sigma \Phi_{12} \\
	& + (2\bar\alpha - \bar\pi  - 2\beta - 2\tau)\Phi_{20} \\
	& + 2(\epsilon+ \rho - \bar\rho)\Phi_{21} + \bar\kappa \Phi_{22} . 
\end{split}
\eeqa

\section{Perturbation equations of massless spin particles}
\label{app:pert_eom_spin-s}

\subsection{Perturbation field equations in NP formalism}

We derive and summarize the field perturbation equations of all spin-weighted massless fields in Petrov type-D spacetime in the NP formalism. By consolidating the Rarita-Schwinger fields, this can be viewed as a generalization of those in spherical, symmetric spacetime~\cite{Arbey:2021jif}.

\subsubsection{Neutrino equations}
The Dirac equations in the NP formalism become~\cite{Chandrasekhar:1976ap}
\beqa
\begin{split}
&  (\delta'-\alpha + \pi + i e A\cdot \bar{m})\chi_0  - (D - \rho + \epsilon + i e A\cdot l) \chi_1 \\
& = i \mu_e \eta_0, \\
& (D'-\gamma + \mu + i e A\cdot n) \chi_0  - (\delta - \tau + \beta + i e A \cdot m )\chi_1 \\
& =  i \mu_e \eta_1, \\
& + (\delta - \bar\alpha + \bar\pi + i e A \cdot m) \eta_0 - (D - \bar\rho + \bar\epsilon + ie A \cdot l)\eta_1 \\
& = - i \mu_e \chi_0, \\
& (D' - \bar\gamma + \bar\mu + i e A \cdot n)\eta_0 - (\delta' - \bar\tau + \bar\beta + i e A \cdot m ) \eta_1 \\
& = -i \mu_e \chi_1, \\ 
\end{split}
\eeqa
where we have denoted two components of the spinor $\chi_A=(\chi_0,\chi_1)=\epsilon_{AB}P^B$ and $\eta_{A'}=(\eta_0,\eta_1)=\epsilon_{A'B'}Q^{B'}$, $\mu_e=m_e/\sqrt{2}$ are Dirac fermion mass. For the neutrino case, $e=0$ and $\mu_e=0$, the Dirac equations are decoupled as
\beqa
\begin{split}
& -(D - \rho + \epsilon) \chi_1 + (\delta'-\alpha + \pi)\chi_0 = 0, \\
& (D'-\gamma + \mu) \chi_0 - (\delta - \tau + \beta)\chi_1 = 0, \\
& -(D - \bar\rho + \bar\epsilon)\eta_1 + (\delta - \bar\alpha + \bar\pi) \eta_0 = 0, \\
& (D' - \bar\gamma + \bar\mu )\eta_0 - (\delta' - \bar\tau + \bar\beta ) \eta_1 = 0. \\ 
\end{split}
\eeqa
Since $D=\bar{D}$ and $D'=\bar{D}'$, the third and fourth equations are just the complex conjugates of the first and second equations. For the neutral case, by identifying $\bar\eta=\chi$, we obtain the massless Dirac spin-$1/2$ field equation in terms of differential operators and the spin coefficients as
\beqa
\begin{split}
& (D - \rho + \epsilon) \chi_1 - (\delta'-\alpha + \pi)\chi_0 = 0, \\
& (D'-\gamma + \mu) \chi_0 - (\delta - \tau + \beta)\chi_1 = 0. \\
\end{split} \label{eom_neutrino}
\eeqa
They are the components of the Weyl equation, i.e., the field equations for $s=1/2$ fields $\chi_A$.
By imposing the operators
\beqa
  [\delta - \bar\alpha - \tau + \bar\pi], \quad [ D + \bar\epsilon - \rho - \bar\rho ] ,
\eeqa
respectively, upon the first and second equations in Eq.~(\ref{eom_neutrino}), we have
\beqa
\begin{split}
& [\delta - \bar\alpha - \tau + \bar\pi](D + \epsilon -\rho)\chi_1 \\
& = [\delta - \bar\alpha - \tau + \bar\pi](\delta'-\alpha+\pi)\chi_0  , \\
& [ D + \bar\epsilon - \rho - \bar\rho ](\delta  + \beta - \tau ) \chi_1 \\
& = [ D + \bar\epsilon - \rho - \bar\rho ](D'-\gamma+\mu)\chi_0  . \\
\end{split}
\eeqa
By using the first identity in Eq.~(\ref{ID_delta-D}) with $p=-1$ and $q=-1$, we have
\beqa
\begin{split}
& [D + \bar\epsilon - \rho -\bar\rho](\delta + \beta -\tau) \\
& - [\delta - \bar\alpha + \bar\pi -\tau](D + \epsilon - \rho) = 0. \\
\end{split}
\eeqa
By subtracting the two Weyl equations on both sides, we get the Dirac equation for $\chi_0$ as
\beqa
\begin{split}
& [ (\delta - \bar\alpha - \tau + \bar\pi)(\delta'+\pi-\alpha) \\
& - ( D + \bar\epsilon - \rho - \bar\rho )(D'+\mu-\gamma) ] \chi_0  =  0 .  \\
\end{split} \label{eom_chi_0}
\eeqa
When $\epsilon=\tau=\pi=0$ and $\bar\rho=\rho$, the decoupled ordinary differential equation recovers that in spherical, symmetric spacetime~\cite{Arbey:2021jif}.
Alternatively, by imposing the operators
\beqa
[D' -\bar\gamma + \mu + \bar\mu ] , \quad [\delta'+\bar\beta-\bar\tau+\pi],
\eeqa
respectively, upon the first and second equations in Eq.~(\ref{eom_neutrino}), we have
\beqa
\begin{split}
& [D' -\bar\gamma + \mu + \bar\mu ](D + \epsilon -\rho)\chi_1 \\
& = [D' -\bar\gamma + \mu + \bar\mu ](\delta'-\alpha+\pi)\chi_0  , \\
& [\delta'+\bar\beta-\bar\tau+\pi](\delta  + \beta - \tau ) \chi_1  \\
& = [\delta'+\bar\beta-\bar\tau+\pi](D'-\gamma+\mu)\chi_0  . \\
\end{split}
\eeqa
By using the second identity in Eq.~(\ref{ID_delta-D}) with $p=-1$ and $q=-1$
\beqa
\begin{split}
& (D' -\bar\gamma + \mu + \bar\mu )(\delta'- \alpha + \pi) \\
& - (\delta'+\bar\beta-\bar\tau+\pi)(D'-\gamma+\mu) = 0 . \\
\end{split}
\eeqa
By subtracting the two Weyl equations on both sides, we have
\beqa
\begin{split}
& [ (D' -\bar\gamma + \mu + \bar\mu )(D + \epsilon -\rho) \\
& - ( \delta'+\bar\beta-\bar\tau+\pi )(\delta  + \beta - \tau ) ] \chi_1 = 0 .
\end{split}
\label{eom_chi_1}
\eeqa

\subsubsection{Maxwell equations}

The components of the Maxwell equation, i.e., the field equations for $s=1$ fields $(\phi_0,\phi_1,\phi_2)$ in terms of differential operator and spin coefficients in NP formalism are
\beqa
\begin{split}
& (D - 2\rho) \phi_1  = (\delta' + \pi - 2\alpha )\phi_0 - \kappa \phi_2 + 2\pi J_l , \\
& (D' + \mu - 2\gamma ) \phi_0  = ( \delta - 2\tau ) \phi_1 + \sigma \phi_2 - 2\pi J_m , \\
& (D - \rho + 2\epsilon) \phi_2  = - \lambda \phi_0 + ( \delta' + 2 \pi ) \phi_1  + 2\pi J_{\bar{m}} , \\
& (D' +2\mu ) \phi_1  = \nu \phi_0 + (\delta + 2\beta-\tau)\phi_2 - 2\pi J_n ,
\end{split}
\eeqa
where the differential operators can also be expressed as $\delta'=\bar\delta$ and $D'=\Delta$. The current sources are $J_l = J_\mu l^\mu$, $J_m=J^\mu m_\mu$, etc., with $J_\mu$ the $4$-current density. We have used the notation in Eq.~(\ref{NPS_prime}).
By considering Eq.~(\ref{SCs_KTN}), when $\kappa=\sigma = 0 =\nu = \lambda$, the first and third equations are coupled, and become first-order differential equations involving $\phi_0$ and $\phi_1$ only. The second and fourth equations become one involving $\phi_1$ and $\phi_2$ only. In this case, Maxwell's equation becomes~\cite{Teukolsky:1973ha}
\beqa
\begin{split}
& (D - 2\rho) \phi_1 = (\delta' - 2\alpha + \pi )\phi_0  + 2\pi J_l, \\
& (D'+\mu-2\gamma)\phi_0 = ( \delta - 2\tau ) \phi_1 - 2\pi J_m , \\
& (D -\rho + 2\epsilon ) \phi_2 =  (\delta' + 2\pi ) \phi_1 + 2\pi J_{\bar{m}} , \\
& (D' + 2 \mu)\phi_1  =  (\delta + 2\beta-\tau)\phi_2 - 2\pi J_n .
\end{split}
\eeqa
%
By imposing two operators
\beqa
[\delta-\beta-\bar\alpha-2\tau + \bar\pi], \quad [D-\epsilon+\bar\epsilon-2\rho-\bar\rho],
\eeqa
upon the first and second equations, respectively,
\beqa
\begin{split}
& [\delta-\beta-\bar\alpha-2\tau + \bar\pi] (D - 2\rho) \phi_1 \\
& = [\delta-\beta-\bar\alpha-2\tau + \bar\pi](\delta' - 2\alpha + \pi )\phi_0  \\
& + 2\pi [\delta-\beta-\bar\alpha-2\tau + \bar\pi] J_l ,\\
& [D-\epsilon+\bar\epsilon-2\rho-\bar\rho](D'+\mu-2\gamma)\phi_0 \\
& = [D-\epsilon+\bar\epsilon-2\rho-\bar\rho]( \delta - 2\tau ) \phi_1 \\
& - 2\pi [D-\epsilon+\bar\epsilon-2\rho-\bar\rho] J_m ,
\end{split}
\eeqa
and by using the first equation in Eq.~(\ref{ID_delta-D}) with $p=0$ and $q=-2$, i.e.,
\beqa
\begin{split}
& [D-\epsilon + \bar\epsilon - 2 \rho -\bar\rho](\delta - 2 \tau) \\
& - [\delta - \beta - \bar\alpha + \bar\pi - 2 \tau](D  - 2 \rho) = 0 ,
\end{split} 
\eeqa
and combining the two equations together, the terms in $\phi_1$ disappear, leaving a decoupled equation for $\phi_0$ as
\beqa
\begin{split}
& [ (D-\epsilon+\bar\epsilon-2\rho-\bar\rho)(D'+\mu-2\gamma) \\
& - (\delta-\beta-\bar\alpha-2\tau + \bar\pi)(\delta' - 2\alpha + \pi ) ] \phi_0  
= 2\pi J_0  ,  \\
\end{split} \label{eom_phi_0}
\eeqa
where the external source is
\beqa
\begin{split}
J_0 & \equiv  (\delta-\beta-\bar\alpha-2\tau + \bar\pi) J_l \\
& - (D-\epsilon+\bar\epsilon-2\rho-\bar\rho) J_m  .  \\
\end{split}
\eeqa
Similarly, by imposing two operators:
\beqa
[D'+\gamma-\bar\gamma+2\mu+\bar\mu], \quad [\delta'+\alpha+\bar\beta + 2\pi - \bar\tau]
\eeqa
upon the third and fourth equations, respectively, we obtain
\beqa
\begin{split}
	& [D'+\gamma-\bar\gamma+2\mu+\bar\mu](D -\rho + 2\epsilon ) \phi_2 \\
	& =  [D'+\gamma-\bar\gamma+2\mu+\bar\mu](\delta' + 2\pi ) \phi_1 \\
	& + 2\pi [D'+\gamma-\bar\gamma+2\mu+\bar\mu] J_{\bar{m}} , \\
& [\delta'+\alpha+\bar\beta + 2\pi - \bar\tau](D' + 2 \mu)\phi_1  \\
& =  [\delta'+\alpha+\bar\beta + 2\pi - \bar\tau](\delta + 2\beta-\tau)\phi_2 \\
& - 2\pi [\delta'+\alpha+\bar\beta + 2\pi - \bar\tau] J_n ,
\end{split}
\eeqa
and by using the second identity in Eq.~(\ref{ID_delta-D}) with $p=0$ and $q=-2$, i.e.,
\beqa
\begin{split}
& [D' + \gamma - \bar\gamma +2 \mu + \bar\mu](\delta'+ 2 \pi) \\
& - [\delta' + \alpha + \bar\beta-\bar\tau + 2 \pi](D'+2\mu) = 0 . \\
\end{split} 
\eeqa
Combing the two equations together, the terms in $\phi_1$ disappear, leaving a decoupled equation for $\phi_2$ as
\beqa
\begin{split}
& [ (D'+\gamma-\bar\gamma+2\mu+\bar\mu)(D -\rho + 2\epsilon ) \\
& -  (\delta'+\alpha+\bar\beta + 2\pi - \bar\tau)(\delta + 2\beta-\tau) ] \phi_2
= 2\pi J_2 . \\
\end{split} \label{eom_phi_2}
\eeqa
where the source is
\beqa
\begin{split}
J_2 & \equiv (D'+\gamma-\bar\gamma+2\mu+\bar\mu) J_{\bar{m}} \\
& - (\delta'+\alpha+\bar\beta + 2\pi - \bar\tau) J_n . \\
\end{split}
\eeqa
Alternatively, by imposing two operators:
\beqa
[D' - \gamma-\bar\gamma + \mu + \bar\mu ], \quad [\delta'-\alpha + \bar\beta-\bar\tau + \pi], 
\eeqa
upon the first and third equations, we have
\beqa
\begin{split}
& [D' - \gamma-\bar\gamma + \mu + \bar\mu ](D - 2\rho) \phi_1 \\
& = [D' - \gamma-\bar\gamma + \mu + \bar\mu ](\delta' - 2\alpha + \pi )\phi_0  \\
& + 2\pi [D' - \gamma-\bar\gamma + \mu + \bar\mu ] J_l, \\
& [\delta'-\alpha + \bar\beta-\bar\tau + \pi](D'+\mu-2\gamma)\phi_0 \\
& = [\delta'-\alpha + \bar\beta-\bar\tau + \pi]( \delta - 2\tau ) \phi_1 \\
& - 2\pi [\delta'-\alpha + \bar\beta-\bar\tau + \pi] J_m , \\
\end{split}
\eeqa
by using the second equation in Eq.~(\ref{ID_delta-D}) with $p=-2$ and $q=-1$.
\beqa
\begin{split}
& [D'-\gamma-\bar\gamma + \mu + \bar\mu ](\delta'-2\alpha+\pi) \\
& - [\delta'-\alpha+\bar\beta-\bar\tau+\pi](D'-2\gamma+\mu) = 0 , \\
\end{split}
\eeqa
Combing the two equations, eliminating the terms in $\phi_0$, and we obtain a decoupled equation for $\phi_1$ as 
\beqa
\begin{split}
& [ (D' - \gamma-\bar\gamma + \mu + \bar\mu )(D - 2\rho)  \\
& - (\delta'-\alpha + \bar\beta-\bar\tau + \pi)( \delta - 2\tau )  ] \phi_1 \\
& =  2\pi [ ( D' - \gamma-\bar\gamma + \mu + \bar\mu ) J_l \\
& - ( \delta'-\alpha + \bar\beta-\bar\tau + \pi) J_m  ] . \\
\end{split}
\eeqa
We can also impose two operators:
\beqa
[\delta + \beta + \bar\beta' - \bar\tau' - \tau], \quad [D+\epsilon + \bar\epsilon - \rho -\bar\rho], 
\eeqa
upon the third and fourth equations, we have
\beqa
\begin{split}
& [\delta + \beta + \bar\beta' - \bar\tau' - \tau](D -\rho + 2\epsilon ) \phi_2 \\
& =  [\delta + \beta + \bar\beta' - \bar\tau' - \tau](\delta' + 2\pi ) \phi_1 \\
& + 2\pi [\delta + \beta + \bar\beta' - \bar\tau' - \tau] J_{\bar{m}} , \\
& [D+\epsilon + \bar\epsilon - \rho -\bar\rho](D' + 2 \mu)\phi_1  \\
& =  [D+\epsilon + \bar\epsilon - \rho -\bar\rho](\delta + 2\beta-\tau)\phi_2 \\
& - 2\pi [D+\epsilon + \bar\epsilon - \rho -\bar\rho] J_n .
\end{split}
\eeqa
By using the first identity in Eq.~(\ref{ID_delta-D}) with $p=-2$ and $q=-1$ as
\beqa
\begin{split}
& [D+\epsilon + \bar\epsilon - \rho -\bar\rho](\delta + 2 \beta - \tau) \\
& - [\delta + \beta + \bar\beta' - \bar\tau' - \tau](D + 2 \epsilon - \rho) = 0 , \\
\end{split}
\eeqa
by combining the two equations and eliminating the terms in $\phi_0$, we obtain a decoupled equation for $\phi_1$ as
\beqa
\begin{split}
	& [ (D+\epsilon + \bar\epsilon - \rho -\bar\rho)(D' + 2 \mu) \\
	&  - (\delta + \beta + \bar\beta' - \bar\tau' - \tau)(\delta' + 2\pi ) ] \phi_1 
	= 2\pi J_1 , 
\end{split} \label{eom_phi_1}
\eeqa
where the source is
\beqa
\begin{split}
J_1 & \equiv ( \delta + \beta + \bar\beta' - \bar\tau' - \tau ) J_{\bar{m}} \\
& - ( D+\epsilon + \bar\epsilon - \rho -\bar\rho ) J_n . \\
\end{split}
\eeqa
For a spherically symmetric static metric with NP quantities in Eq.~(\ref{NPS_SSS}), the Maxwell equations reduce to~\cite{Arbey:2021jif} 
\beqa
\begin{split}
& (D - 2 \rho)\phi_1  = (\delta' + 2\beta' )\phi_0  , \\
& (D'+2\epsilon'-\rho')\phi_0  = \delta \phi_1  . \\
\end{split}
\eeqa
The coupled first-order differential equations can be reformulated into a pair of decoupled second-order differential equations. By applying $\delta$ (since $\tau=0=\pi$, $\beta=-\alpha$) to the first equation and $D-3\rho$ (since $\bar\rho=\rho$ and $\epsilon=\bar\epsilon=0$) to the third one, adding together two equations, we obtain
\beqa
\begin{split}
& [\delta (\delta' - 2\alpha + \pi ) - (D-3\rho)(D'-2\gamma+\mu)]\phi_0 \\
& = [ \delta (D - 2 \rho) - (D-3\rho) (\delta - 2\tau ) ]\phi_1 . \\
\end{split}
\eeqa
In this case, the field equation for $\phi_1$ in Eq.~(\ref{eom_phi_0}) becomes
\beqa
\begin{split}
& [ (D -3\rho)(D'+\mu-2\gamma) - \delta(\delta' - 2\alpha ) ] \phi_0  \\
& =  2\pi [ \delta J_l - (D - 3\rho  ) J_m ]  .  \\
\end{split} \label{eom_phi_0_SS}
\eeqa
Similarly, by applying $D'+3\mu$ (since $\bar\mu=\mu$) and $\delta'$ (since $\beta=-\alpha$ and $\pi=0=\tau$) to the second and fourth equations and adding together two equations, we obtain the differentiation equation for $\phi_2$.

\subsubsection{Rarita-Schwinger equation}

The spin-$3/2$ particle satisfy the Rarita-Schwinger (RS) equation in the Newman-Penrose form as~\cite{Arbey:2021jif}
\beqa
\begin{split}
	(D-\epsilon-3\rho)H_{001} - (\delta'-3\alpha+\pi)H_{000} - \Psi_2 \psi_{000} =0, \\
	(\delta-\beta-3\tau)H_{001} - (D'-3\gamma+\mu)H_{000} - \Psi_2 \psi_{001} = 0 ,  \\
\end{split} ~
\eeqa
where $\Psi_2$ is the non-vanishing background gravitational component. $H_{000}$ is a combination of spinor components as
\beqa
\begin{split}
H_{000} & = (\delta-2\beta-\bar\alpha+\bar\pi)\psi_{000} - (D-2\epsilon+\bar\epsilon-\bar\rho)\psi_{001}. \\
\end{split} ~ \label{H000}
\eeqa
By using the identity in Eq.~(\ref{ID_delta-D}) with $p=1$ and $q=-3$,
\beqa
\begin{split}
& [D-2\epsilon + \bar\epsilon - 3 \rho -\bar\rho](\delta - \beta - 3 \tau) \\
& - [\delta - 2\beta - \bar\alpha + \bar\pi - 3 \tau](D-\epsilon - 3 \rho) = 0 . \\
\end{split}
\eeqa
By imposing the following operators upon the first and second RS equations above, respectively,
\beqa
[\delta - 2\beta - \bar\alpha + \bar\pi - 3 \tau], \quad [D-2\epsilon + \bar\epsilon - 3 \rho -\bar\rho], 
\eeqa
we obtain
\beqa
\begin{split}
	& [\delta - 2\beta - \bar\alpha + \bar\pi - 3 \tau](D-\epsilon-3\rho)H_{001} \\
	& - [\delta - 2\beta - \bar\alpha + \bar\pi - 3 \tau](\delta'-3\alpha+\pi)H_{000} \\
	& - [\delta - 2\beta - \bar\alpha + \bar\pi - 3 \tau]\Psi_2 \psi_{000} =0, \\
	& [D-2\epsilon + \bar\epsilon - 3 \rho -\bar\rho](\delta-\beta-3\tau)H_{001} \\
	& - [D-2\epsilon + \bar\epsilon - 3 \rho -\bar\rho](D'-3\gamma+\mu)H_{000} \\
	& - [D-2\epsilon + \bar\epsilon - 3 \rho -\bar\rho]\Psi_2 \psi_{001} = 0 . 
\end{split}
\eeqa
By subtracting the two equations, we obtain
\beqa
\begin{split}
& [ (D-2\epsilon + \bar\epsilon - 3 \rho -\bar\rho)(D'-3\gamma+\mu) \\
& - (\delta - 2\beta - \bar\alpha + \bar\pi - 3 \tau)(\delta'-3\alpha+\pi) ] H_{000} \\
& =  [\delta - 2\beta - \bar\alpha + \bar\pi - 3 \tau] (\Psi_2 \psi_{000}) \\
& - [D-2\epsilon + \bar\epsilon - 3 \rho -\bar\rho] (\Psi_2 \psi_{001})  \\
& =  \Psi_2 (\delta  - 2\beta - \bar\alpha + \bar\pi - 3 \tau) \psi_{000}     + \psi_{001} \delta \Psi_2  \\
& - \Psi_2 ( D  - 2\epsilon + \bar\epsilon - 3 \rho -\bar\rho) \psi_{001} - \psi_{001} D \Psi_2
 = \Psi_2 H_{000} . 
\end{split} \label{eom_H_0}
\eeqa
where we have used the Bianchi identities in Eq.~(\ref{BI_Psi_2}). $H_{000}$ is already defined in Eq.~(\ref{H000}).

\subsubsection{Newman-Penrose equations}

The field equations of gravitation in the NP formalism are nothing but the first and third NP Bianchi identities in Eq.~(\ref{BIs}) as~\cite{Teukolsky:1973ha}
\beqa
\begin{split}
& (\bar\delta - 4 \alpha + \pi )  \Psi_0 - (D - 2\epsilon-4\rho) \Psi_1 - 3  \kappa \Psi_2 \\
& = (\delta+\bar\pi-2\bar\alpha-2\beta)\Phi_{00} - (D-2\epsilon-2\bar\rho)\Phi_{01} + 2  \sigma \Phi_{10} \\
& - 2  \kappa \Phi_{11} -  {\bar\kappa} \Phi_{02}, \\
& (D'-4\gamma+\mu) \Psi_0 - (\delta - 4 \tau - 2\beta) \Psi_1 - 3  \sigma \Psi_2 \\
& = (\delta + 2 \bar\pi - 2 \beta) \Phi_{01} - (D-2\epsilon+2\bar\epsilon-\bar\rho)\Phi_{02} - \bar\lambda \Phi_{00} \\
& + 2  \sigma \Phi_{11} - 2  \kappa \Phi_{12} ,  \\
& (D-\rho-\bar\rho-3\epsilon	+ \bar\epsilon)  \sigma - (\delta-\tau + \bar\pi-\bar\alpha - 3 \beta)  \kappa =  \Psi_0 . \\
\end{split} \label{eoms_Psi_1st}
\eeqa
where $\Phi_{\mu\nu}$ are Ricci tensors defined in Eq.~(\ref{Rij}). At leading order, $\Psi_{0,1,3,4}=0$ and $\Phi_{\mu\nu} = 0$ in vacuum spacetime.

In linear homogeneous equations up to first-order perturbation, there are six perturbation equations, of which four come from the eight NP Bianchi identities and two come from the $18$ NP Ricci identities. The first three are:
\beqa
\begin{split}
& (\bar\delta - 4 \alpha + \pi ) \hat\Psi_0 - (D - 2\epsilon-4\rho)\hat\Psi_1 =  3 \hat\kappa \Psi_2  \\
& + \frac{\kappa_N}{2} [ (\delta+\bar\pi-2\bar\alpha-2\beta) \hat{T}_{ll} - (D-2\epsilon-2\bar\rho) \hat{T}_{lm}  ] , \\
& (D'-4\gamma+\mu)\hat\Psi_0 - (\delta - 4 \tau - 2\beta)\hat\Psi_1 = 3 \hat\sigma \Psi_2 \\
& + \frac{\kappa_N}{2} [ (\delta + 2 \bar\pi - 2 \beta) \hat{T}_{lm} - (D-2\epsilon+2\bar\epsilon-\bar\rho) \hat{T}_{mm} ]  ,  \\
& (D-\rho-\bar\rho-3\epsilon	+ \bar\epsilon) \hat\sigma - (\delta-\tau + \bar\pi-\bar\alpha - 3 \beta) \hat\kappa = \hat\Psi_0, \\
\end{split} \label{eoms_Psi_1st}
\eeqa
By using the exchange symmetry and $\hat\Psi_4 \leftrightarrow \hat\Psi_0 $, and $\hat\Psi_1 \leftrightarrow \hat\Psi_3 $, we have the last three equations as
\beqa
\begin{split}
& (\delta + 4 \beta - \tau ) \hat\Psi_4 - (D' + 2 \gamma + 4\mu)\hat\Psi_3 =  - 3 \hat\nu \Psi_2  \\
& + \frac{\kappa_N}{2} [ (\delta' - \bar\tau +2\bar\beta+2\alpha) \hat{T}_{nn} - (D'+2\gamma+2\bar\mu) \hat{T}_{n\bar{m}}  ] , \\
& (D+4\epsilon-\rho)\hat\Psi_4 - (\delta' + 4 \pi + 2\alpha)\hat\Psi_3 = - 3 \hat\lambda \Psi_2 \\
& + \frac{\kappa_N}{2} [ (\delta' - 2 \bar\tau + 2 \alpha) \hat{T}_{n\bar{m}} - (D'+2\gamma-2\bar\gamma'+\bar\mu) \hat{T}_{\bar{m}\bar{m}} ]  ,  \\
& - (D'+\mu+\bar\mu+3\gamma - \bar\gamma) \hat\lambda \\
& + (\delta'+\pi - \bar\tau + \bar\beta + 3 \alpha) \hat\nu = \hat\Psi_4. \\
\end{split} \label{eoms_Psi_2nd}
\eeqa
These are gravitational wave perturbation equations with a source.
The first three equations couple $\hat\Psi_0$ and $\hat\Psi_1$; the last three equations couple $\hat\Psi_3$ and $\hat\Psi_4$, where $\hat\Psi_{0,1,3,4}$ are the perturbed components of the Weyl tensor. $\Psi_2$ is the only non-vanishing background component. It satisfies
\beqa
D\Psi_2 = 3\rho \Psi_2, \quad \delta \Psi_2 = 3\tau \Psi_2 . \label{BI_Psi_2}
\eeqa
According to the exchange symmetry in Eq.~(\ref{sym_exc}), we have
\beqa
D' \Psi_2 =  - 3 \mu \Psi_2, \quad \delta \Psi_2 =  - 3 \pi \Psi_2 .
\eeqa
Thus, the third and sixth equations in Eqs.~(\ref{eoms_Psi_1st}) and (\ref{eoms_Psi_2nd}), respectively, can be rewritten as
\beqa
\begin{split}
& (D-4\rho-\bar\rho-3\epsilon	+ \bar\epsilon) \Psi_2 \hat\sigma \\
& - (\delta-4\tau + \bar\pi-\bar\alpha - 3 \beta) \Psi_2 \hat\kappa  = \hat\Psi_0 \Psi_2 , \\
& (D'+4\mu+\bar\mu+3\gamma - \bar\gamma) \Psi_2 \hat\lambda \\
& - (\delta'+4\pi - \bar\tau + \bar\beta + 3 \alpha) \Psi_2 \hat\nu = -  \hat\Psi_4  \Psi_2 .
\end{split}
 \label{eom_BI}
\eeqa
Note that the differential operators act upon $\Psi_2$ first.
Quantities with hat refer to perturbed NP scalars, e.g., $(\hat\kappa,\hat\sigma,\hat\lambda,\hat\nu)$ are the perturbed values of $(\kappa, \sigma,\lambda,\nu)$, and all other quantities are unperturbed.
By observing the first two equations in Eq.~(\ref{eoms_Psi_1st}), we may choose $p=2,q=-4$ so that the first identity in Eq.~(\ref{ID_delta-D}) becomes
\beqa
\begin{split}
& [D- 3 \epsilon + \bar\epsilon - 4 \rho -\bar\rho](\delta - 2 \beta - 4 \tau) \\
& - [\delta - 3 \beta - \bar\alpha + \bar\pi - 4 \tau](D - 2\epsilon - 4 \rho) = 0 . \\
\end{split}
\eeqa
By imposing the differential operators
\beqa
[\delta - 3 \beta - \bar\alpha + \bar\pi - 4 \tau], \quad [D- 3 \epsilon + \bar\epsilon - 4 \rho -\bar\rho] ,
\eeqa
respectively, upon the first and second equations in Eq.~(\ref{eoms_Psi_1st}), they become
\beqa
\begin{split}
& [\delta - 3 \beta - \bar\alpha + \bar\pi - 4 \tau](\bar\delta - 4 \alpha + \pi ) \hat\Psi_0 \\
& - [\delta - 3 \beta - \bar\alpha + \bar\pi - 4 \tau](D - 2\epsilon-4\rho)\hat\Psi_1 \\
& =  3 [\delta - 3 \beta - \bar\alpha + \bar\pi - 4 \tau]\hat\kappa \Psi_2  + \frac{\kappa_N}{2} [\delta - 3 \beta - \bar\alpha + \bar\pi - 4 \tau] \\
& \times [ (\delta+\bar\pi-2\bar\alpha-2\beta) \hat{T}_{ll} - (D-2\epsilon-2\bar\rho) \hat{T}_{lm}  ] , \\
& [D- 3 \epsilon + \bar\epsilon - 4 \rho -\bar\rho] (D'-4\gamma+\mu)\hat\Psi_0 \\
& - [D- 3 \epsilon + \bar\epsilon - 4 \rho -\bar\rho](\delta - 4 \tau - 2\beta)\hat\Psi_1 \\
& = 3 [D- 3 \epsilon + \bar\epsilon - 4 \rho -\bar\rho] \hat\sigma \Psi_2 + \frac{\kappa_N}{2} [D- 3 \epsilon + \bar\epsilon - 4 \rho -\bar\rho] \\
& \times [ (\delta + 2 \bar\pi - 2 \beta) \hat{T}_{lm} - (D-2\epsilon+2\bar\epsilon-\bar\rho) \hat{T}_{mm} ]  .  \\
\end{split}
\eeqa
By subtraction both sides, the $\Psi_1$ is decoupled, and the equation becomes
\beqa
\begin{split}
& [ ( \delta - 3 \beta - \bar\alpha + \bar\pi - 4 \tau)(\bar\delta - 4 \alpha + \pi ) \\
&  - (D- 3 \epsilon + \bar\epsilon - 4 \rho -\bar\rho)(D'-4\gamma+\mu) ]\hat\Psi_0  , \\
& = 3 [( \delta - 3 \beta - \bar\alpha + \bar\pi - 4 \tau) \hat\kappa  \\
& -  3 (D- 3 \epsilon + \bar\epsilon - 4 \rho -\bar\rho) \hat\sigma ] \Psi_2  \\
& + \frac{\kappa_N}{2} [\delta - 3 \beta - \bar\alpha + \bar\pi - 4 \tau] \\
& \times [ (\delta+\bar\pi-2\bar\alpha-2\beta) \hat{T}_{ll} - (D-2\epsilon-2\bar\rho) \hat{T}_{lm}  ] \\
& - \frac{\kappa_N}{2} [D- 3 \epsilon + \bar\epsilon - 4 \rho -\bar\rho] \\
& \times [ (\delta + 2 \bar\pi - 2 \beta) \hat{T}_{lm} - (D-2\epsilon+2\bar\epsilon-\bar\rho) \hat{T}_{mm} ] , \\
\end{split}
\eeqa
where $\tau'=-\pi$ and $\beta'=-\alpha$.
When combined with the Eq.~(\ref{eom_BI}), it becomes
\beqa
\begin{split}
& [ ( \delta - 3 \beta - \bar\alpha + \bar\pi - 4 \tau)(\bar\delta - 4 \alpha + \pi )  \\
& - (D- 3 \epsilon + \bar\epsilon - 4 \rho -\bar\rho)(D'-4\gamma+\mu) + 3  \Psi_2 ]\hat\Psi_0 \\
& = \frac{\kappa_N}{2} T_0 , \\
\end{split} \label{eom_Psi_0}
\eeqa
where the matter sources are
\beqa
\begin{split}
T_0 & \equiv   (\delta - 3 \beta - \bar\alpha + \bar\pi - 4 \tau) [ (\delta+\bar\pi-2\bar\alpha-2\beta) \hat{T}_{ll} \\
& - (D-2\epsilon-2\bar\rho) \hat{T}_{lm}  ] \\
& - ( D- 3 \epsilon + \bar\epsilon - 4 \rho -\bar\rho) [ (\delta + 2 \bar\pi - 2 \beta) \hat{T}_{lm} \\
& - (D-2\epsilon+2\bar\epsilon-\bar\rho) \hat{T}_{mm} ] . \\
\end{split}
\eeqa
For Petrov type-D spacetime, $\sigma=\lambda=\kappa=\nu=0$. The right-hand side is vanishing; the equation is a second-order differential equation for $\Psi_0$.
By observing the fourth and fifth equations in Eq.~(\ref{eoms_Psi_2nd}), we may choose $p=2,q=-4$, so that the first identity in Eq.~(\ref{ID_delta-D}) becomes
\beqa
\begin{split}
& [D'+3\gamma-\bar\gamma + 4 \mu + \bar\mu](\delta'+2\alpha+4\pi) \\
& -[\delta'+3\alpha + \bar\beta-\bar\tau+4\pi](D'+2\gamma+4\mu) = 0. \\
\end{split}
\eeqa
By imposing the differential operators
\beqa
 [\delta'+3\alpha + \bar\beta-\bar\tau+4\pi], \quad [D'+3\gamma-\bar\gamma + 4 \mu + \bar\mu ], 
\eeqa
respectively, upon the fourth and fifth equations in Eq.~(\ref{eoms_Psi_2nd}),
\beqa
\begin{split}
& [\delta'+3\alpha + \bar\beta-\bar\tau+4\pi](\delta + 4 \beta - \tau ) \hat\Psi_4 \\
& - [\delta'+3\alpha + \bar\beta-\bar\tau+4\pi](D' + 2 \gamma + 4\mu)\hat\Psi_3 \\
& =  - 3 \hat\nu [\delta'+3\alpha + \bar\beta-\bar\tau+4\pi] \Psi_2 \\
& + \frac{\kappa_N}{2} [\delta'+3\alpha + \bar\beta-\bar\tau+4\pi]\\
& \times [ (\delta' - \bar\tau +2\bar\beta+2\alpha) \hat{T}_{nn} - (D'+2\gamma+2\bar\mu) \hat{T}_{n\bar{m}}  ] , \\
& [D'+3\gamma-\bar\gamma + 4 \mu + \bar\mu ](D+4\epsilon-\rho)\hat\Psi_4 \\
& - [D'+3\gamma-\bar\gamma + 4 \mu + \bar\mu ](\delta' + 4 \pi + 2\alpha)\hat\Psi_3 \\
& = - 3 \hat\lambda [D'+3\gamma-\bar\gamma + 4 \mu + \bar\mu ] \Psi_2 \\
& + \frac{\kappa_N}{2} [D'+3\gamma-\bar\gamma + 4 \mu + \bar\mu ] \\
& \times [ (\delta' - 2 \bar\tau + 2 \alpha) \hat{T}_{n\bar{m}} - (D'+2\gamma-2\bar\gamma'+\bar\mu) \hat{T}_{\bar{m}\bar{m}} ]  ,  \\
\end{split} 
\eeqa
By subtraction both sides, the $\Psi_3$ is decoupled, and the equation becomes
\beqa
\begin{split}
& \big[ [\delta'+3\alpha + \bar\beta-\bar\tau+4\pi](\delta + 4\beta - \tau)  \\
& - [D'+3\gamma-\bar\gamma + 4 \mu + \bar\mu ](D+4\epsilon-\rho) \big] \hat\Psi_4\\
& = \big[ - 3 \hat\nu ( \delta'+3\alpha + \bar\beta-\bar\tau+4\pi ) \\
& + 3 \hat\lambda ( D'+3\gamma-\bar\gamma + 4 \mu + \bar\mu ) \big] \Psi_2 \\ 
& = \frac{\kappa_N}{2} \big[ [\delta'+3\alpha + \bar\beta-\bar\tau+4\pi]\\
& \times [ (\delta' - \bar\tau +2\bar\beta+2\alpha) \hat{T}_{nn} - (D'+2\gamma+2\bar\mu) \hat{T}_{n\bar{m}}  ]  \\
& - [D'+3\gamma-\bar\gamma + 4 \mu + \bar\mu ] \\
& \times [ (\delta' - 2 \bar\tau + 2 \alpha) \hat{T}_{n\bar{m}} - (D'+2\gamma-2\bar\gamma'+\bar\mu) \hat{T}_{\bar{m}\bar{m}} ]  \big] , \\
\end{split}
\eeqa
where we used the notation in the first row of Eq.~(\ref{NPS_prime}). Now, combing with the sixth equation in Eq.~(\ref{eom_BI}), the equation becomes
\beqa
\begin{split}
&  [ (\delta'+3\alpha + \bar\beta-\bar\tau+4\pi)(\delta + 4\beta - \tau) \\
& - (D'+3\gamma-\bar\gamma + 4 \mu + \bar\mu )(D+4\epsilon-\rho) + 3 \Psi_2 ] \hat\Psi_4 \\
& = \frac{\kappa_N}{2} T_4 , \\ 
\end{split} \label{eom_Psi_4}
\eeqa
where the perturbations of the matter source are
\beqa
\begin{split}
	T_4 & =   [\delta'+3\alpha + \bar\beta-\bar\tau+4\pi][ (\delta' - \bar\tau +2\bar\beta+2\alpha) \hat{T}_{nn} \\
	& - (D'+2\gamma+2\bar\mu) \hat{T}_{n\bar{m}}  ]  \\
& -  [D'+3\gamma-\bar\gamma + 4 \mu + \bar\mu ] [ (\delta' - 2 \bar\tau + 2 \alpha) \hat{T}_{n\bar{m}} \\
& - (D'+2\gamma-2\bar\gamma'+\bar\mu) \hat{T}_{\bar{m}\bar{m}} ] 
\end{split} 
\eeqa
For Petrov type-D spacetime, $\sigma=\lambda=\kappa=\nu=0$. The right-hand side is vanishing; the equation is a second-order differential equation for $\Psi_4$. Therefore, according to Eqs.~(\ref{eom_Psi_0}) and (\ref{eom_Psi_4}), we obtain two decoupled Teukolsky differential equations for $\Psi_{0}$ and $\Psi_{4}$, as
\beqa
\begin{split}
& [ (\delta - 3 \beta - \bar\alpha + \bar\pi - 4 \tau)(\bar\delta - 4 \alpha + \pi ) \\
& - (D- 3 \epsilon + \bar\epsilon - 4 \rho -\bar\rho)(D'-4\gamma+\mu) + 3  \Psi_2 ] \Psi_0 = 0 , \\
& [ (D'+3\gamma-\bar\gamma + 4 \mu + \bar\mu )(D+4\epsilon-\rho)  \\
& - (\bar\delta +3\alpha+\bar\beta-\bar\tau+4\pi)(\delta + 4\beta - \tau)  - 3  \Psi_2 ] \Psi_4 = 0 , \\
\end{split} \label{eom_s=2}
\eeqa
where $\delta'=\bar\delta$. They can be written as a single master equation that applies to the sourceless perturbation equations for scalar, two-compoent neutrino, electromagnetic fields, and gravitational fields.

\subsection{All spin massless field equations}

In summary, we have obtained all the perturbative equations for massless spin particles, including netrino, Maxwell, Rarita-Schwinger, and gravitational field. We can reorganize Eqs.~(\ref{eom_chi_0}), (\ref{eom_chi_1}), (\ref{eom_phi_0}), (\ref{eom_phi_1}), (\ref{eom_H_0}) and (\ref{eom_s=2}) into a unified generalized first-order perturbation equation for all massless spin-$s$ particles as below:
\begin{widetext}
\beqa
\begin{split}
& \big[ [ D-(2s-1)\epsilon+\bar\epsilon-2s\rho-\bar\rho ] (D'-2s\gamma+\mu)  - [\delta + \bar\pi - \bar\alpha - (2s-1)\beta - 2s \tau](\delta'+\pi -2s\alpha) \\
& - (2s-1)(s-1)\Psi_2 \big] \hat\Phi_s  = 4\pi T_s , \\
& \big[ [ D'+(2s-1)\gamma-\bar\gamma+2s\mu+\bar\mu ] (D+2s\epsilon-\rho)  - [\delta' - \bar\tau + \bar\beta + (2s-1)\alpha + 2s \pi](\delta-\tau + 2s \beta) \\
& - (2s-1)(s-1)\Psi_2 \big] \hat\Phi_{-s} = 4\pi T_{-s} ,  \\
\end{split} \label{s-Teukolsky master equation}
\eeqa
where the spin states $\Phi_s=(\phi, \chi_0, \varphi_0, \psi_0, \Psi_0)$ includes arbitrary spins: $s=0, 1/2, 1, 3/2, 2$. Alternatively, $\Phi_{-s}=(\phi, \chi_1, \varphi_2, \psi_3, \Psi_4)$ for spin-weights $-s = 0, -1/2, -1, -3/2, -2$, respectively. $T_{\pm s}$ denotes the corresponding source, which is vanishing for fields in vacuum spacetime.
It's worthy of emphasizing that the second set of equivalent master perturbations equations for $\Phi_{-s}\propto \Delta^s \Psi_s$ is different but contains the same information as the first set equation for $\Psi_s$. The second equation can be obtained due to the exchange symmetry $l\leftrightarrow n$ and $m\leftrightarrow \bar{m}$ in Eq.~(\ref{sym_exc}).

\subsection{Teukolsky master equations in axial symmetric spacetime for all spin}

As a check for all spin-$s$ field equations in Eq.~(\ref{s-Teukolsky master equation}), let's consider a generic stationary axis-symmetric metric in Eq.~(\ref{ds2_generic}). We can choose the tetrad in Eq.~(\ref{lnm_fgh}), and then the corresponding spin coefficients (or optical scalars) are
\beqa
\begin{split}
& \kappa = \nu = \sigma = \lambda = \tau = \pi = 0, \\
& \rho = - \frac{h'}{2h}\sqrt\frac{g}{f} +  \frac{1}{2h} \frac{i \partial_\theta w }{\sin\theta}, \quad \mu = -\frac{h'}{4h}\sqrt{fg} + \frac{f}{4h} \frac{i \partial_\theta w}{\sin\theta} , \\
& \alpha = - \beta = - \frac{\cot\theta}{2\sqrt{2h}}, \quad \gamma = \frac{f'}{4}\sqrt\frac{g}{f} + \frac{f}{8h} \frac{i\partial_\theta w}{\sin\theta}, \quad \epsilon = \frac{i\partial_\theta w}{4h \sin\theta}, \\
\end{split}
\eeqa
where the prime denotes the derivative along the radial direction.
The Teukolsky master equations are
\beqa
\begin{split}
& \bigg( - \frac{h}{f} + \frac{w^2}{\sin^2\theta} \bigg) \frac{\partial^2}{\partial t^2}  \Phi_s  + s \bigg[ \sqrt\frac{g}{f}\bigg( \frac{f'}{f}h - h' \bigg) + 2i\frac{ \partial_\theta w -  w \cot\theta}{\sin\theta} \bigg] \partial_t \Phi_s - 2 \frac{w}{\sin^2\theta} \frac{\partial^2}{\partial t \partial \phi } \Phi_s \\
& + gh \bigg[ \partial_r^2 \Phi_s + \bigg(  \frac{g'}{2g} + (s+1/2)\frac{f'}{f} + (s+1)\frac{h'}{h} - i s \sqrt\frac{f}{g} \frac{1}{h}\frac{\partial_\theta w}{\sin\theta} \bigg) \partial_r \Phi_s \bigg]\\
& +  \frac{1}{\sin\theta} \frac{\partial}{\partial \theta} \bigg( \sin\theta \frac{\partial}{\partial \theta} \Phi_s \bigg) + 2is\frac{\cot\theta}{\sin\theta} \partial_\phi \Phi_s + \frac{1}{\sin^2\theta}  \frac{\partial^2}{\partial \phi^2}   \Phi_s   + (s - s^2 \cot^2\theta) \Phi_s - \frac{2s+1}{12}gh \bigg[ \frac{4(s+1)}{gh} \\
&  - (5s+2)\frac{f'}{f}\frac{h'}{h} - (1+s) \bigg( \frac{f'}{f} \frac{g'}{g} + 2 \frac{f''}{f} - \frac{f'^2}{f^2}  \bigg) + (1-2s)\frac{h'^2}{h^2} + (s-2)\bigg( \frac{g'}{g}\frac{h'}{h} + 2 \frac{h''}{h} \bigg)  \bigg] \Phi_s \\
& + \bigg[ - \frac{i\csc\theta \partial_\theta w}{2} \sqrt\frac{g}{f} \bigg( s(s+1)f' + s^2 f \frac{h'}{h} \bigg) + \frac{s^2-1}{12\sin^2\theta}\frac{f}{h} (\partial_\theta w)^2 \bigg] \Phi_s = 0.
\end{split} \label{eom_Teukolsky-master_asym_static}
\eeqa
\end{widetext}
This is a generalization of the Teukolsky master equations for all massless spin particles in axially symmetric static spacetime.
When $w=0$, it recovers that in the spherical, symmetric spacetime~\cite{Arbey:2021jif}.

\end{CJK}

\bibliography{references}

\end{document}